\documentclass[10pt]{article}
\usepackage[ansinew]{inputenc}
\usepackage{fancyhdr}

\usepackage{amssymb}
\usepackage{tipa}
\usepackage[ansinew]{inputenc}
\usepackage{fancyhdr}
\usepackage[all]{xy}
\usepackage{sectsty} 
\usepackage[normalem]{ulem}
\allsectionsfont{\large}
\usepackage{amsmath}
\usepackage{mathenv}
\usepackage{dsfont}

\usepackage{graphicx}
\usepackage{color}
\usepackage[colorlinks=true, citecolor=blue, linkcolor=blue, urlcolor=blue]{hyperref}
\usepackage[active,new,noold,marker]{xrcs}
\usepackage[active]{srcltx} 

\title{\begin{flushright} \vspace{-70pt}\textit{\footnotesize Computer Science, Engineering \& \\ \vspace{-10pt}  General Sciences, Phase: Theor. Proj.} \\
\textit{}\\\vspace{-20pt}
\textit{\footnotesize Rep. on Comp. Phys.}$\,$\footnotesize \textbf{\textit{Vol.}}\textit{ }\textbf{1}, \emph{Ver.} 1, 1--51\\ \vspace{2pt} \footnotesize Report Posted and Published on 01 Oct 2007
\\ --------------------------------------------------------------- \end{flushright} \vspace{12pt} \Large  \textbf{The Theory} {\emph{of}} \textbf{Unified Relativity}\emph{ for } \\ \emph{a}\textbf { Biovielectroluminescence Phenomenon} {\emph{via} \textbf{Fly's Visual and Imaging System}}\\
} 

\begin{document}


\definecolor{MyPurple}{rgb}{0.4,0.08,0.45}

\definecolor{MyBrown}{rgb}{0.6,0.4,0}

\date{}
\maketitle
\begin{center}
\vspace{-50pt}
\long\def\symbolfootnote[#1]#2{\begingroup%
\def\thefootnote{\fnsymbol{footnote}}\footnote[#1]{#2}\endgroup}
\smallskip
By Philip Baback Alipour \(^{1}\,^{,}\, ^{2}\,^{,}\) \noindent\symbolfootnote[1]{Author for correspondence (\htmladdnormallink {\textcolor{blue}{philipbaback\_orbsix@msn.com}}{mailto:philipbaback_orbsix@msn.com}).} $^{,}$ \noindent\symbolfootnote[2]{This paper merely represents the theoretical phase of biovielectroluminescence phenomenon, where the practical phase demonstrating this phenomenon requests a healthy grant from some academic body, hence to research the theory as being submitted to ArXiv.org on a realistic level in the future.}
\end{center}

\begin{center}
\textit{1-Computer Science, Physics and Biology, Personalized Research Project,}

\textit{Elm Tree Farm, Wallingfen Lane, Newport, Brough, HU15 1RF, UK}

\smallskip
\textit{2- Computer Sciences `Departmental Character Reference Support',}

\textit{University of Lincoln, Brayford Pool, Lincoln, LN6 7TS, UK}

\end{center}

\noindent \textbf{Abstract ----- }\small {The elucidation upon fly's neuronal patterns as a link to computer graphics and human vision, is investigated for the phenomenon by propounding \emph{a unified theory of Einstein's two known relativities}. It is conclusive that flies could contribute a certain amount of neuromatrices indicating an imagery function based on a visual-computational system of \emph{n}-artificial ommatidia into \emph{computer graphics} and \emph{visual superimposition}. The visual system involves the time aspect, whereas flies possess faster pulses compared to humans' visual ability due to charge \emph{Q}'s \textbf{E}-field state on an active fly's eye surface. This behaviour can be tested on a dissected fly specimen at its \emph{ommatidia}. Electro-optical contacts and electrodes are wired through the flesh at the dorsal to the fly's anterior, forming organic emitter layer to stimulate light emission, thereby to a computer circuit adjacent to artificial compound eyes. The next step is applying a threshold voltage with secondary voltages to the circuit denoting an array of essential electrodes for bit switch. As a result, dormant pulses of biochemical-electrical in some parts of the circuit versus active pulses of biovielectro-luminescence relative to the specimen's environment are recorded. The outcome matrix, is deemed to possess a construction of red-green-blue (RGB) and time radicals expressing the space-time problem in consumption, allocating it into computational algorithms, enhancing the technology far beyond. The obtained formula is \(v_{r,g,b}\:t_{\mathrm{cons}}\) which generates a consumed distance of $x$ in form of function \(\mathrm{cons}(x)\), denoting circuital travel between data source/sink, for pixel data and \emph{bendable wavelengths}. Once `image logic' is in place, incorporating this point of graphical acceleration permits one to enhance graphics and optimize immensely central processing, data transmissions between memory and computer visual system. The phenomenon can be mainly used in holography, ${360^{\rm o}}$ viewing/display, 3D scanning techniques, military and medicine, a robust and cheap substitution for e.g., flash photography, pre-motion pattern analysis, real-time rendering and LCDs.}

\begin{center}
\vspace{6pt}
\smallskip
\end{center}

\section{Introduction and project objectives}
\label{section1}
\vspace{1pt}
\markboth{}{A UR Theory and Biovielectroluminescence}

\normalsize It would not be erroneously speaking about how best image sensors could improve by incorporating biochemical and organic material~\cite[vii]{06-Wiki}. In this case, it is aimed for a biovielectrical composition of some inactive tissue of an insect with the involvement of its compound eyes' neurons and thus, processed by a computer system on the organic scale where the dead body is preserved. The result would be a phenomenon called the biovielectro-luminescence phenomenon.

The term `biovielectro-luminescence' herein pronounced as, \textcolor{blue}{\textipa{b\=i$'$\=o-v\=i-$\breve{\mathrm{i}}$-l$\breve{\mathrm{e}}$k"tr\=o-l\=u"m@-n$\breve{\mathrm{e}}$s$'$@ns}}, consists of acronyms, \emph{bio-}, for \emph{biological}, \emph{vi-} or \emph{vie-}, for compound noun \emph{visual image/input}, whereas \emph{Biovie-}, represents \emph{visually utilized biological system} or simply, \emph{biological vision}.

By definition, this part of the phrase i.e. \emph{Biovie-}, represents applications-\{supporting from/engaging with/competeing for\}-the perception of the visually utilized body here, a `biological system' observing/recording events from its world. The next compound of the word is the noun `electroluminescence' which is well-defined to be expressed as: `an optical phenomenon and electrical phenomenon where a material emits light in response to an electric current passed through it, or to a strong electric field'~\cite{06-Wiki}. Prior to the phenomenon's performance, the experiment chiefly concentrates on utilizing its components of source with inexpensive organic and non-organic compounds.\\

The project objectives are:

\begin{enumerate}
\item [{1-}] Synchronizing a 2D-space-time planar view to 5D-vision in a state of a maximally 4D-human vision, now adapted to an \emph{n}-dimensional perception. ----------- Subject to all of the upcoming Sections, and Major to \S\ref{section2} and \S\ref{section3}.

\item [{2-}]Training a learning AI circuit as a perceptron in terms of a neuromatrix, practically in aim of emphasizing the actual best outcome and efficiency factor on the pulses' qualitative and quantitative radicals concerning the biovielectroluminescence phenomenon. ----------- Major to \S\ref{section4}.

\item [{3-}]Computing the unifying theory of Einstein's relativities, special and general (SR and GR), in terms of the time-dilation prospect and time-data synchronization on a Newtonian observer's clock, in relevance to `image logic'~\cite[e.g., the latter Ref.'s \S1]{01-Alipour,{02-Alipour}}. ----------- Major to \S\ref{section2} and \S\ref{section3}.

\item [{4-}]Pattern recognition upon motion-stretch analysis. Mobile and stationary objects in place to the scene's field relative to pre/post-motion capture and all sequences of incoming data, computed and enhanced in accordance with the collected image data. ----------- Major to \S\ref{section2}, \S\ref{section3} and \S\ref{section5}.

\item [{5-}] Chemical against physical pulses of electronics as: bioelectrical-chemical and biovielectrical pulses in investigation assigned to the experimenter when introduced to the project putting into prospect this question: \emph{which is the best product combination of pulses of either nature to be in action} ? ----------- Subject to \S\ref{section4} and \S\ref{section5}.

\item [{6-}] Demonstrate the traversed distances in terms of space-time in consumption as \emph{bendable wavelengths} in the spectra exhibited by biovielectroluminescence phenomenon, in combination with Objectives \#1 and \#3. ----------- Major to \S\ref{section2} and \S\ref{section3}.

\item [{7-}] By presenting an algorithm, make relative comparisons between the notion of pixel velocity using its displacement parameter $x_{\mathrm{cons}}$, and rendering systems in computer graphics improving e.g., pixel displacement mapping~\cite{52-Kautz} and image reconstruction techniques~\cite{51-Asma et al.}, in combination with Objectives \#4 and \#6. ----------- Subject to \S\ref{section2}, \S\ref{section3}, Major to \S\ref{section5} and future reports.

\item [{8-}] Conclude the project based on powerful deductions on Objectives \#3 and \#6, \emph{the theory of unified relativity} (UR), as a provable theory and empirically correct in \emph{n}-dimensional physical systems and application. ----------- Major to \S\ref{section5}, \S\ref{section6}, Appendix I and future empirical observations under laboratorial conditions.

\end{enumerate}

\noindent General notations, expressions and their abbreviations relevant to the above-mentioned objectives of this project, are submitted in Appendix III. \\

Using methods of reverse engineering in artificial intelligence (AI), and having the fact that Lee's team managed to fabricate biologically inspired artificial compound eyes~\cite{03-Lee}, an experimental setup could take place demonstrating the visual ability of a fly, data-point anticipation of future events before occurrence via rapid pulsing of light, e.g., biovielectroluminescence. Furthermore, it is understood from the above-mentioned composition, a phenomenon to be emerged which is in contrast with an optical and electrical phenomenon in the class of electroluminescence (EL), e.g., semiconductor electroluminescent devices such as light emitting diodes (LEDs). That is, the phenomenon to be in response to an electric current passing through organic and non-organic materials under the affect of \textbf{E}-field of charge \emph{Q} in the presence of time varying \textbf{B}-field. This response expectation of the phenomenon is comprehensively comparable to those characteristics of organic light-emitting diodes (OLEDs). As a result, the following points would be the robust/privileged accomplishment of this project: \textbf{i-} advancing the scope and frequency range of imaging devices and CCDs in receiving far-end light pulses; \textbf{ii-} conjecturing future captured frames of motion, in motion pictures industry; \textbf{iii-} surveillance imaging systems detecting an act of subject before it happens; \textbf{iv-} general and advanced application use in medical and physical sciences. One of the other aspects is that, \textbf{v-} this phenomenon supports new unique techniques on restricted integration of semiconductors.

As envisioned in point \#iii, security systems can become more advanced in surveilling images and detecting movements in terms of \emph{pre-motioned} or by definition: \emph{the pre-kinematical or unoccured motion of any body-form in an arbitrary scene of some arbitrary visual recording device}. For instance, CCTVs incorporating biovielectroluminescent hardware in their system. The kinematics concept can merged into dynamics when events are studied in more detail allowing complex variables accept values of a specimen for e.g., electro-microscopy, STM techniques~\cite{05-Biro},~\cite[i]{06-Wiki}, even subatomic levels of some atom-ion interactions in form of audio and visual. The anticipatory behaviour in capturing pictures of the scenes before they occur, compared with nowadays imaging systems' technology, is objectively illustrated in Fig.$\,3.1$, in recognition of Fig.$\,2.3$, \S\ref{section2}.

\section{Principles, methods and formulae to the conceptual investigation by illustration}
\label{section2}

During the conceptual investigation made upon project objectives, a result of pixel imaging with colour velocities is obtained i.e., pixel velocity $v_{r,g,b}$, and a consumed distance describing the phenomenon of biovielectroluminescence in terms of \emph{bendable wavelengths} illustrated in Fig.$\:2.3$. Usage of the symbol $\lambda_{\mathrm{bend}}$ in the future equations, e.g., Eq.$\:(2.2\mathrm{a})$, is solely for this purpose establishing connexion between $x$ as classical distance and $x_{\mathrm{cons}}$ as $x$ in consumption, traversed pre-kinematically. The index $r, g, b$ in $v_{r,g,b}$ represents red-green-blue of approximately 16.7 million colour spectra, denoting a state of pixel transition from point A to point B, where point B is the point of imagery and electrical-analogal pulse's database (see Fig.$\,2.1$). An intersection of $x_{\mathrm{cons}}$ and a time fallen into consumption as $t_{\mathrm{cons}}$, and probable space-time Gaussian curvature states of $k_i$ to $k_n$, forms proportionalities of certain synchronized space-time equations in discrete packets, pertinent to one's vision seeing things through the other(s) vision (predicate 2.1c, Eqs.$\,$2.5, 2.6). This in visual dimensional state representation, signifies Heisenberg's uncertainty principle into unification of relativities in an \emph{n}-dimensional space-time equations and thus, quantum's entanglement principle. For now, let this `quantum entanglement' be regarded into the visual aspect and superimposition of \emph{n}-dimensional images and not the `matter' itself, expressed in some superposition phenomenon.

The intersection of space and time parameters in consumption mentioned earlier, shall exhibit a \emph{dimensional motion-stretch} subject to Eq.$\,(2.6)$ of this Section. The parameter of classical velocity in quantitative terms to pixels, is represented via an observable consumed distance of a scene in terms of
\vspace{-17pt}
\begin{flushleft}
\[\left. \mathrm{cons}\;(x) = {x_{\mathrm{cons}}  = v_{r,g,b} t_{\mathrm{cons}} } \right. \left| \begin{array}{l} \; x_{\mathrm{cons}}  = \mathop {\lim }\limits_{\Delta x \to [0,\infty]} |x|{\rm  }, \; {\rm  }t_{\mathrm{cons}}  < t \equiv f(t)= t \ + \\ f(\backepsilon\epsilon) \; \pm \; \epsilon \ , \end{array} \right. \;  \;
\; \; \; \; \; \; \; \; \; \; \; \; \; \; \; \; \; \; \; \; \; \; \; \; \; \; \; \; \; \; \; \; \; \; \; \; \;\]
\[\mathrm{where} \ f(\backepsilon\epsilon) = \; \mathop {\lim }\limits_{\Delta \epsilon \to ]0,1[\rightleftarrows]0,1[\leftarrow\Delta\backepsilon}\backepsilon\epsilon =  \epsilon - \backepsilon \;=\;  \backepsilon - \epsilon = 0\;, \; \; \; \; \; \; \; \; \; \; \; \; \; \; \; \; \; \; \; \; \; \; \; \; \; \ \:  (2.1\mathrm{a})\]
\[ \mathrm{whereas}, \; \mathrm{if} \;  \int_\mathrm{\; A}^\mathrm{\; B} \mathrm{d}t_{\mathrm{cons}} = \int \mathrm{d}t \;, \; {\rm or} \; \; t_{\mathrm{cons}}\geq t\equiv f(t)= t + \epsilon\;| \; 0<\epsilon<1 \, \mathrm{s}\; \; \mathrm{\; thence},\; \; \; \; \; \; \; \; \; \; \; \; \; \; \; \; \; \; \; \; \; \; \; \; \; \; \; \; \; \; \; \; \; \; \; \; \; \; \; \; \; \; \; \; \; \; \; \; \; \; \; \; \; \; \; \; \; \; \; \; \; \; \; \; \; \; \; \; \; \] \[ \left. \therefore \int_\mathrm{\; A}^\mathrm{\; B}{\mathrm{d}x_{\mathrm{cons}}  = v_{r,g,b}(t_{\mathrm{cons}} - t) \approxeq 0 \; , \; \{\forall  x_{\mathrm{cons}}=0^{+}::\mathrm{F}\rightarrow x = vt ::\mathrm{T}\}\Leftrightarrow \mathrm{T} .} \right.
\;   \vspace{-6pt} \] $  \ \ \ \ \ \ \ \ \ \ \ \ \ \ \ \ \ \ \ \ \ \ \ \ \ \ \ \ \ \ \ \ \ \ \ \ \ \ \ \ \ \ \ \ \ \ \ \ \ \ \ \ \ \ \ \ \ \ \ \ \ \ \ \ \ \ \ \ \ \ \ \ \ \ \ \ \ \ \ \ \ \ \ \ \ \ \ \ \ \ \ \ \ \ \ \ (2.1\mathrm{b})$
\end{flushleft}
\vspace{-6pt}
\long\def\symbolfootnote[#1]#2{\begingroup%
\def\thefootnote{\fnsymbol{footnote}}\footnote[#1]{#2}\endgroup}

\noindent Let by conventional limit to functional Eq.$\:(2.1\mathrm{a})$ on some body displacement, distance change $\Delta x$ for `a geometrically pre-travelled future distance' as consumed distance $x_{\mathrm{cons}}$ approach 0 or $\infty$, but not an absolute number of either the latter or former in the denoted interval on $\Delta x$. In other words, $x_{\mathrm{cons}}$ generates values of near zero or near infinity for the $\pm$ side of $x$ in modulus $|x|$, in sign of (2.1b). One could mathematically interpret $x_{\mathrm{cons}}$ as:

\vspace{4pt}
\noindent ----------- \textbf{i-} \emph{A deformed timelike worldline}, satisfies conditions of $\mathrm{d}s^{2}>0$ in the context of general theory of relativity (GR) in the realization of, Minkowski and Einstein's contribution on the principle of relativity.\noindent\symbolfootnote[1]{ The axiom stated on p.$\,$80 and, Eqs.$\:$(20 to 46) of pp.$\,$ 131-143, \S2 \S9, Chaps. II and VII, Ref.~\cite{19-Einstein et al.}.} That is, incorporating $x_{\mathrm{cons}}$ in any formulae in conjunction with the future Eqs.$\,(2.5)$, (2.6), which aims to satisfy the Geodesics concept in GR via the subsequent predicate for \emph{the theory of unified relativity} (UR), subject to \S\ref{section3}. Hence for now, write upon this orientated statement interpretation, a deformed timelike worldline predicate
\vspace{-28pt}
\begin{flushleft}
\[\left\{\forall x_{\mathrm{cons}}\in \textmd{g}(f(s)) \left| f(s)=\mathrm{ d}s^{2}<0 , \; \textmd{g}(s) = \mathrm{d}s^{2}>0 \;, \; \mathrm{for } \; \textmd{g}(f(s)): f(s)\mapsto \textmd{g}(s)= \ \begin{array}{l}
\\ \vspace{-2pt}\frac{}{}
\end{array}\;  \right.\right.\]\end{flushleft}\vspace{-20pt}
\begin{flushleft}
\[\left.{\mathop {\lim }\limits_{\Delta x \to [0,\infty]} |x| \wedge \{{\mathrm{d}s^{2}<0}\Leftrightarrow v<c\}\cap\{{\mathrm{d}s^{2}>0}\Leftrightarrow v>c\} \equiv \frac {x_{\mathrm{cons}} \times \{\mathrm{d}s^{2}<0\}} {\{\mathrm{d}s^{2}>0\}} \longrightarrow }\right.\]
\[\left.{ \frac {v_{k} \times (\mathrm{d}x \in \{\mathrm{d}s^{2}<0\} \times \mathrm{d}y\in \{\ v_{r,g,b}\})\: \mathrm{cos}\:\theta}{(c \: \mathrm{d}t)^2\in \{\mathrm{d}s^{2}>0\}} =x_{k} \; , \; \mathrm{d}s^{2} = \: \emph{g}_{\mu\nu} \: \mathrm{d}x^\mu \: \mathrm{d}x^\nu}\right\} \; , \; \; \; \; \; \; \; \; \; \; \; \; \; \; \; \; \; \; \; \; \; \; \; \; \; \; \; \; \; \; \; \;   \]
\end{flushleft} \begin{flushright}\vspace{-2pt}
(2.1c)\end{flushright}\vspace{-6pt}
\noindent and hence obtains \[\therefore  x_{k}\equiv \sqrt{\mid \mathrm{d}s_{\mathrm{cons}}^{2} \mid} \; . \]\begin{flushright}\vspace{-20pt}
(2.1d)\end{flushright}

\noindent  By means of which, $\mathrm{d}s_{\mathrm{cons}}$ corresponds to events of consumed coordinates of classical length $x$ and space-time interval $\mathrm{d}s$, in terms of time in consumption or $t_{\mathrm{cons}}$, from the set form on time-like space-like ratio
\[\{\mathrm{d}s^{2}>0\Leftrightarrow v<c\}:\{\mathrm{d}s^{2}<0\Leftrightarrow v>c\}\:,\]

\noindent is in regard to the degree of revolution, $\mathrm{cos}\:\theta$, for those pre-travelled coordinated events. These pre-travelled events in the context of electrodynamics could be expressed in virtue of predictable loci of lengths in some observable \emph{n}-dimensional space-time-form system (\emph{n}-DST). Out of the total possible spacelike events for curvature state \emph{k}, one can determine in the denominator of the articulated equivalence say, $(c \: \mathrm{d}t)^2$ as an element of set $\{{\mathrm{d}}s^{2}>0\}$, in every calculation. This interpretation implies correspondingly to the timelike events in the numerator part of the equivalence and thus, future equations when vectors are being used for calculations on magnitudes and angular responses of predictable loci. The inequalities $v<c$ and $v>c$ where $v$ appearing as velocity and $c$ as the speed of light, correspond to scenarios of timelike and spacelike, respectively. ------------

\vspace{6pt}
\noindent ------------ \textbf{ii- }Equivalently, \emph{a graded spacelike worldline}, where this type of worldline functions reversibly compared with interp.$\,$\textbf{i}. That is when numerator contains vectors of $x$ in set $\{\mathrm{d}s^{2}\}$ conceiving $\emph{g}_{\mu\nu}\: \mathrm{d}x^\mu \: \mathrm{d}x^\nu$, and then $y$ in set $\{v_{r,g,b}\}$, reflecting coefficients $\emph{g}_{i}$ in denominator, thus describing the curvature state $k$ of space-time and its deviation from a Euclidean nature in the context of GR. These coefficients are, in general, functions of $t, x, y$ and $z$ for which all $\emph{g}_{i}=1$~\cite{47-Krane}. The Minkowski metric from Euclidean metric in three-space $\mathbb{R}^3$~\cite[i and ii]{42-Wolfram}, is given in terms of line element $\mathrm{d}s^{2} = \emph{g}_{\mu\nu} \: \mathrm{d}x^\mu \: \mathrm{d}x^\nu$, which is processed in Eqs.$\,$(2.6) and (3.15). ------------

\vspace{7pt}
The two interpretations outlined above, \textbf{i} and \textbf{ii}, in a simplistic manner, function in a course which define all formulaic components in regard to space-time interval $s$, as $k_i$ or degrees of Gaussian curvature in the context of SR. This formulaic representation is subject to Eq.$\,$(2.6), in an antiderivative form reflecting relativistic generation of Minkowski metric giving the proper time $\mathrm{d}\tau^2$ via arbitrary constant $\mathfrak{c}$ (see also, Ref.~\cite[ii]{42-Wolfram}). An example is provided at the end of this Section where one could study events between observable systems in both contexts of relativities, SR and GR in the form of UR. It is crucial to signify the submitted interpretations incorporate Gödel logic and deductive fuzzy logic~\cite{40-Behounek}, hence the notions of entropy between systems are in place. These notions in particular, define visual and scope ability between some \emph{n}-DST system, in the mappability of spontaneous changes or transformations of the system based on energy state and pixel velocity.

\noindent ------------ Pay attention to the use of logic symbols e.g., $\wedge$ and $\cap$ for the entropic behaviour in Pred.$\,$(2.1c). ------------ As we shall confront shortly, these visually physical systems of transformations, may refer to an observable energy $E$ product to pixel velocity $v_{r,g,b}$ or, $Ev_{r,g,b}$ (relation 3.5e, \S\ref{section3}). However, the time fallen into consumption problem, in ratio terms to the visual system between two non-subatomic and sub-atomic bodies, fixate conditions of time factor $\dot{t}^{2} \ne \dot{t}^{2} $ as a `free variable' satisfying global time scenarios that obey Russell's Paradox, discussed by Alipour~\cite[relations~(4.1) to (4.6)]{01-Alipour,{02-Alipour}}.

Let in continue to relation (2.1b) by conventional limit on event prediction of the same body, consumed time $t_{\mathrm{cons}}$ between two space-time points A and B by definition, represent a completely used up time $t$, which if it occurs, $x_{\mathrm{cons}} = 0$, according to (2.1b) satisfying time scenario $t_{\mathrm{cons}} = t$ via $t$'s functional equation $f(t)= t + \epsilon$. Assume $t$ on its own, ordinarily represent a Newtonian time frame standard shown on a clock of a human observer, observing events between points A and B on earth without duality. So, a condition satisfying scenarios of $t_{\mathrm{cons}} > t$ via the same function of time, forms a greater time increment $t_{\mathrm{cons}} > t+\epsilon$, and not $t_{\mathrm{cons}} < t-\epsilon$ (aim of objectives \#3 and \#4, \S\ref{section1}). This is due to the result remaining zero and not $\epsilon$-dependant according to the equivalence stipulated on $t$ in functional Eq.$\,$(2.1a), or better to say: `no distance is being consumed and just dimension $x$ is being travelled interdependent with $\epsilon$'. Ergo, it is said that, the laws of time tick as $t + \epsilon$, where $0<\epsilon<1$ being some small time increment, are bounded to classical physics when $t_{\mathrm{cons}}\geq t$, overriding `$t_{\mathrm{cons}} < t-\epsilon$' against absolute value $|t|\geq0$ as non-negativity, thus destroying the overall possibility of $x_{\mathrm{cons}}$, deduced by the righthand predicate in integral relation (2.1b). The latter is usual to occur, based on the currently well-known theories. That is, since the paradox of infinitesimal points of time conserve the problem of past-to-future events interconnected to some current event occurrence in our universe, difficult, but possible to tell the current event's future or past outcome based on Eq.$\,$(2.1a) satisfying the current theory.\noindent\symbolfootnote[1]{ Science tells us the minimum amount of time that can be measured is called Planck Time. This is around $10^{-43}$ seconds. Below that length of time there cannot be said to have a future or past~\cite[xi]{06-Wiki}.} To compute $x_{\mathrm{cons}}$, we firstly should contemplate the means of an experimental setup, observation and thus perspicacity; \emph{Firstly}, the installation process of essential components between points A and B of the experiment. \emph{Secondly}, biological surgical operation in removing some organs from the insect and from there, observational operations studying bio-organic events inclusive to physical electro-optical analyses relevant to biovielectroluminescence phenomenon. \emph{Finally}, data collection of the future Eqs.$\,$(2.2), (2.4), and relation (2.3), complying/aligning with UR theory, \S\ref{section3}.
\vspace{6pt}

The above-mentioned tasks for the pragmatics phase and analysis, commence as follows:

\vspace{6pt}
It is the optical waveguide fibers, organic substrate and electrodes installation to be taken into consideration first hand. This requires a surgical operation on the insect, i.e., dissecting it prior to the optical waveguide installation. Subsequent to optical fibers installation which is after the growth of pillar-carbon nanotubes (CNTs), (see the next paragraph), techniques of dark field microscopy such as Rheinberg illumination, would assist the experimenter in studying biovielectro-luminescence with relevant light sensors according to Figs. 2.1 and 2.3\emph{a}.

From Fig.$\,$2.2\emph{b}, kernel to the concept of Step IV of Fig.$\,$2.1, it is of \S\ref{section3}, hypothesized that, the fly's eye does possess a time varying \textbf{B}-field indicating the phenomenon of sooner predictability of event occurrences compared with a human's eye (as single-aperture eyes) understanding the 4th dimension. That is, the `time' dimension during capturing pictures of a scene in one's memory state. As an example, this is shown by a few depicted magnetic field lines in combination with polarization as the transverse light phenomenon, described by the time varying \textbf{B}-field with the probable presence of charge \emph{Q} with an \textbf{E}-field property of a situation of electrodynamics in this picture. Charge \emph{Q}, is deemed to satisfy Eqs. of \S\ref{section2}, defining proper time $\tau$-property for vision (experienced by the particle $q$ and the visually utilized body in its own rest frame), in respect to a consumed distance and time in the context of Einstein's SR. Fig.$\,$2.2\emph{c} is the theoretical representation of the wiring construct, pre and post single-voltage application concerning charge quantity \emph{Q}.

In Fig.$\,$2.1, during experimentation, a neuron sodium, potassium or calcium ion (Na$^{+}$, K$^{+}$ or Ca$^{2+}$)-feeder in aim of constructing `voltage-gated ion channels' in the fly's body, is deemed to be used before single-voltage application, hence to study the electrical and field effect property produced via time-varying \textbf{B}-filed~\cite{41-Bezanilla}. The feeding course is feasted upon the channels installing a new artificial circuit that once ported the brain and ventral nerve cord (VNC) comprised of about 1,500 neurons, discussed by Shwartz~\cite{24-Shwartz}.
\smallskip
\vspace{62.5mm}
\begin{flushleft}
\includegraphics[width=22mm, viewport= 0 0 10 65]{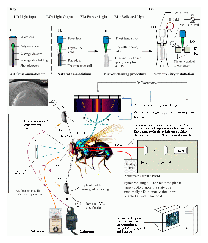}\\
\end{flushleft}

\noindent{\footnotesize{\textbf{Figure 2.1.}  The base anatomy, finite state and diagrammatic representation of the project from point `A', as source of pulse(s) to point `B', promoting the analysis stage, advanced vision detection and storage of pulses(s). \emph{Upper-left corner}: a cross section with the spherical arrangement of artificial ommatidia consisting of microlenses, polymer cones, and waveguide arrays (Luke Lee photo, courtesy of Science magazine); Parts of its anatomy is shown in View I. In the artificial ommatidium, light impinges onto a microlens and thereby coupled with polymer cones and waveguides and then guided to the end of the waveguide; A natural ommatidium however, consists of a facet lens, a crystalline cone, and photoreceptor cells with a wave-guiding rhabdom, Ref.~\cite{03-Lee}. The microsuctioning procedure on the natural ommatidium (the real fly specimen), is carried out preparing it for optical/electical/organic circuitry installation and fabrication process in accordance with Steps II through IV. In connection with Step IV, see \S10.6.1 and \S10.6.2, and for the electro-optic property~\cite{12-Senior}; See~\cite{49-Borchardt} for the organic emitter property, pertinent to the current author's design modifications and adaptation.}
\bigskip

\normalsize

\vspace{62.5mm}
\begin{flushleft}
\includegraphics[width=27mm, viewport= 0 0 10 43]{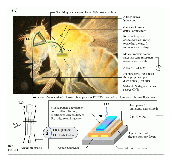}\\
\includegraphics[width=27mm, viewport= 0 0 10 21]{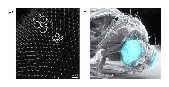}
\end{flushleft}
\vspace{-2pt}
\noindent{\footnotesize{\textbf{Figure 2.2. }(\emph{a}) Cross section installation and insertion technique of optical contacts including, monochromatic top-emitting OLED~\cite{54-Tan et al.} array, and stalactite/stalagmite angular positioning of pillar-CNTs in the fruit fly's body. The contacts connect from RGB filter to fly's anterior for the diagram illustrated in Fig.$\,$2.1. The anatomy of an element of the array as a TOLED with a monochromatic light output is also shown. The given techniques are here to assist ways of biovielectroluminescence propagation against energy loss from its source. The fruit fly's 800 unit eyes or ommatidia~\cite{06-Wiki}, compared with the 4,000 image-forming elements of a typical housefly, holds good for basic visual investigations between fruit fly's visual ability and e.g., dragonfly's sophisticated visual system. However, dragonfly's thousands of unit eyes adaptation as the artificial apparatuses $\mathbf{A}'$ and $\mathbf{A''}$ of part (\emph{a}), shall be the most privileged; (\emph{b}) The fly's eye under electro-microscope; (\emph{c}) The hypothetical biovielectroluminescence effect off of the fly's corneal surface shown in a \htmladdnormallink{14 frame graphical image sequence}{http://mojanzygon.netfirms.com/ug013a25.exe} (\htmladdnormallink{530 KB}{http://mojanzygon.netfirms.com/ug013a25.exe}). ----------- Pic.$\,$(\emph{a}), the fruit fly photo, courtesy of \htmladdnormallink{T. Grant}{http://microscopy-uk.net/}; Pic.$\,$(\emph{b}), courtesy of \htmladdnormallink{\emph{Dept. Biol.}, University of Wisconsin}{http://images.google.co.uk/imgres?imgurl=http://www.uwec.edu/biology/Facilities/FacilitiesPhotos/FlyEye.jpg&imgrefurl=http://www.uwec.edu/biology/Facilities/Facilities.htm}; Pic.$\,$(\emph{c}), courtesy of \htmladdnormallink{\emph{Dept. of Insects}, Chicago Field Museum}{http://www.localinfinities.com/infinitesimals/eye.jpg}. Photo alterations including the 14 frame graphical animation, are done by the current author. -----------}
\bigskip

\normalsize This is merely a preliminary step of approach on simulating a neuromatrix system once the growth of CNTs, incorporation of organic emitters and single-mode optical fibers are thus populated based on electrolysis as being obtained in this account; examples are given by Wang \emph{et al.}~\cite{15-Wang}; Kaushik \emph{et al.}~\cite{32-Kaushik et al.}, respectively. In addition, in Fig.$\,$2.2\emph{a}, multimode optical waveguides for reflected light loss problem (RL) adjacent to the electro-optics, is taken into consideration. The method for dissecting an adult fly's brain using dissecting microscope and special forceps accompanied with other experimental tools is guided by Schaner \& Sniffen~\cite{36-Sniffen}. The objective of the ionic feeding method is to assess the \textbf{E} and \textbf{B}-filed generation, allowing the measurement process to proceed for the fly's corneal surface. The material under the influence of the electric field shall be concentrated between the fly's eye and its \emph{lower dorsal}.

A microsuctioning procedure indicated in Fig.$\,$2.1, is after the ionic approach and data investigation on the field issue on the eye surface. Hence to make appropriate data signal comparisons between the biological concept versus organically-artificial concept, where the latter is fabricated within the insect's body. The suction made, is to take out different parts of ommatidium except the lens and crystalline cone for a reverse projection light process, contrasted to their regular task which is focusing light into the eye. This is done by installing a discrete array of miniaturized glass (or plastic) optical microlenses at the ommatidia to the organic emitter, thus acting passive for light input (LI), and focused for light output (LO), between metal electrodes for the array of electro-optic switch (see Fig.$\,$2.2). The microlenses' mechanism could resemble micro-Fresnel lens characteristics. For the electro-optic deportment, it is to install the new artificial circuit in form of grown pillar-shaped CNTs aligned along the direction perpendicular to the substrate surface (p.$\:$751, Ref.~\cite{15-Wang}).

These CNTs should be inclined to the hemispherical angle of the insect's eyes within the specimen's body. The circuitry installation and fabrication process, aim to exhibit \textbf{E}-field emission (pp.$\,$751-752, Ref.~\cite{15-Wang}. See also, Ref.~\cite{37-Katayama}), on the ommatidium surface, whilst the CNTs adjacent to microlens and crystalline cones remain intact to the outer layer of the insect's eye. The positioning and formation of pillar-CNTs inclusive to their substrate, are in state of stalactite/stalagmite geometry (Fig.$\,$2.2\emph{a}). This type of installation after surgery, is to permit light emission on the principles of biovielectroluminescence energy state from Eq.$\,$(5.4), \S\ref{section5}, pertinent to Eq.$\,$(5.3) and governed by Fowler-Nordheim equation~\cite[x]{06-Wiki}. Subsequently, artificial ommatidia collection apparatuses $\mathrm{A}'$ and $\mathrm{A}''$, receive and detect the reflected photons off of the subject(s) in the scene, ready for post-step signal imaging and pixel data processing.

\long\def\symbolfootnote[#1]#2{\begingroup%
\def\thefootnote{\fnsymbol{footnote}}\footnote[#1]{#2}\endgroup}

In this illustration, Fig.$\,$2.1, a mono-MUX element out of total MUXs as multi bit-multiplexer, is considered for the CCD and other electrical channels. The MUX is combined with a demultiplexer (DEMUX) adjacent to analogue and optical signal carriers of wavelength-division multiplexing (WDM), discussed by Stallings~\cite{30-Stallings}; Senior~\cite{31-Senior}. Contemplating the WDM method makes possible to establish a bidirectional communication between points A and B, whilst maintaining all analogue/digital signals' conversion to the computer unit. A memory system (preferably a read-only memory (ROM) type in case of preventing data modifications), in the circuit, is for both CCD colour result e.g., 24-bit technology and pulses from the Calyptrate Diptera fly vision,\noindent\symbolfootnote[1]{Other types of flies e.g., honeybees and dragonflies with apposition compound eyes~\cite{18-Gaten}, for specially-purposed imaging applications vsversus abstract imaging systems~\cite{22-Grossberg}, such as faster movement detection with high resolution relevant to the project, is permissable instead~\cite{06-Wiki}. The more lenses the compound eye bears, the higher the resolution of the image.} in combination. The computer unit thereby accesses memory location (address) to categorically store data as colour signals against combined signals of both former and latter types. A projector projects light onto a photographic film slot (under dark room conditions) as an original camera negative (OCN) process, then developed firstly into colour pictures and in closure, picture sequence or motion picture. All is recorded in discrete intervals. The next step to do is making relevant comparisons between the recorded events in discrete intervals, simulating the same ongoing actual events assigned to the object situated within the experimental scenery setup, Fig.$\,$3.1, \S\ref{section3}. This setup is arranged by installing two units of artificial ommatidia collection apparatuses, $\mathrm{A}'$ and $\mathrm{A}''$ symmetrically.

Luminance density and contrast per time sequence, is also recorded during this action whilst single-voltage application is uniformly distributed onto the fly's eye surface, recording \textbf{E}-field and \textbf{B}-field events by electromagnetic sensors. A matrix of an RGB angle calculator is to deal with lens dispersion problems dependent to photodetectors. Thus the calculator's task is to correct secondary performance of chromatic abberation problems in a variant scale due to the way of microlenses distribution and the polymer replication process discussed by Lee's team~\cite{03-Lee}. Alleviating problems such as angular acceptance, polymer waveguide modes, light propagation and other essential optical measurements, are via optical sectioning technique (p.$\,$560, Ref.~\cite{03-Lee}).

As hinted in Fig.$\,$2.1, a matrix calculator is installed on the bottom of the optical data receiver/projector section, above the relevant optical waveguides, so to virtually mix imagery angles off of the elements of the scene per microlens of the ommatidium collection when necessary. This image matrix technique is to eliminate blur and maintain faster capture speed on fly's vision which is experimentally measured by a light-meter relative to biovielectroluminescence effect, before and after the large collection of artificial ommatidia (here, e.g., dragonfly's 30,000 ommatidia). The result would be producing properly exposed discrete sequence of images in unification.

The matrix mainly supports methods of multi-channeled optical imaging system in form of a calculator of matrix of RGB angle $\alpha_{r,g,b}$, computing image pixels in favour of an installed virtual filter onto a computer software program as a colour filter, discussed in \S\ref{section4} and \S\ref{section5}. The formation of polymer waveguides self-aligned with microlenses by Lee~\cite{03-Lee}, is considered for the means of comparisons whereas repositioning the optical channels could be realized by reconfiguring microlenses and thus secondary corrections made upon the RGB matrix program. Its specification is premised in terms of a cyclic data flow of \emph{virtual/real variable lens and image matrix technique} (or, `RGB matrix program'), a self-built reconfigured optical sectioning technique on angles of focus. Executing this technique would be in form of series of consecutive images from different depths of field of the lens using a programmable memory chip reacting to matrix points that are sensitive to chromatic abberation. Alternatively, immiscible (non-mixing) fluids of different refractive index (optical properties) of variable-focused lens (convergent to divergent and back) by Royal Philips Electronics~\cite{34-Philips}, could be designated for the `real matrix'. The implementation process dedicated to the values of matrix points of light refraction, is assigned to its program. The convenience to this alternative is, no need for mechanical parts, and just benefiting from the above-mentioned involved \textbf{E}-field, adjusting the microlenses shape as a practical solution against the `virtual matrix'(see also, Ref.\cite{33-Ren}).

For instance, the technique used in the matrix's construct of different light wavelengths, can give an auto-focus zooming behaviour used in modern videocams as an extra function devoted to artificial waveguides and ommatidia polymer microlenses. Moreover, for the two `artificial ommatidia collection apparatuses' $\mathrm{A}'$ and $\mathrm{A}''$, the received left, right and center images are recombined, thus to detect and compute precise 3D motion and relative time factors of the moving object in the scene. This extra function would also include image correction based on sufficient software complexity and data for image post-processing without delay (refer to \S\ref{section5} or see general literature on this term e.g., Ref.~\cite{06-Wiki}).

The operation mechanism having quicker frames of image capture akin to a fly's visual cortex encoding visual stimulus, also possess motion detection from the scene compared with a simple one-eyed function. This mechanism subsequently unifies all objects and their motion capture in an average point, stretched from left to right of the scene. This would be given in terms of \emph{motion-stretch } $x_{k_{i,...,n}<0}$, on curvature $k<0$ in terms of its coefficient $k_i<0$ as generalized in Fig.$\,$3.1, \S\ref{section3}, pertinent to Eq.$\,$(2.5). An example of averaging the captured image in the RGB matrix from the scene positions \textsf{L}, as left, \textsf{C}, as centre and \textsf{R}, as right, is given in \S\ref{section5}, governing the proposed statements, concept and hypothesis of \S\ref{section3} as well.

Since the azimuthal filed of view of a fly is almost $360^{\mathrm{o}}$~\cite{23-Orghidan} and possesses two large spherical eyes covering an approximate $360^{\mathrm{o}}$ vision, one cold observe and adapt the very imaging sensors of a system equipped with low cost normal lenses (traditional cameras), into $360^{\mathrm{o}}$. This is considered for the development stage, a substitution for expensive fish eye lenses with a wide field-of-view (FOV) over $90^{\mathrm{o}}$ imagining ability~\cite{06-Wiki}. In reason, fish eye lenses are not economical in mass production for surveillance imaging needs and photography.

In continue, governing sequences of images for a common 4,000 image-forming elements of a typical housefly in adjacency with/relative to photodetection of the artificial compound eyes in Fig.$\,$2.3 (e.g., a 30,300 dragonfly of Ref.~\cite{03-Lee}), is prompted to acquire synchronization of a time dilation product from SR to UR. No matter how its resultants remain minuscule, the general time dilation product remains significant in anticipating future events of an electrodynamical system. This time dealation product, as one of the main objectives denoted in Fig.$\,$2.1, plugs in all pixel data in synchronization to the captured events based on time dilation resultants given in Tab.$\,$3.1, \S\ref{section3}, for an insect relative to human. One could ponder, from an insect's \emph{visual rate ability} in the category of \emph{speed and interrogation of incoming sequences of image frames}, adjacent to artificially inspired insect's \emph{visual resolution and angular ability}~\cite{03-Lee}, altogether fulfil a sequential output of Hi-Resolution images, efficiently conjure up 100\%. This claimed efficiency herein states: the efficiency of 100\%$^+$ or even more\emph{ neurovisual parallel impairments}, hereon seeing more than one parallel world to another linking to yours. This is compared with a human's vision and cognitive perception (say, its usage in AI), once being adapted under real environmental circumstances. So, space-time curvatures and pre-kinematical anticipation of objects in motion, could be promoted and captured as vital data to the system in the proposed project (see explanations over Fig.$\,$3.1, \S\ref{section3}).

In Fig.$\,$2.3, it is shown that curvature states of $k$ for any occupiable unit circle of space and time can be deformed when a wave is propagated in space. A perfect example is the propagation of the electromagnetic wave such that, time and space are being dedicated to \textbf{E} and \textbf{B}-fields and not just space. In other words, the conventional performance of how the \textbf{E}-field, the \textbf{B}-field and the wave vector inter-relate as the electromagnetic wave propagates through space could be reconsidered especially for propagation of electromagnetic wave  in space-time varying dielectrics and plasmas. This is hypothesized by Fante in 1972~\cite{44-Fante}; visualized by Whites in 1998~\cite{45-Whites}. In White's illustration over lossless media (Fig.$\,$2.2\emph{a}), in his plot for a wave with amplitude (Vm$^{-1}$), $\mathbf{E}_{\mathrm{m_{plus}}}=1$, the $\mathbf{E}_{\mathrm{x_{plus}}}=1$ is shown as a function of $z$ (meter) at three different time values: $t = 0$, $t = \mathrm{T}/4$ and $t = \mathrm{T}/2$, where $\mathrm{T}$ is the period of one time cycle.

Ergo, applications concerning the study of space into space-time occupation by some moving object in scene dissipating energy (dynamical), verify the relativeness of the object to biovielectroluminescence holding an electromagnetic property and its time-varying media. The phenomenon too represents a function of electric field that varies in space and time as a wave in respect to unoccured $k$-events. The time-varying media can be framed as the constituents of the creature's eye, the corneal layer.

In the UR theory, it is perceived that\emph{ bendable wavelengths}, do obey such characteristics especially for pre-kinematical events of some relativistic occupation of space and time, whereas there must at least exist one light source and a media relative to multiple photo-receivers and their detectors. In this case, artificial apparatuses $\mathbf{A}'$ and $\mathbf{A}''$ of Figs.$\,$2.2 and 3.1. For the wave propagation's unit circle $C_1$, relative to a medium in motion occupying space, the formulae on its pre-kinematics occurrence of some $k$-curvature as `\emph{curvature n-state propagations or motion-stretch}' like the one illustrated in Fig.$\,$3.1, suggests \\
\vspace{-6pt}
\[ \ \ \ \ \ \ \ \ \ \  \ \ \ \ \ \ \ \ \ \ \ \ \ \ \ \ x_{k} \equiv \lambda_{\mathrm{bend}} =\frac{v_w t}{\nu t_{\mathrm{cons}}}\ \ \mathrm{against} \ \ \lambda= \frac{v_w}{\nu} \;, \ \ \ \ \ \ \ \ \ \ \ \ \ \ \ \ (2.2\mathrm{a})\]
\noindent conditionally. That is, wavelength $\lambda$, possessing $v_w$ which is the propagation velocity of the waves $\lambda$ and $\lambda_{\mathrm{bend}}$, corroborating with bendable distances travelled by the object pre-kinematically against the former i.e., the spontaneous $\lambda$ in favour of electric field \textbf{E} in conventional Maxwell equations in consideration of discrete structure of the electromagnetic energy

\begin{center}\vspace{-18pt}
\[E=h\nu \; \mid \; h \approx 6.626068...\times 10^{-34} \ \mathrm{kg}\:\mathrm{m}^{2}\: \mathrm{s}^{-1}\;,\]
\end{center}\begin{flushright}\vspace{-22pt}
(2.2b)\end{flushright}
\vspace{2pt}
\noindent  where $h$ is Planck's constant, and $\nu$ is the frequency of the photon $h\nu$, in favour of direction $x$ as illustrated in the following diagram.

\bigskip
\vspace{62.5mm}
\begin{flushleft}
\includegraphics[width=17.6mm, viewport= 0 0 10 22]{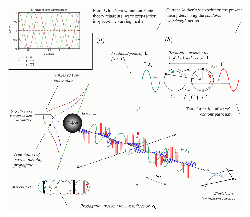}\\
\end{flushleft}

\noindent{\footnotesize{\textbf{Figure 2.3.} The \emph{bendable wavelength }(\emph{b}), compared with a conventional wavelength (\emph{a}), from a classical electromagnetic wave propagation into a wave-in-consumption, respectively; (\emph{b}) middle portion, courtesy of Fu-Kwun Hwang, \htmladdnormallink{NTNUJAVA Virtual Physics Laboratory}{http://www.phy.ntnu.edu.tw/ntnujava/index.php?topic=35}, sever from author's alterations prior to modifications on the remaining parts describing \emph{bendable wavelengths} inclusive to \emph{transformation of wave's unit circle paradigm} or equally, \emph{transformation of wave's contour paradigm}.}
\bigskip

\normalsize
The unit-unit circle $C_1$ during shape and coordinate transformation on the propagating wave, appreciates the formality of \emph{transformation of wave's contour paradigm}. That is doable via the unit circle (contour) transformation function $T(C)$. The function is specified in Fig.$\,$2.3, transforming contour $C_1$ into $C_2$ in specialization of pre-kinematical or, unoccured future events' motion as curvatures of $x$ in states of $k$. These states of curvature could be expressed as e.g., negative curvature sequence $k_{i,...,n}<0$, occupied by the object within the scene posed by the incoming-propagating electromagnetic wave.

These types of curvatures defining the concept of pre-kinematics of an object, permit the system to anticipate future incoming motion(s) as `time prediction' linked to the object by a proper algorithm, i.e., \emph{pattern recognition of pre-motion}. In other words, the negatively curved distance $x_{k<0}$ denoting the previously-mentioned geometric stretch for a coordinate transformation of discrete area $\mathbf{A}$ in the scene, against the obtained formulation of $v_{r,g,b}\:t_{\mathrm{cons}}$ in Eq.$\,$(2.1). This behaviour generates a ratio of interval $s$ regular to Euclidean space versus consumed distances in a course of spatial scalar product $\mathbf{x}\cdot \mathbf{y}$ for Eq.$\,$(2.5). One distance is of $x_\mathrm{C}$ as `regular', denoting circuital travel between data source and destination, such that $x_{\mathrm{cons}}$, is of a consumed distance for data as pixels' information in Eqs.$\,$(2.1) to (2.5). The other distance relating to $x_{k<0}$, is of coordinate system, $x^{\mu}$, in an \emph{n}-dimensional Euclidian space~\cite{06-Wiki} in the context of SR, which involves the object, scene and the time-varying field altogether in the course of \emph{dynamics-into-kinematics}. By this notion, one could convert any form of \emph{velocity of an object in motion within some observable scene} into pixel velocity $v_{r,g,b}$ on screen or, \emph{xy} for the set of scene's object picture-elements (displayable pixels after point of destination here, point B) in constance to the promised ratio of the involved distances, or, \[
\frac{{x_{k < 0} }}{{x_{\mathrm{cons}} }} = \frac{{\left( {vt} \right)_{k < 0} }}{{v_{r,g,b} t_{\mathrm{cons}} }}
\; \; \; \; \; \; \; \; \; \; \; \; \; \; \; \; \; \; \; \; \; \; \; \; \; \; \; \; \; \; \; \; \; \; \; \; \; \; \; \; \; \; \; \; \; \; \; \; \; \; \; \; \; \; \; \; \; \; \; \; \; \; \; \; \; \; \; \; \; \; \; \; \; \; \; \; \; \; \; \; \; \; \; \; \; \; (2.2\mathrm{c})\]
\noindent whereby
\[
v_{r,g,b}  \leftrightarrow v_{k < 0} {\; \; \rm iff \; }f\left( {v_{r,g,b} } \right):v_{r,g,b}  \to \frac{{xy_{r,g,b} }}{t},\; \; x_{k < 0}  \ne x_{\mathrm{cons}}
\; \; \; \; \; \; \; \; \; \; \; \; \; \; \; \; \; \; \; \; \; \; \; \; \; \;  (2.3)\]
\noindent hence
\[
v_{k < 0}  = v_{r,g,b} {\: \rm o \:}\frac{{xy_{r,g,b} }}{t}, \; v_{k < 0}  \ne v_{r,g,b} {\: \rm o \:}\frac{{x_{\mathrm{cons}} }}{t}, \; f\left( {v_{r,g,b} } \right) \in \left\{ {\frac{{xy_{r,g,b} }}{t},\frac{{x_{\mathrm{cons}} }}{t}} \right\}.
\; \; \; \;\]
\begin{flushright} \vspace{-6pt}
(2.4a)\end{flushright}
\vspace{-2pt}
Thus, $ v_{r,g,b}$  is always in ratio to \(x_{\mathrm{cons}} t_{\mathrm{cons}}^{ - 1}\), or \(v_{r,g,b} :x_{\mathrm{cons}} t_{\mathrm{cons}}^{ - 1}\) for Eq.$\,$(2.2c), unless described in terms of \(v_{r,g,b} :xy_{r,g,b} t^{ - 1}\) which is equal to \(v_{k < 0}\) in Eq.$\,$(2.4a), correspondingly. One can compatibly merge this concept of anticipation of unoccured events into phase velocity $v_\mathrm{p}$, \emph{virtually} appearing as $v_\mathrm{p}>c$ in relativistic relations for energy $E$ and momentum $p$,~\cite[xvii]{06-Wiki}. In this case, the evaluating phase of matter-waves does not necessarily have to exceed $c$ for this sort of curvature group pre-geometric determination, and just a mere contemplation of its \emph{virtual establishment}. That is, suppose the condition `$>c$', has precedently/already occurred. In fact, since all phases of group velocities $v_{g_{1,...,n}}$ are in place for a transformation into pixel velocity $v_{r,g,b}$, it is assumed a consumed distance with curvature $k$ is already travelled, once pixel velocity is multiplied by $t_{\mathrm{cons}}$. Let this anticipation of unoccured events be applicable in form of function, $\mathfrak{A}$. Ergo, the matter-wave phase relativistic relation originated from de Broglie's quantum relations, is merged into
\[ \mathrm{if} \; \forall \ v_\mathrm{p} \stackrel{\exists}{\longrightarrow} \frak{A}(v_\mathrm{p}) \; \mathrm{then} \; v_\mathrm{p} = \frac {\frak{A}(\omega)}{\frak{A}(k)} = \frac{\frak{A}(E)}{\frak{A}(p)} \equiv \frac{\gamma m c^2 \longrightarrow m\lambda_{\mathrm{bend}}\nu c}{\gamma m v\longrightarrow mv_{k_{i,...,n}}} \ , \]
\noindent generating
\[ \ \ \ \ \ \ \ \ \therefore |v_\mathrm{p}| \equiv \frac{c^2}{v}\longrightarrow \frac {t_{\mathrm{cons}}c^2}{x_{\mathrm{cons}}} = \frac{c}{\beta} \stackrel {\mathrm{dis}}{\sim}  c_{\mathrm{cons}} = \frac{c^2}{v_{r,g,b}}   \; , \ \ v_{r,g,b} \ll \{ c ,v \} \ \ \ \ \ \ \ \ (2.4\mathrm{b}) \]

\noindent where $E$ is the total energy of the particle (i.e. rest energy plus kinetic energy in kinematic sense), $p$ the momentum, $\gamma$ the Lorentz factor, $c$ the speed of light, and $\beta$ the velocity of a consumed or anticipated fraction of $c$, which is now distributed into $c_{\mathrm{cons}}$ (denoted by symbol, $\stackrel {\mathrm{dis}}{\sim}$) representing pixel velocity $v_{r,g,b}$. The variable $v$ can either be taken as the velocity of the particle or the group velocity of the corresponding matter-wave in consumption (as the anticipated matter-wave). Since the particle velocity $v < c$ for a massive particle according to SR, phase velocity of matter-waves would now exceed $c$ in terms of its consumed or anticipated version distributed into pixel velocity $v_{r,g,b}$, or

\[\because c-c_{\mathrm{cons}} \equiv v_{r,g,b} \ll \{ c ,v \} \;,\]
\[ \ \ \ \ \ \ \ \ \ \ \ \ \  \ \ \ \ \ \ \ \ \ \ \ \ \ \ \ \ \ \ \therefore |v_{\mathrm{p}}| > c_{\mathrm{cons}} \stackrel {\mathrm{dis}}{\sim} v_{r,g,b}\ \ \ \mathrm{or},  \ \ |v_{\mathrm{p}}| \geq |c|\: , \ \ \ \ \ \ \ \ \ \ \ \ \  \ \ \ \ \ (2.4\mathrm{c}) \]
\noindent or
\[\ \ \ \ \ \ \ \ \ \ \ \ \ \ \ \ \ \ \ \ \ \ \ \ \ \  \ \ \ \ \ \ \ \ \ \therefore \ |v_{\mathrm{p}}| \geq |\xi c| \ \ \mathrm{where,} \ \: \xi \geq1 \; , \ \ \ \ \ \ \ \ \ \ \ \ \ \ \ \ \ \ \ \ \ \ \ (2.4\mathrm{d}) \]
\noindent explicitly
\[\ \ \ \ \ \ \ \ \ \ \ \ \ \ \ \ \ \ \ \ \ \ \therefore \ v_{\mathrm{p}} = |v_{\mathrm{p}}| + c = \xi c + c \ , \ \ \ \ \ \ \ \ \ \ \ \ \ \ \ \ \ \ \ \ \ \ \]
\[\ \ \ \ \ \ \ \ \ \ \ \ \ \ \ \ \ \ \ \ \ \ \ \ \ \ \ \ \ \ \ \ \ \ \ \ \  \therefore \ v_{\mathrm{p}} \gg c \  \ \mathrm{and}, \ \ \nexists\ v_{\mathrm{p}} > c \ . \  \ \ \ \ \ \ \ \ \ \ \ \ \ \ \ \ \ \ \ \ \ \ \ \ \ \ \ (2.4\mathrm{e}) \]
\noindent As we can see, $v_{\mathrm{p}}$ via $|v_{\mathrm{p}}|$ in (2.4e), approaches further over the $c$-limit (or, values of $\xi c + c \: $) when the particle velocity is in the relativistic range of anticipatory variables of events of space and time. One may call this pase velocity between bars, $|$, as $|v_{\mathrm{p}}|$, the \emph{anticipated} or \emph{virtual part of phase velocity} $v_\mathrm{p}$. Each `group velocity' to the argumentum mentioned above, is defined by the following equation and a line integral over a unit circle
\[ v_\mathrm{g} \equiv \frac{\partial \omega}{\partial \textbf{\emph{k}}} \; , \ \mathrm{if} \; \oint_{C} f ({\textbf{\emph{k}}})\; \mathrm{d}k = \frac {2\pi \jmath}{2\pi f(\lambda)} = \frac {\jmath}{\lambda \leftrightarrow \lambda_{\mathrm{bend}}} \stackrel{\forall \textbf{\emph{k}}}{ \longrightarrow}  \exists \, f(\tilde{\nu})\equiv \tilde{\nu}_{C}=\frac {\jmath}{\lambda \leftrightarrow \lambda_{\mathrm{bend}}} \]
\[ \mathrm{where}, \  \ \jmath = \sqrt{-1} \, , \ \ \ \ \ \ \ \ \ \ \ \ \ \ \ \ \ \ \ \ \ \ \ \ \ \  \ \ \ \ \ \ \ \ \ \ \ \ \ \ \ \ \ \ \ \ \ \ \ \ \ \ \ \ \ \ \ \ \ \ \ \ \ \ \ \ \ \ \ \ \ \ \ \ \ \ \ \ \ \ \ \ \ \ \ \  \ \ \ \ \ \ \   (2.4\mathrm{f}) \]
\noindent $\omega$ is the wave's angular frequency and $\textbf{\emph{k}}$ is the angular wavenumber~\cite[xvii]{06-Wiki}.

Let $\tilde{\nu}$ represent spectroscopic wavenumber function, thus $\tilde{\nu}$ remains as a wavenumber from a unit circle $C_{i}$ closing an integral over $\textbf{\emph{k}}$ in form of $\tilde{\nu}_{C}$. The latter is said to correspond to reciprocal meters (or m$^{-1}$) of unoccured curvatures of $k$ in their sequence form $x_{k_{i,...,n}}^{-1}$ for all Gaussian unoccured curvature states. For the incorporated line integral, see the concept in virtue of~\cite[xxi]{06-Wiki}.

In relevance to project Objectives \#4, \#6, we firstly consider a 2D-object to be recognizable in front of a photographic film positioned onto a light reflector, so to record and compute pixel-velocity when in motion. Secondly, we remove the film and use an actual 3D-object identical to the 2D-object only with a depth dimension onto a new film. Subsequently, we compute the object's motion and pre-motion (stationary moments/mobile moments) by capturing frames of the scene, accumulating them, thus ready for the analysis stage after both films are developed. Now, the motion-stretch on curvature $k$ is preceded via the following equation benefiting from composite function equation $v_{k < 0}  = v_{r,g,b} {\: \rm o \:}{xy_{r,g,b} }t^{-1}$, as in Eq.$\,$(2.4a), promoting time in the context of SR to UR,
\[x_{k<0} \mapsto g(f(x_{k<0})) \equiv x_{k_{i,...,n}<0} = \frac{{v_{r,g,b} \int_{\mathbf{|A|}}{{\rm d} x_i {\rm d} y_i } }}{{c^2 \int {{\rm d}t_i}}} = \frac{{v_{k<0} \: |\mathbf{x}|_{x^{\mu},x_\mathrm{C}} |\mathbf{y}|_{r,g,b} \cos \theta  }}{{c^2 u\left( t \right)}}\; \; \; \; \; \;  \]
\[\mathrm{where},\; \; u\left( t \right) = 1{\mathop{\; \rm s}\nolimits}. \; \; \; \; \; \; \; \; \; \; \; \; \; \; \; \; \; \; \; \; \; \; \; \; \; \; \; \; \; \; \; \; \; \; \; \; \; \; \; \; \; \; \; \; \; \; \; \; \; \; \; \; \; \; \; \; \; \; \; \; \; \; \; \; \; \; \; \; \; \; \; \; \; \; \; \; \; \; \; \; \; \; \; \; \; \; \; \;(2.5)\]
\vspace{-2pt}

\noindent Whereas for all cases special to Subhyp.$\,$3.1.2., \S\ref{section3}, involving total Gaussian curvature (surface integral) of geodesic triangles for a one-dimensional motion stretch in the context of GR to UR,
\vspace{-14pt}
\begin{flushleft}
\[x_{K_{i,...,n}}=\frac{{x_{\mathrm{cons}}  \int_{\mathbf{|A|}} {{\rm d} x_i {\rm d} y_i } }}{{
s^{2}\rightarrow \int 2s \;\mathrm{d}s \rightarrow{\emph{g}_{i,...,n}(c \int {\rm d}t_i}})^2} = \frac{{v_{K_{i,...,n}} |\mathbf{x}|_{x^{\mu},\lambda_{\mathrm{bend}}} |\mathbf{y}|_{r,g,b} \cos \theta  }}{\emph{g}_{n}\!\mid \!\emph{g}_{i}\ {c^2 \; u\left( t \right)'\!\mid \! u(t)}} \] \[ =\frac{{x_{\mathrm{cons}}\overrightarrow{xy}}}{2\emph{g}_{n+1}(c^2\:tt_P \mid u(t))} \; ,\; \int 2s \;\mathrm{d}s = \emph{g}_{\mu \nu } \mathrm{d}x^\mu  \mathrm{d}x^\nu + \mathfrak{c}\left|\mathfrak{c}\rightarrow\mathrm{d}\tau^2 \right., \; \mathrm{where} \; \; \; \; \; \; \; \; \; \; \; \; \] \[ \ \ \; \; \; \; \; \mathrm{d}\tau^2 = \eta_{\mu\nu}dx^\mu dx^\nu \:,\  u\left( t \right)'\in[tt_P,t(u(t))],\; \;\sum\limits_{i = 1}^3 {\theta _i }  = \pi  + \int {\int_T {K{\rm d}A} } \;. \; \; \; \; \; \; \; \;  (2.6)\]
\end{flushleft}\begin{flushright} \vspace{-12pt}
\end{flushright}

\noindent  This motion stretch framing $x_{K_{i,...,n}}$, is of future product relation (3.7), component $\mathcal{V}_\mathfrak{V}s^{-2}$, acting one-dimensional satisfying an observation-scope function $scope\left( o \right)$, on \emph{infinitesimal points of length and time} for at least one non-subatomic body relative to channeled in/out quantum body. It also carries out $t_P  = \sqrt {\frac{{\hbar G}}{{c^5 }}}  \approx 5.39121\left( {40} \right) \times 10^{ -44} {\rm s}$ via consumed time $t_{\mathrm{cons}}$ from Eq.$\,$(2.2), which is too inclusively symmetric to unit of time function $u(t)$, between space-time past and future events. Planck time $t_P$, is the time it would take a photon travelling at the speed of light in a vacuum to cross a distance equal to the Planck length $l_P  = \sqrt {\frac{{\hbar G}}{{c^3 }}}  \approx 1.61624\left( {12} \right)\times 10^{ -35} {\mathop{\rm m}\nolimits}$. In the Planck time and length unit relations, $\hbar$ stands for Dirac's constant, Planck's constant divided by $2\pi$, $G$ represents Newton's gravitational constant, $c$ is the speed of light in a vacuum and $t_P$, is in seconds~\cite[xi]{06-Wiki}. This type of measurement involving the smallest time step of quantum mechanics here, Planck time $t_P$, inclusively symmetric and relative to unit of time interval, 1 s from classical mechanics, applies to cases of \emph{non-subatomic bodies channeling quantum bodies}, subject to Subhyp.$\,$3.1.2. The multi-dimensional form covering an observation scope of more than four dimensions, benefits from the righthand component of product of observation scope in product relation (3.7), that is, $\mathcal{V}_\mathfrak{V}s^2$. This form representing higher dimensions of non-Euclidean states for Eq.$\,$(2.6) in one's observation scope and its perception of colour, shape and motion (subjects of~\cite[xii]{06-Wiki}) channeling-in quantum bodies, can be expressed in terms of a `Sample problem' exemplified in Appendix I, and Eqs.$\,$(2.5), (2.6). The basics are subject to the subsequent Section, in conjunction with Subhyp.$\,$3.1.2 and Fig.$\,$3.1.

\section{The theory of unified relativity}
\label{section3}

\noindent Consider the next two Tables containing assumptions that could be tested as facts based on one's approach (see e.g., Refs.~\cite{{06-Wiki,{26-Starman},{27-Kippers},{28-Schlessingerman}}}), \\

\vspace{2pt}
\footnotesize
\hspace{-8pt}--------------------------------------------------------------------------------------------------------------------\\
\begin{tabular}{p{1in}p{1.1in}p{2.1in}}
\textit{Typical housefly} & \textit{Typical human} & \textit{Mean comparison of range by ratio in \%             } \\
--------------------------  \vspace{1pt}Adult size: 5-7 mm = 0.005-0.007 m & --------------------------\vspace{1pt} Adult size: 1.5-1.8 m & ---------------------------------------------------\vspace{1pt}$100\%-0.003636\%\approx 99.636\%$ \\
\vspace{1pt}Adult mass: 12 mg = 0.000 012 kg & \vspace{1pt} Adult mass: 61-70 kg & \vspace{1pt}100\%$-$1.832061$×10^{-5}\%\approx$99.9999817\% \\
\vspace{1pt}Lifespan: 7 days = 604800 s & \vspace{1pt} Lifespan: 40-75 yr = $1.262277\times10^{9}$-$2.366769\times10^{9}$ s & \vspace{1pt}$100\%-0.03333107\%\approx 99.9666689\%$
\\
\end{tabular}
\begin{center}
\hspace{-4pt}---------------------------------------------------------------------------------------------------------------------
\end{center}
\vspace{-3pt}
\noindent{\footnotesize{\textbf{Table 3.1.} Body size, mass and decay parameters of a typical housefly and human including their value comparisons (see also p.$\,$3 of Ref.~\cite{26-Starman}).
\bigskip

\normalsize The comparisons made between the exemplified bodies in Tab.$\,$3.1, are measured in percent, where 100\%, is the default assumption made for the mentioned parameters given to a human in average. For instance, (61+70)/2 kg for 100\% thus computing \emph{p}\% for the fly's mass by ratio, which gives $\approx 1.83206107×10^{-5}\%$, resulting 99.9999817\%, for the comparison of bodies' mass. Note that, in this table, the mean value of human lifespan, is based on an average point of the founding made upon `life expectancy' between years, 1955-2005~\cite{06-Wiki}. The size proportionality between the two bodies in Tab.$\,$3.1, eventually relates to the vision problem on space-time curvatures in respect to body size. Thus, the occupied space by either bodies' mass despite of their lifespan, measured in terms of volume, shall lead to product resultant ${( \frac{ \cal{V}_{\mathfrak{V}} }{s^{2}}, {\cal{V}_\frak{V}}s^{2} )}$ and $[2\pi \partial \vec{r} \ \; 2n\pi \partial \vec{s}^{\; 2n-1}]$. These `product resultants' are signified by the forthcoming concept and solution details, leading to relations (3.7), (3.8), prompting events using methods of discrete mathematics, tensors and systematic symmetrical uncertainties for space-time dimensions in form of partial differentiation, respectively.

In content, Tab.$\,$3.2 remains significant when events are studied for either bodies' visual ability based on value comparisons. For instance, when trying values of relativistic gamma $\gamma$ in rows \#1-3, and studying the notion of length contraction in SR, supposing $\gamma$ approaches a Lorentz factor of 1, indicating a man's body nearing a non-relativistic state (almost stopping nearby), the view of his/her body gets pregnantly gigantic to the fly's vision. (Let the notion of `man' in UR, represent `human' for either gender male/female in expressions)
\vspace{-18pt}
\begin{flushleft}
\noindent
\end{flushleft}
\footnotesize \hspace{8pt}---------------------------------------------------------------------------------------------------------------------\\
\hspace{6pt}\begin{tabular}{p{4.6in}}
\textit{Bodies' visual observation in motion relative to each other, total traverse and occupation of space-time based on the new UR theory} \\
---------------------------------------------------------------------------------------------------------------------\\ \vspace{2pt}$\begin{array}{l} {{\rm {\mathcal B}}\left(\ell \right)_{{\rm fly}} \ll {\rm {\mathcal B}}\left(\ell \right)_{\rm man} ,{\rm if\; {\mathcal B}}\left(\ell \right)\to \ell \wedge \left({\rm {\mathfrak V}}\wedge {\rm {\mathcal V}}\right),} \\ {\therefore {\rm {\mathcal V}}_{{\rm {\mathfrak V}}} \cdot \ell _{{\rm fly}}^{-1} \ne {\rm {\mathcal V}}_{{\rm {\mathfrak V}}} \cdot \ell _{{\rm man}}^{-1} ,} \\ {\therefore f\left(x_{{\rm fly}\left(o\right)} \right)<\frac{\ell _{{\rm man}} \gamma }{\mathop{\lim }\limits_{\delta \gamma \to 1} \gamma \to 1} \ge \ell _{{\rm man}} \gamma =\ell '_{{\rm man}} \to \int _{>1}^{1}\int \ell _{{\rm man}} \gamma \mathrm{d}\gamma \mathrm{d}\ell   ,{\rm \; }\gamma =\frac{1}{\sqrt{1-v^{2} c^{-2} } } >1} \\ {} \end{array}$         \newline \newline $\begin{array}{l} {M_{{\rm fly}} \ll M_{{\rm man}} ,{\rm if\; }M\wedge \left({\rm {\mathfrak V}}\wedge {\rm {\mathcal V}}\right),} \\ {\therefore M_{{\rm fly}} \cdot {\rm {\mathcal V}}_{{\rm {\mathfrak V}}}^{-1} \ne M_{{\rm man}} \cdot {\rm {\mathcal V}}_{{\rm {\mathfrak V}}}^{-1} ,} \\ {\therefore \rho _{{\rm fly}\left(o\right)} \gg \rho _{{\rm man}\left(o\right)} }\; ,   \\ \because M_{\rm man} \gg M_{\rm fly}\;. {} \end{array}$  $\left. {\underline{\begin{array}{l} {M\buildrel\wedge\over= {\rm body\; mass\; \; \; \; \; \; \; \; ,\; }{\rm {\mathcal B}}\left(\ell \right)\buildrel\wedge\over= {\rm body \;size\; function}} \\ {\mathcal B}\buildrel\wedge\over= {\rm body\; state;\; body} \\ {\rho_{\mathcal{B}(o)}\buildrel\wedge\over= {\rm \; observation \; density\; of \; relative \; body} \; \; } \\ {{\rm {\mathcal V}}\buildrel\wedge\over= {\rm body\; volume\; \; \; \; \; \; ,\; }\ell \buildrel\wedge\over= {\rm body\; size\; or\; length}} \\ {{\rm {\mathcal V}}_{{\rm {\mathfrak V}}} \buildrel\wedge\over= {\rm body\; volume\; utilized\; with\; visual\; system\; {\mathfrak V}}} \end{array}}}  \!\right| $\newline     \\
$\begin{array}{l} {T_{{\rm fly}} \ll T_{{\rm man}} ,} \\ {\left(vT\right)_{{\rm fly}} \ne \left(vT\right)_{{\rm man}} \vee \infty } \\ {{\rm if\; }T\wedge \forall x_{{\rm {\mathfrak V}}} \in \prod _{i}^{\infty }scope\left(o\right) ,} \\ {\therefore v_{{\rm fly}\left(o\right)} >v_{{\rm man}\left(o\right)} } \end{array}$         $\left. {\underline{\begin{array}{l} {T\buildrel\wedge\over= {\rm body\; lifetime\; \; \; \; ,} \; v\buildrel\wedge\over= {\rm velocity}} \\ {o\buildrel\wedge\over= {\rm observation\; \; \; \; \; \; \; ,\; }\; x\buildrel\wedge\over= {\rm distance}} \\ {scope\left(o\right)\buildrel\wedge\over= {\rm observation\; scope\; function}} \end{array}}}  \!\right| $\newline                     \newline  \\---------------------------------------------------------------------------------------------------------------------
\end{tabular}

\vspace{4pt}
\noindent{\footnotesize{\textbf{Table 3.2.} The comparison on observation parameters that are generalized from the new unified relativity in the following notes. This table has inheritance relationship with Tab.$\,$3.1.}\\

\normalsize In other words, `length expansion', and not other scenarios indicating `length contraction' and `proper length' of the dimensions of the body with mass $M_{\mathrm{man}}$ and size $\ell_{\mathrm{man}}$, for the relative body with mass $M_{\mathrm{fly}}$ and size $\ell_{\mathrm{fly}}$, paradoxically. Another factor to consider is of the deduction made in rows \#4-7. For instance, assume the division, $70 \: {\rm kg}/1.5\: \rm mm^{3}$ for $\rho_{\mathcal{B}(o)}$ as $\rho _{{\rm fly}\left(o\right)}$, where the denominator denotes an example of fly's vicinity as an occupiable space for its body and observation, relative to a man approaching it. For now, ignore precise relativistic measurements and just contemplate speculative relativity. Of course in this frame of observation, objects in terms of their space-time curvature such as the man approaching the fly, appear denser than they seem defining the states of observation density $\rho _{{\rm fly}\left(o\right)}$. For the man being observed, regardless of its exaggerated and expanded shape, the division's result $4.66666667 × 10^{10} \: \rm kg \: \rm m^{-3}$, is due to the \emph{observed curves and distance of approach} as being a lot closer than they seem to the fly's visual interpretation. Hence the densest or `heaviest compared to body size' bending curves of space and time, is the density of observation state on the human subject. That is why, it is most presumptuously correct for the small body with visually well-equipped spectacles, fast motion detection of a body approaching it in the scene, becomes extruded in many ways (angles of body approach), and densest in terms of observation (not the actual density of the body). Accordingly, the content of Tab.$\,$3.2, is subject to the subsequent notes and consequently, Major to \S\ref{section5}.

Note that, for one's way of interpreting `decay' here, should not be mistaken with `particle decay' described as e.g., `an elementary particle transforming into other elementary particles' in particle physics. In fact, since the subjects in the scene are of non-subatomic type, this type of decay would therefore relate to the lifespan of a body and not its afterlife-decomposition (if deemed biological). On the contrary, if such bodies are utilized with subatomic systems, such as the proficiency of a fly's visual system and lenses mapped to a man's visual system artificially/organically/biosynthetically, the study of particle transformation into other particles becomes relevant in practice.\\

\noindent \textbf{Concept and theorem 3.1.} \emph{Imagine a visual system tends to see the human world from the fly's eyes. Compared to a human, imagine a man to remain as a New York City's skyscraper relative to the size and dimensions of the fly's body. Thus, one could envisage the objects within the scene to appear in the context of GR, involving gravity familiar from Newtonian gravitation studying curvature, whilst for the trajectory of a moving particle e.g. specific charge $q$, appearers on the fly's eye surface to the scene's environment. It as at this point, space-time events connect with the calculations made on proper time $\tau$ in the context of SR, remaining ourselves with the following self-evident postulates} \\

\noindent From Einstein's SR theory published in 1905~\cite{25-Einstein}, \\

\noindent \textbf{Statement 3.1. }\emph{The laws of physics are the same for all non-accelerating observers}. \\

\noindent \textbf{Statement 3.2. }\emph{Space and time form a 4-dimensional continuum}. \\

\noindent and from Einstein's GR theory~\cite{35-Einstein},\\

\noindent \textbf{Statement 3.3. }\emph{The global Lorentz covariance of special relativity becomes a local Lorentz covariance in the presence of matter. The presence of matter `curves' space-time, and this curvature affects the path of free particles (and even the path of light)}.\\

Having in context the following consequences from SR's Postulate 3.1, and the postulate made on the speed of light $c$, as a universal constant in~\cite{55-Prosper} (doubtably ?, due to e.g., the experiment conducted by Mojahedi discovering a loophole in Einstein's law~\cite{43-Mojahedi}),\\

\noindent \textbf{Consequences of SR: }\emph{`Events that may be simultaneous for one observer can occur at different times for another. This leads to length contraction and time dilation, the slowing down of time in a moving frame. Every observer has her own personal time, caller proper time. That is, the time measured by a clock at the observer's location. Two observers, initially the same age as given by their proper times, could have different ages when they meet again after travelling along different space-time paths.'}\\

\noindent one could therefore by associating the global Lorentz covariance stated in Statement 3.3, and based upon the above-propounded Postulates 3.1, 3.2, and the Concept 3.1, paradoxically hypothesize \\

\noindent \textbf{The non-subatomic bodies' unifying hypothesis 3.1.}\emph{ For two non-sub atomic bodies occupying space with a great difference in body-size having diverse physical properties utilized with a visual system, relative to other bodies occupying space in our universe, observe events differently. No matter the laws of physics describing their course of motion, decay and time, the relativity for these bodies are unified in one and only one frame of synchronistic observation. That is, allowing one body to observe an event as same as the other.}\\

\noindent The following proof by deduction which interactively uses the basis of \emph{deduction theorem} from mathematical logic, leads to the current hypothesis into two forms of Subhypotheses:\\

\noindent \textbf{The non-subatomic bodies' unifying subhypothesis 3.1.1.}\emph{ Non-subatomic bodies with relative size to other bodies, the visually utilized small body observing events should thus have a spacelike difference, prior to a timelike difference compared with a visually utilized giant body from one space-time plane to another.}\vspace{8pt}

\noindent \textbf{Proof by deduction. }\emph{It is provable by deduction to Subhyp.$\,$3.1.1, to put in place an assumption for non-subatomic bodies with relative size to other bodies, positioned in a non-Euclidian system. One body, utilized with a visual tool and function, if, significantly smaller than the remaining bodies in motion relative to still bodies occupying either system of geometry, Euclidian/non-Euclidean, the utilized body thus has spacelike difference $\Delta s^{2}<0$. This is prime/prior to timelike difference $\Delta s^{2}>0$ in observing events compared to a significantly giant body in the same space possessing time $t$ applied to a time dilation, now of space-time interval $s$, for the same observable two events in planes of simultaneity.}\\

\noindent Ergo, in unification comes about the subsequent Subhypothesis:\\

\noindent \textbf{The non-subatomic bodies channeling quantum bodies' unifying subhypothesis 3.1.2.}\emph{ Non-subatomic bodies channeling out quantum bodies for bodies who observe the emission, all bodies relative to this quantum frame possess visual time difference subsequent to their motion, and last to their occupiable space. This occupiable space is the promising space reserved for the observer in motion/still, and remains intact to the individual's visual time differentiations.}

\noindent \textbf{Proof by deduction. }\emph{The visually utilized and significantly small non- subatomic bodies in the unifying Hyp.$\,$3.1, likely possess spacelike or past-timelike which define themselves in a geodesics states of GR; such that, all subatomic bodies in the context of non-gravitational state i.e. quantum state, are significant and greater in their force's relative strength, $10^{36}$. Assuming an exemplar to this comparison of states is, a photon being emitted from the visually utilized small body, compared to a device measuring the difference of time between two events for all of the visually utilized bodies, measures the unit of time interval, $1 \: \mathrm{s}$, relative to Planck time $t_P$. For the small body in this frame, the body deals with curves of space with relative strength of `$\:1\!$' for gravitation, while all systems of geometry hold in existence to their occupants, the \{smallest bodies and giant bodies of the same frame$\,$\}, engage with the three stronger forces against the weak force of gravitation in unification. These visual involvements and time anticipations are perpetual to a state of space-time change and differentiation, in regard to the body observing events in an n-dimensional space-time geometry. That is being relative to firstly, an occupiable space, secondly, motion and thirdly, visual time difference as follows}:\\

\noindent Here goes an axiomatic predicate for all measurable non-subatomic bodies in our universe,
\[\left\{\forall \: body(\mathcal{B})  \in sys^{\mathcal{U}} \; \left| \; small \: body \: \ll  \: big \: body \: \ll giant \: body \; \equiv \mathcal{B}^{>0}\ll\ll\mathcal{B^{<\infty}} \right. \right\}  \]
\[, \; \mathcal{B}\rightarrow m,\mathcal{V},\mathfrak{V} \; \; \; \; \; \; \; \; \; \; \; \; \; \; \; \; \; \; \; \; \; \; \; \; \; \; \; \; \; \; \; \; \; \; \; \; \; \; \; \; \; \; \; \; \; \; \; \; \; \; \; \; \; \; \; \; \; \; \; \; \; \; \; \; \; \; \; \; \; \; \; \; \; \; \; \; \; \; \; \; \; \; \; \; \; \; \; \; \; \; \; \; \; \; (3.1) \]
now incorporating the previous axiomatic predicate into
\[\tau _{\mathcal{B}^{>0}}  \equiv \sum {\Delta t_{sys} }  = t_{B^{<\infty}}  - t_{\mathcal{B}^{>0}}  = t_{B^{<\infty}}  - \int\limits_{a < total{\rm \:  }t}^{b \le total{\rm\:  }t} {t_{B^{<\infty}} } \; \mathrm{d} t \; \; \; \; \; \; \; \; \; \; \; \; \; \; \; \; \; \; \; \; \; \; \; \; \; \; \; (3.2)\]
\noindent it is deducible in the realization of Pred.$\,$(3.1), relation $\,$(2.1b), to some curvature state $k$ satisfying Eq.$\,$(2.1d) of \S\ref{section2}, for body ${\mathcal{B}^{>0}}$'s proper time $\tau$, ensues
\[\therefore \tau _{\mathcal{B}^{>0}} = n\: {\rm s}  - \mathop  \sum \limits_{i > 0}^{j \lessapprox 1} i\left( {n} \right) \: \mathrm{s}\equiv n \; \mathrm{s}  + \mathrm{cons} \; {\rm s}\; , \; i,j \in\mathbb{ R}^{+} \wedge \stackrel{\propto}{=} \epsilon \; , \; t \in sys^{\mathcal{U}} \; \;  \; \; \; \; \; \; \; \; \; \; \; \; \; \; (3.3)\] \vspace{-6pt}

\noindent where body function $body(\mathcal{B})$ in Pred.$\,$(3.1), specifies $x$, as an arbitrary matter within the non-subatomic spatial system powered by universe or, $sys^{\mathcal{U}}$ in question, implies to mass $m$ and volume $\mathcal{V}$ (taking up space), utilized with a visual system $\mathfrak{V}$, for the involved bodies. This could be called \emph{the theory of unified relativity} (UR) or, `\emph{the special and general theories of relativity in unification}'. One could thereby for two bodies with an approximate body-size with similar physical properties in this unified relativity, and using geometric congruency between shapes for their inertial frames, articulate, \vspace{-3pt}
\[\tau _{\mathcal{B}^{<\infty}\cong\mathcal{B'}^{<\infty}}= n  - \mathop  \sum \limits_{i > 0}^{j \ll 1} i\left( {n } \right)\; \mathrm{s} \approx  0\;{\mathop {\mathrm{cons}}\nolimits} \; \mathrm{s} ,\; \mathrm{iff} \;\mathcal{B,B'}\rightarrow \{m,\mathcal{V},\mathfrak{V}\cong \wedge \neq m',\mathcal{V}',\mathfrak{V}'\}\; \; \; \; \; \; \; \; \; \; \; \; \; \; \; \; \; \; \; \]
\begin{flushright}\vspace{-15pt}
(3.4)\end{flushright}
\vspace{-3pt}
Perceivably, from the following conceptual solution and all inertial frames for accelerative/non-accelerative bodies down to a point of their vector constituent (unit vector), would still appreciate Newton's 1st and 2nd laws of motion regardless of the effect of the force on the motion of bodies (i.e. rigid body dynamics, Ref.~\cite{06-Wiki}). Thus acceleration in broader scale of measurements is the product, which is in this case `irrelevant'. In other words, \emph{the equivalence of all inertial frames despite of the quality of Newton force involvement, holds good in the unifying theory of relativity once a solution to concept 3.1, the electrodynamics--on the kinematics of the rigid body} (p.$\,$2, accompanying \S1, \S2, of Ref.~\cite{25-Einstein})\emph{, is tried out for all Lorentzian type against Galilean type transformations in classical mechanics $\ldots$}\\

\noindent \textbf{Solution to concept 3.1.} Since we already know the Lorentz force equation must be modified according to SR solely when particle speeds approach the speed of light~\cite[xiii]{06-Wiki}, cases of `luminosity' concerning the Proof on Subhyp.$\,$3.1.2, the amount of energy radiated from the fly's eye surface constituent per unit time, suggests
\[
\frac{{\mathrm{d}\left( {\gamma m{\bf v}} \right)}}{{\mathrm{d}t}} = {\bf F} = q\left( {{\bf E} + {\bf v} \times {\bf B}} \right),
\]
\noindent where \vspace{-8pt}\[
\ \ \ \ \ \ \ \ \ \ \ \ \ \ \ \ \ \ \ \ \ \ \ \  \ \ \ \ \ \ \ \ \ \ \ \ \ \ \ \ \gamma \mathop  = \limits^{{\rm def}} \frac{1}{{\sqrt {1 - \left| {\bf v} \right|^2 c^{ - 2} } }} \ ,
 \ \ \ \ \ \ \ \ \ \ \ \ \ \ \ \ \ \ \  \ \ \ \ \ \ \ \  \ \ \ \ \ \ \ \ (3.5\mathrm{a})\] \\
\noindent is called the Lorentz factor and $c$ is the speed of light in a vacuum. \textbf{F} is the force in newtons; \textbf{E} is the electric field in volts per meter; \textbf{B} is the magnetic field in teslas; $q$ is the electric charge of the particle in coulombs; $\mathbf{v}$ is the instantaneous velocity of the particle in meters per second and `$\times$' is the cross product. This relativistic form is identical to the conventional expression of the Lorentz force if the momentum form of Newton's law, $F= {\rm d}p \: {\rm d}t^{-1}$, is used, and the momentum $p$ is assumed to be $p = \gamma mv$. The change of energy due to the electric and magnetic fields, in relativistic form, is simply
\[ \ \ \ \ \ \ \ \ \ \ \ \ \ \ \ \ \ \ \ \ \ \ \ \ \ \ \ \ \ \ \ \ \ \ \ \ \ \ \ \
\frac{{\mathrm{d}\left( {\gamma mc^2 } \right)}}{{\mathrm{d}t}} = q{\bf E} \cdot {\bf v}. \ \ \ \ \ \ \ \ \ \ \ \ \ \ \ \ \ \ \ \ \ \ \ \ \ \ \ \ \  (3.5\mathrm{b})
\]

\vspace{-2pt}
\noindent The \emph{change in energy} depends only on the electric field, and not on the magnetic field. Thereby, one via charge equation \(Q=It\) can speculate upon the physical bodies within the scene under observation:\\

\vspace{2pt}
Let \(Q=It\) apply for specific charge $q$, time $t$ at a given instant in path $P$ of a timelike varying field for some expected radiation (non-ionizing) from the retarded scalar (timelike) proving biovielectroluminescence to appear as
\begin{flushleft}
\[\forall \: q \in Q, \: \exists \: \mathrm{d}t \in \tau \left| \tau = \int_P \sqrt {\mathrm{d}t^2 - \mathrm{d}x^2c^{-2} - \mathrm{d}y^2c^{-2} - \mathrm{d}z^2c^{-2}} \ , \right.\]
\[\therefore t_{q} \equiv \int \mathrm{d}t \; \mathrm{for} \; q \; \mathrm{of} \; x^{2} \;,\; t_{q} = \frac {\tau}{\sqrt{1-{v_{q}^2}{c^{-2}}}} \ ,  \]
\[ \therefore t_{q} \equiv \int \mathrm{d}t \; \mathrm{for} \; x_{q}^{2} \rightarrow ds^{2} \; \mathrm{of} \; \emph{g}_{\mu\nu} \; \mathrm{d}x^\mu \; \mathrm{d}x^\nu \ , \] \[  \ \ \ \ \ \ \ \ \ \ \ \ \ \ \ \ \ \ \ \ \ \ \ \ \ \ \ \ \ \ \ \ \because \mathrm{d}s^{2} = \; \emph{g}_{\mu\nu} \; \mathrm{d}x^\mu \; \mathrm{d}x^\nu\ \; \mathrm{for} \;  \mathcal{B^{<\infty}}\gg\gg\mathcal{B}^{>0}\; . \ \ \ \ \ \ \ \ \ \ \ \ \ \ \ \ \ \ \ \ \ \  (3.5\mathrm{c}) \]
\end{flushleft}
\vspace{3pt}
\noindent which relates to derivation principle on any space-time for the timelike paths travelled by giant bodies here, the invariant interval $\mathrm{d}s$ between events with an incremental coordinate separation $\mathrm{d}x^{\mu}$, becomes proper time $\tau$ based on $\tau \stackrel{\mathrm{def}}{=}s$, in the context of GR~\cite[xiv]{06-Wiki}. While having giant body ${\mathcal{B}^{<\infty}}$ in picture, small bodies of ${\mathcal{B}^{>0}}$ relative to ${\mathcal{B}^{<\infty}}$, are too in this picture, thus, ${\mathcal{B}^{<\infty}}$ described in the context of SR satisfies moments of $t_q$. Ergo, lightlike (null) path followed by an expected probabilistic photon $h\nu$ relative to particle with specific charge $q$, both the latter and former are channeled by non-subatomic body ${\mathcal{B}^{>0}}$. Hence, passage of proper time to timelike relative to spacelike are in simultaneity for the separated events (recall Pred.$\,$2.1c, Eq.$\,$2.1d, \S\ref{section2}). Thence proving thus far
\[\forall \: q\wedge h\nu \in E \wedge \textbf{F}, \: \exists \: \frac {(y \vee z)_{h\nu}}{t_{h\nu}}=c \wedge t_{q} \in \tau \left| \tau = \int_P \sqrt {\mathrm{d}t^2 - \frac {\mathrm{d}x^2}{c^{2}} - \frac{\mathrm{d}((z\vee y)t_{h\nu})^{2}}{\mathrm{d}(y\vee z)^2}} \right. \]

\[E_q=m_{q}c^2={\gamma m_{0}c^2 } \rightarrow \left\{q{\bf E} \cdot {\bf v}\right\}t_{q} \mathbf{ }\;,\; E_{h\nu}=h\nu \rightarrow \left\{q{\bf E} \cdot {\bf v}\right\}t_{h\nu}| t_{h\nu} =\frac{ct_{\mathrm{cons}}t_q}{\lambda} \]

\noindent whereon for all of these components, a conventional proof formulating a\emph{ path integral in consumption} suffices the current concept. Let this path integral in limit over time be $P\rightarrow \mathrm{cons}$, formal to the involvement of time phase $t_{h\nu}$ for either coordinate $y$ or $z$ on photon $h\nu$ separated in space-time events. Whilst the latter in separation being relative to $x$ on charge $q$ for charge time $t_q$, now derives
\[\therefore \tau = \int_{P\rightarrow {\mathrm{cons}}} \sqrt {t_q^2 - \frac {x_q^2}{c^{2}} -
\frac{{\lambda ^2 t_{h\nu }^2 }}{{c^2 t_{\mathrm{cons}}^2 }}- \frac{t_q{v_{r,g,b}^2 }}{{c^2 }}
}\ , \ \ \ \ \ \ \ \ \ \ \ \ \ \ \ \ \ \ \ \ \ \ \ \ \ \ \ \ \ \ \ \ \ \ \ \ (3.5\mathrm{d}) \]

\noindent which holds good with a pixel velocity $v_{r,g,b}$ from Eq.$\,$(2.1a), \S\ref{section1}, indicating a displacement field on charge $q$ between points of area $A$ covering experimental volume points, from A to B (observe Fig.$\,$3.1). The displacement is relative to the area of \emph{biovielectroluminescence phenomenon's emergence} say, from ${\cal{B}^{>\mathrm{0}}}$ as its vector form, $\mathbf{A}_{\cal{B}^{>\mathrm{0}}}$, involving time phase $t_{h\nu}$ relative to $\mathcal{B^{<\infty}}$, establishing
\[
\frac {\mathbf{F}\mathbf{A}_{\cal{B}^{>\mathrm{0}}}}{t_{\mathrm{cons}}}= Ev_{r,g,b}=h\nu v_{r,g,b}  = \nu _L  \cdot Q \; {\mathop{\rm rel}\nolimits} \left( {A_{\cal{B}^{>\mathrm{0}}} } \right){\cal V} = \frac{{\rm d}({\gamma VI)\cal V}}{{\rm d}{A_{\cal{B}^{>\mathrm{0}}} }} = \left\{ {q{\bf E} \cdot \bf{v}} \right\}x_{\mathrm{cons}}
\]
\[\ \ \ \ \ \ \ \ \ \ \  \ = q{\bf E} \cdot {\bf v\cdot v}_{r,g,b} t_{\mathrm{cons}} \;  {\rm in \;   J\:}{\mathop{\rm m}\nolimits} {\mathop{\rm s}\nolimits} ^{ - 1} \; . \ \ \ \ \ \ \ \ \ \ \ \ \ \ \ \ \ \ \ \ \ \ \ \ \ \ \ \ \ \ \ \ \ \ \ \ \ \ \ \ \ \ \ \ \ \ \ \ \ \ \ (3.5\mathrm{e}) \]

\noindent It is written that, charge $q$ travelling in the line of coordinate $(x,t)$ for a timelike property $s^2>0$ from Pred.$\,$(2.1c), \S\ref{section2}, defines energy $E$ to product with pixel velocity $v_{r,g,b}$, or, $Ev_{r,g,b}$ for biovielectroluminescence $\beta_\nu$. This product formation is due to the relative coordinate transformation of charge $q$ to photon $h\nu$ of $\beta_\nu$, and concludes in \S\ref{section5} commencing with Eq.$\:(5.1)$. In that equation, $\nu _L$ is luminance frequency for observable luminance $L$ off of the body's image data.
\vspace{-4pt}
\bigskip
\vspace{4pt}
\begin{flushleft}
\includegraphics[width=24mm, viewport= 0 0 10 50]{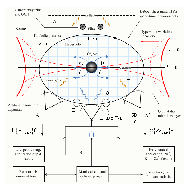}\\
\end{flushleft}

\noindent{\footnotesize{\textbf{Figure 3.1.}  A diagrammatic representation of the angular vision of the fly focused onto the field of objects or scene while electromagnetic phenomena take over, conducting relative $\lambda$ and $\lambda_{\mathrm{bend}}$ measurements. The fly's two large spherical eyes give an angular vision $\approx 360^{\rm o}$. The inclusion of $\beta_\nu$, expands vision over parabolic world-lines. The theoretical study of the scene suggests the observation of space-time curves pertinent to Hyp.$\,$3.1 satisfying Postulate 3.3.}
\smallskip
\vspace{6pt}
\normalsize

One aspect coming about on these elementary components of proof, defines one's visual ability in terms of its scope of observation based on the relativity, qualitative and quantitative factors such as: magnitude, function and dimensions of the involved time-varying magnetic field in space-time geometry. The aspect encompassing volumetric operations on one's vision and scope of observation, relevant to relation (3.5e) for provable Eq.$\:(5.4)$ on biovielectroluminescence $\beta_\nu$, is presented as follows:

\vspace{12pt}
Based on Fig.$\,$2.2\emph{c}, `fly's eye surface constituent', could geometrically in the large represent a hemispherical surface area $A$ divided by two, which is equivalent to the surface of a top of a cylinder, now included within the concept of space-time curvatures in the context of GR for the approximate Riemannian geometry. The overall geometry is defined by the following relation which too benefits from the Pythagorean theorem, Reiman sum, and methods of integration such as the shell method, herein
\[\overline {A_\mathfrak{V} } s^2  = \frac{{4\pi r^2 }}{2} \cdot \mathop {\lim }\limits_{\delta x \to 0} \sum\limits_i^n {s_i^2 \delta x}  = 2\pi \int_a^b {r^2 } {\rm d}\: x\bigcap\nolimits_{ - \infty }^{ + \infty } {s_i^2 } \left| {f\left( r \right)\bigcap\nolimits_{ - \infty }^{ + \infty } {f\left( s \right)_i }  = \left( {\vec r,\vec s} \right)_i} \right.\] \vspace{-20pt}
\begin{flushright}(3.6a)\end{flushright}
\noindent where $\vec r$, represents radius vector and $\vec s$, represents space-time vector relevant to Cartesian spatial coordinates $x,y,z$ including time as a vector (`time' being directional as `time vector' Ref.~\cite{20-David}), $\vec t$, inclusively. Distance $x$ for radius $r$ in terms of defining the geometry of the vision's scope in a two-dimensional space-time interval, interval $s$ quantifies this new distance `$x$', in Cartesian coordinates $(x,y,z,t)$, such that,
\vspace{-2pt}
\[ s^{2} = c^{2}t^{2} - r^{2}\]
\vspace{-2pt}
where $c$ is the speed of light; the differences of the space and time coordinates of the two events are denoted by $r$ and $t$, respectively, where
\[\ \ \ \ \ \ \ \ \ \ \ \ \ \ \ \ \ \ \ \ \ \forall r^{2}\in r_{\mathfrak{V}}^{2} \left| r^{2}= \int 2x \mathrm{d}x \rightarrow x^{2}+ y^{2}+ z^{2} \cap \{x^{2} + \mathfrak{c}\}\right. .     \ \ \ \ \ \ \ \ \ \ \ \ (3.6\mathrm{b})\]

\noindent Note that the choice of signs in the above follows the Landau-Lifshitz spacelike convention. Compare this with Ref.~\cite{06-Wiki}.

Let $r_{\mathfrak{V}}^{2}$ represent the radius-time interval of vision's observation scope of an observer, and $r$, remain as radius of revolution for all volumetric and surface integrations on the space-time coordinate system. The indefinite integral, $\int 2x \mathrm{d}x$, transforming dimensions of $r^{2}$ into $r_{\mathfrak{V}}^{2}$, holds an antiderivative position for the vision's imaginary scope later known by its function, `observation-scope function' $scope\left( o \right)$; where $\mathfrak{c}$ is any constant of zero, positive or negative which might suggest in the vision's scope to indicate a metric position of expanding dimensions more than four otherwise, merely one-dimension for the product form of $scope\left( o \right)$ in relation (3.7).

The above-mentioned imaginary scope of vision is determinable via $\jmath$ for interval $s$, where $\jmath$ is the imaginary unit equal to $\sqrt{-1}$ satisfying vision's imaginary coordinate $\jmath ct$ in $\{x^{2} + \mathfrak{c}\}$. The result would be of a space-time, synchronized versus asynchronous distances of observation-scope, definite to vectors $\vec r$ and $\vec s$ within the extended dimensions, i.e. a maximum of \emph{n}-dimensions $\geq 4$. This dimensional extension occurs whilst evaluating scenery objects is in a limited manner e.g., one-to-two-dimensional for the initiated observation. The result defining types of observation scope in functionality must satisfy the outcome of the general matrix form given in the forthcoming product relation (3.7).

\begin{flushleft}
\noindent Now let, \(
\forall \left( {x,y} \right)\left| {x \in X \wedge y \in Y,x_{i \to n}  \in X \wedge y_{i \to n}  \in Y::true,} \right.
\) then\\
\vspace{2pt}
\[
\left\{ {\forall \: Y \subseteq X,{\rm  }\exists \: scope\left( o \right)\left| \begin{array}{l}
 X \times Y = \left( {\left\{ {2\pi \int {r^2 }  {\rm d}  x} \right\}_i ,{\rm  }\left\{ {2\pi \int {r^2 }  {\rm d}  x} \right\}_i } \right){\rm  for } \\
 scope{\rm  }\left( {\overline {\; A_\mathfrak{V} } s^2 _{1 \le i \to n} ,\overline {A_\mathfrak{V}} s^2 _{1 \le i < \infty } } \right) \equiv scope\left( o \right) \\
 \end{array} \right.} \right\}, \; \; \; \; \; \; \; \; \; \; \; \; \; \; \; \; \; \;
\] \[\mathrm{if} \; \left\{ {\forall f\left( s \right),\textmd{g}\left( {A_\mathfrak{V} } \right) \in \prod\limits_{i = 1}^\infty  {scope\left( o \right)} \left| \begin{array}{l}
 \mathop {\min }\limits_{\delta x \to  \pm \infty }  \to f\left( s \right)_i  = x_i \wedge \left\{ {\mathop {\min }\limits_{\delta x \to  \pm \infty } \left( \mathbf{o} \right),} \right. \\
 \ \ \ \ \  \left. {\mathop {\lim }\limits_{\delta x \to  \pm \infty } \left( \mathbf{o} \right)} \right\} \to f\left( s \right)_i  = y_i  \in \left( {\emph{\textbf{C}}_f \textmd{g}} \right)_{m \times n} , \\
 \mathop {\lim }\limits_{i \to n}  \to \textmd{g}\left( {A_\mathfrak{V} } \right)_i  = \left( {x,y} \right) \in \left( {\emph{\textbf{C}}_f \textmd{g}} \right)_{m \times n} \; , \\
 \end{array} \right.}  \right. \; \; \; \; \; \; \; \; \; \; \; \; \; \; \; \; \; \; \; \; \; \; \; \; \; \; \; \; \; \; \; \; \; \; \;\]\vspace{4pt}
\end{flushleft}
\vspace{-12pt}
\noindent for $\textmd{g}{\mathop{\:\rm o \:}\nolimits} f:X \to Z$, then for the scope's geometry satisfying $A_\mathfrak{V}$ and $s$, in their function composition via composition operator, $\emph{\textbf{C}}_f \textmd{g}$, generating a matrix product form, $\left({\emph{\textbf{C}}_f \textmd{g}} \right)_{m \times n}$, one can thereby deduce
\vspace{-4pt}
\[\prod\limits_{i = 1}^\infty  {scope\left( o \right)}  = x \mapsto \left( {\emph{\textbf{C}}_f \textmd{g}} \right)_{m \times n}  = x \mapsto \textmd{g}\left( {f\left( x,y \right)} \right)_{m \times n}  = \left[ {\begin{array}{*{20}c}
   {\mathop {\lim }\limits_{i \to n} \partial(t) }  \\
   {\mathop {\lim }\limits_{i \to n} \partial(t) }  \\
\end{array}} \right]*\left[ {2\pi r^2 \mathfrak{V}_i } \right] \] \\ \vspace{-40pt}
\begin{flushleft}
\[\; \; \; \; \; \; \; \; \; \; \; \;  \mathbf{o} \left[ {\begin{array}{*{20}c}
   {\mathop {\min }\limits_{\delta x \to  \pm \infty } \partial(t) }  \\
   {\mathop {\lim }\limits_{\delta x \to  \pm \infty } \partial(t) }  \\
\end{array}} \right]*\left[ {\sum\limits_i {s_i^2 \delta x} } \right] \; \mathrm{where,} {\mathop {\min }\limits_{\delta x \to  \pm \infty }f\left( s \right)_i } \in \; \mathop {\lim }\limits_{\scriptstyle \delta x \to 0 \hfill \atop
  \scriptstyle n \to \infty  \hfill} \sum\limits_{\mathop i\limits_{n = 0} }^{n - 1} {s_i^2 \delta x} \]
\end{flushleft}
\[\mathrm{and},\; \mathbf{o} \; \mathrm{for}\; \{x \mapsto \textmd{g}\left( {f\left( x \right)} \right)_{m \times n}\} \equiv \left\{ {\bigcup\nolimits_a^b {f\left( x,y,z \right)}  \cap f\left( x \right),\bigcup\nolimits_a^b {f\left(x,y,z \right)}  \cup f\left( x \right)} \right\},\] wherein continue,
\vspace{-4pt}
\[\therefore \prod\limits_{i = 1}^\infty  {scope\left( o \right)}  = \left( {\bigcup\nolimits_{\mathfrak{V}_i }^{\mathfrak{V}_n } {2\pi \int_a^b {r^2 }  {\rm d}  x} \bigcap\nolimits_{ - \infty }^{ + \infty } {s_i^2 } {\rm \; \; , \; \; }\bigcup\nolimits_{\mathfrak{V}_i }^{\mathfrak{V}_n } {2\pi \int_a^b {r^2 }  {\rm d}  x\bigcup\nolimits_{ - \infty }^{ + \infty } {s_i^2 } } } \right) \; , \; \; \; \; \; \; \; \; \; \; \; \; \; \; \; \; \; \; \; \; \; \; \; \; \; \; \; \; \;\]

\[ \therefore \prod\limits_{i = 1}^\infty  {scope\left( o \right)}= {\left( \frac{ \cal{V}_{\mathfrak{V}} }{s^{2}}, {\cal{V}_\frak{V}} s^{2} \right)}= {(\:\vec r_{ij},\vec s_{ij}\:)}_{m\times n}\; . \ \ \ \ \ \ \ \ \ \ \ \ \ \ \ \ \ \ \ \ \ \ \ \ \ \ \ \ \ \ \ \ \  \ \ \ \ \ (3.7) \]

\noindent In this discrete Cartesian product's \emph{set}-and-\emph{matrix} over function $f(x,y,z)$ with expandable function $f(x)$ to its matrix form $\textmd{g}\left( {f\left( x \right)} \right)_{m \times n}$ via abstract algebraic operator $ \mathbf{o}$, in the course of observation-scope function $scope\left( o \right)$, we commuted and subsequently unified the geometry of the vision's scope with the field's scope. That is, defining the state of vision for the utilized body in space (in this case, the fly), with the field's scope as specified. This commutability of observing and recording events prior to unification, is done by incorporating discrete mathematical operators $\bigcup\nolimits_a^b$ and $\bigcap\nolimits_{a'}^{b'}$ on space-time interval $s$, denoting a symmetrical presence of all vector values in ranges of $\infty$ and $\mathfrak{V}_i$; hence, determining the states of Maxwell continuity equations for the range given on observation vectors. Having this determination of states, concluding a result on consumed time $t_{\mathrm{cons}}$ against time difference $\mathrm{d}\: t$,  shall in isolation derive relation (3.8), expectably. The mathematical commutative law at work satisfying the observation's commutability in coproduct (categorical sum) terms, is relevant to vision and biovielectroluminescence $ \beta _{\gamma }$, which is submitted in \S\ref{section5} inclusively.

In continue, let operator $\mathbf{o}$ for a subfield in a differential field, $scope\left(o\right)^{{'} } $of $scope\left(o\right)$, satisfy scope implications $\mathop{\min }\limits_{\delta x\to \pm \infty } (\mathbf{o})\to f\left(s\right)_{i} =x_{i} $ and $\left\{\mathop{\min }\limits_{\delta x\to \pm \infty }(\mathbf{o}) , \right. $ $ \left.\mathop{\lim }\limits_{\delta x\to \pm \infty }(\mathbf{o}) \right\}\to f\left(s\right)_{i} =y_{i} $ by discrete union and intersection closures $\bigcup\nolimits_a^b$ and $\bigcap\nolimits_{a'}^{b'}$, for the range of differential $\partial \left(u\right)$ in $x_{i} $, and for $\partial \left(u\right)$, which is the range of differential in $y_{i} $ of the matrix, as deduced in relation (3.7). The result of differentiated values is distributive and forms an efficient matrix composition, which is a construct to a database system for the recoded observable events from the scene of observation and event occurrences. Thence, if the product of $scope\left(o\right)^{{'} } $ is a \textit{differential field} and \textit{differentially closed}, then the field of constants is $\mathds{k}=\left\{u\in scope\left(o\right)^{{'} } \left| \: \partial \left(u\right)=0\right. \right\}$, whereon the differential product incorporating Leibniz law thus derives
\[\prod _{i=1}^{\partial \left(u\right)}scope\left(o\right)^{{'} }  =\left(\partial \left(\frac{A_\mathfrak{V}\to A_{\bigcirc } }{r^{2} } r_{i} s_{i} \right)^{2} ,\partial \left(\frac{A_{\bigcirc } \to \sum A_{\bigcirc }  }{r^{2} } r_{i} n_{i} \right)^{2} \right) \] \[\ \ \ \ \ \ \ \ \ \ \ \ \ \ \ \ \ \ =\left(\partial \left(\pi r_{i} \left|s_{i} \right. \right)^{2} ,\partial \left(\pi s_{i} \right)^{2n} \right)=\left[\begin{array}{cc} {2\pi \partial \vec{r}} & {2n\pi \partial \vec{s}^{\; 2n-1} } \end{array}\right] , \] where \[ \left\{\bigcup _{\mathfrak{V}_{i} }^{\mathfrak{V}_{n} }\ni n\in \bigcup _{-\infty }^{+\infty }s_{i}^{2} ,{\mathbb N}  \right\} \stackrel{\subset}{=}\mathop{\lim }\limits_{\infty \to n} f(n) \vec{s}^{\; 2f(n)}\equiv n\eta _{\mu \nu } x^{\mu } x^{\nu } , \ \eta =  \footnotesize \left[\begin{array}{cccc} {-1} & {0} & {0} & {0} \\ {0} & {1} & {0} & {0} \\ {0} & {0} & {1} & {0} \\ {0} & {0} & {0} & {1} \end{array}\right]\:. \normalsize  \]
$\ \ \ \ \ \ \ \ \ \ \ \ \ \ \ \ \ \ \ \ \ \ \ \ \ \ \ \ \ \ \ \ \ \ \ \ \ \ \ \ \ \ \ \ \ \ \ \ \ \ \ \ \ \ \ \ \ \ \ \ \ \ \ \ \ \ \ \ \ \ \ \ \ \ \ \ \ \ \ \ \ \ \ \ \ \ \ \ \ \ \ \ \ \ \ \ \ \ \ \ \ \ \ \ \ \ \ \ \ \ \ \ \ \ \ \ (3.8)$

\noindent The product is of two-dimensional over interval $s$ in the matrix attributes containing one-dimensional spin vectors satisfying some trigonometric function e.g., cosine otherwise sine function, conditionally. The interval $s$ differentially closed restriction, is $r_{i} \left|s_{i} \right. $, which implies to its neighbour (matrix member) containing radius $r$ for that one-dimensional spin. The $A_{\bigcirc } $ is the area of visual transformation $A_\mathfrak{V}\to A_{\bigcirc } $, in discrete slices of world's visual system in form of circles (within a sum form) against contour $C$ in the unitary field limits. Contour $C$ is already understood in Eqs.$\,$(2.2a), (2.4) and (2.5) framing bendable wavelength $\lambda_{\mathrm{bend}}$, satisfying virtual parts of phase velocity $v_\mathrm{p}$. Note that $scope\left(o\right)^{{'} } $, is unitary plus, an element which is of $x$ or $y$ due to $i=1$ for the subfield and sustains all vectors concerning the closure limit, $\mathop{\lim }\limits_{\infty \to n} $, for the Minkowski metric on signatures of $\eta $ for higher/inner dimensions of observation (the four properties in Minkowski space Refs.~\cite[xx, ii]{06-Wiki,{42-Wolfram}}). Imperatively, in the field and scope equations involving GR and SR, we can present both scopes, entire and minimal, into an adjacency matrices inclusive to a 4-by-4 matrix, where the latter is of Minkowski metric defining matrix $\eta $ from SR's space-time concept. This construction of matrix type, depends on the generalization of tuples and attributes of the system's database, recording sequences of events as they occur within the observable environment relative to the non-subatomic bodies in e.g., a coversine state for a giant body's vision.

It is in closure to the products formulated in relations (3.8) and (3.9), the remaining features of ratios of velocity concerning $|v_{{\rm p}}|$ as virtual part of phase velocity $v_\mathrm{p}$, strongly interdependent with the recorded values onto the database system, Appendix II. Future and past events against present, do appreciate the hypotheses and formulaic representations made by the current author, \S4 of~\cite{01-Alipour} or~\cite{02-Alipour}, which explain the role of `Gaussian curvature' in this context (see also Ref.~\cite{19-Einstein et al.}). This also broadens the aspect from a semiconducting viewpoint to the aspects of biovielectroluminescence phenomenon which is subject to \S\ref{section5}.

\section{Artificial neuron configuration with organic inputs and outputs}
\label{section4}
\vspace{6pt}

\normalsize
`Dormant pulses' (see Abstract), or in this case, `$ bias + x_{1} + x_{2} + ... + x_{n}$ inputs', are for a state fallen into abeyance until the moment of threshold and thus, full-activation state. These pulses can be classed as 3rd grade chemical-electrical (ionic feeding stage, \S\ref{section2} and Ref.~\cite{41-Bezanilla}) then, 2nd grade biovielectrical pulses, the latter and former temporarily `in abeyance' which eventually grade to biovielectroluminescence pulses as 1st grade class. The perceptron to fire irrespective of its inputs would be, \textbf{E}-field and \textbf{B}-field dependent, on the Fly's eyes surface where it evidently associates the operations that relate to the organic material of the dissected fly, i.e. spatial visual channels~\cite{07-Braitenberg} compared to a human's visual system. The Sigmoid function concerning activation and saturated states, based on the following curve and condition, emphasizing that for a basic on/off type function, if 0 $> x$ then 0, else if $x \geq$ 0 then 1,
\[\ \ \ \ \ \ \ \ \  \ f(x)=y=\frac{1}{1+e^{-x}}\; , \;e=2.71828\ldots \; , \; f(x)= \left\{ \begin{array}{l}
0 \; \mathrm{if}\ 0 >x
\\ 1 \; \mathrm{if} \ x \geq 0 \; ,
\end{array}\ \ \ \ \ \ \ \ (4.1\mathrm{a}) \right.\]

\noindent plays a crucial role resulting a flat configuration in the architecture which support forms of biovielectroluminescence frequencies and their distribution with respect to pulse population. In other words, the stronger the input (the receiving detector) in terms of \textbf{E}-field and time varying \textbf{B}-field configuration based on the meter's data, the fastest (not even `faster') in higher rate firing rates~\cite{10-Barber}. The logarithmic output and its distribution per output population is thereby formed based on the latter AI interpretation. The logarithmic output and its supplemented measurements would be studied by `Gaussian distribution' satisfying `continuous probability distribution' of sequences of I/O data on electrical signals between non-organic components and organic components of the project, associatively (see, Ref.~\cite[iv]{06-Wiki}). Note that in principle, the output of the perceptron, however, is \emph{always} boolean~\cite{10-Barber} and must not be confused with the general logarithmic algorithm serving the expected normal distribution. Benefiting from the original characterization of normal distribution, its modified version in favour of a basic equation would herein look like this
\[\ \ \ \ \ \ \ \  \mathcal{X} \sim \mathcal{N}(\mu, \sigma^2)\longrightarrow\mathcal{X}_{\beta_\gamma} (\pi)\sim \mathcal{N}(\mu, \sigma^2)^{\mathbf{C}}\left|\;\mathcal{N}(\mu, \sigma^2)_{i}^{\mathbf{CC}} \; , \; i\in \mathbb{N} \; \right. , \ \ \ \ \ \ \ \ (4.1\mathrm{b}) \]

\noindent where $\mathcal{X}$ indicates a random variable of the pulses product, $\pi$, and is normally distributed with mean $µ$ and variance $\sigma^2\geq0$, wherein continue, $\beta_\gamma$ represents a biovielectroluminescence function propounded in \S\ref{section5}. Notation \textbf{C}, indicates the pulses' class whereas for the project, $i$ for \textbf{C} would be $1 \leq i_{\mathbf{C}} \leq3$, since we just investigate 3 classes of the availably-degree-of-biovielectroluminescent pulses in combination. An example to this is given in the extended database exemplar in Appendix II, in the realization of the corresponding product's normal distribution density function
\[\ \ \ \ \ \ \ \  \ \ \ \ \ \ \ \ \ \ \ \ \ \ \ \ \ \ \ \ \ \ \ \  f\left(x\right)=\frac{1}{\sqrt{2\pi } \sigma } e^{-\left(1/2\right)\left[\left(x-\mu \right)/\sigma \right]^{2} } \ \ \ \ \ \ \ \ \ \ \ \ \ \ \ \ \ \ \ \ \ \ \ \  (4.1\mathrm{c}) \]

\noindent where $\sigma$ is the deviation, and $\mu$ is the expected value or `mean' for the occurring distribution. Such type of representation expressing classes entailing their product's abstract-ratio, is submitted in the forthcoming neuromatrix Eqs.$\,$(4.3) and (4.4). The AI concept does not deal with any programming language and solely initiates with the potential worked-out examples of I/O data as perceptron dependent functions. That is, pattern recognition on e.g., Sigmoid function, which studies the objectives of the project on a piece of paper. This is to retrain the computer and its neuromatrix to learn vital information for further data interrogation in comparing data points against potential errors that might deviate the project standards against the gathered information. `Data interrogation' also allows the experimenter to question the computer in terms of: \\

\noindent----------- \emph{Which one is the best product combination satisfying Eqs.$\,$(5.3), (5.4), of organic and nonorganic pulses suggesting convenient biovielectroluminescence frequencies for the objectives given in this project ?} -----------\\ \vspace{-2pt}

After pragmatically responding to this question, the best product choice based on all of the above, would generate a secondary matrix which stores combined images and sub-images of data (subject to \S\ref{section4}) in the most enhanced manner. Reasonably, the images product is clear enough to support more incoming data stream due to the biovielectroluminescence 1st class choice made by the neuromatrix, which was amongst other pulses' product attributes, satisfying appropriate angular values of $\alpha$ for r,g,b and the phenomenon's efficiency.

The AI software program however, incorporates in its body, the codes of programmable variables treated as `global variables' for the multi-channeled optical imaging system. This method is here to associate organic-matter output signals, thus eliminating the barrier problem of `private variables and procedures' embedded into the program's source code. The computer language could be of e.g., an object oriented programming (OOP) language like C++, otherwise, a simpler and faster approach for database instantaneous response and analysis in favour of AI's I/O variables, e.g., Visual Basic (VB) which is an event-driven-programming language and enables rapid application development (RAD)~\cite{06-Wiki}.

Ergo, the neuromatrix concept which consequentially generates neuromatrices configuration, can be formulaically visualized in terms of its symmetrical-sum representation (denoted by an $S$) including the degree of class of pulses,
\[
S(w_i^\mathbf{C}x_i^\mathbf{C}) = \left( \sum\limits_{i = 1}^m {bias}  + w_i^{\mathbf{3}} x_i^{\mathbf{3}} \left| {\sum\limits_{i = 1}^m {bias}  + w_i^{\mathbf{2}} x_i^{\mathbf{2}} } \right.\left| {\sum\limits_{i = 1}^m {bias}  + w_i^{\mathbf{1}} x_i^{\mathbf{1}} } \right ) = \right. \ \ \ \ \ \ \ \ \ \ \] \[ \begin{array}{l} \left(\{w_{1}^{\mathbf{3}}x_{1}^{\mathbf{3}}| w_{1}^{\mathbf{2}}x_{1}^{\mathbf{2}}| w_{1}^{\mathbf{1}}x_{1}^{\mathbf{1}}\}+ \{w_{2}^{\mathbf{3}}x_{2}^{\mathbf{3}}| w_{2}^{\mathbf{2}}x_{2}^{\mathbf{2}}| w_{2}^{\mathbf{1}}x_{2}^{\mathbf{1}}\} + \; \cdots \; + \right.
\\
\left\{w_{n}^{\mathbf{3}}x_{n}^{\mathbf{3}}| w_{n}^{\mathbf{2}}x_{n}^{\mathbf{2}}| w_{n}^{\mathbf{1}}x_{n}^{\mathbf{1}}\} + \{w_{b}^{\mathbf{3}}b^{\mathbf{3}}| w_{b}^{\mathbf{2}}b^{\mathbf{2}}| w_{b}^{\mathbf{1}}b^{\mathbf{1}}\} \right) \geq \left( \theta^{\textbf{3}}|\theta^{\textbf{2}}|\theta^{\textbf{1}}\right)\;, \; b\buildrel\wedge\over= bias , \; x\in \mathbb{R} \; ,
\end{array} \]
\begin{flushright} (4.2) \end{flushright}
\noindent which could also be envisaged in terms of sequences of lists in the context of arrays data structures and algorithms. Note that, the threshold value is defined by $\theta$, and the pulses symmetry of inputs and sum, fires whenever the previous equation is true. Class and grade of the the neuromatrix is constructed explicitly in the form \vspace{-4pt}
\[ \mathcal{N}_{\bf \pi } : = \left( {\left( {w_i^\mathbf{C} x_i^\mathbf{C} } \right)_{\alpha ,\beta } } \right)_{3 \times 1}  = \left[ {\begin{array}{*{20}c}
   {w_i^\mathbf{3} x_i^\mathbf{3} }  \\
   {w_i^\mathbf{2} x_i^\mathbf{2} }  \\
   {w_i^{\bf 1} x_i^{\bf \mathbf{1}} }  \\
\end{array}} \right], \; 1 \le \alpha  \le 3, \; \beta  = 1, \; \; w_i = x_i\delta  \ \ \ \ \ \ \ \ \ \  (4.3\mathrm{a}) \]

\vspace{6pt}
\noindent and its yield efficiency is,
\[  \ \ \ \ \ \ \ \ \ \ \ \ \ \ \ \ \ \ \ \  \ \ \ \ \ \ \ \ \ \  \ \ \ \ \ \ \ \ \ \ \mathcal{Y}\left( {\bf \pi } \right) = \frac{{\left\{ {w_i^{\mathbf{2}} x_i^{\mathbf{2}} ,w_i^{\mathbf{3}} x_i^{\mathbf{3}} } \right\} \times 100\% }}{{w_i^{\mathbf{1}} x_i^{\mathbf{1}} }} \; .  \ \ \ \ \ \ \ \ \ \  \ \ \ \ \ \ \ \ \ \   \ \ \ \ \ \ \ \ \ \ (4.3\mathrm{b})\]

\noindent All of these forms are designated for the neuromatrix of the perceptron and specialized for the above-mentioned pulses including their grades of inputs configuration, superscripted by $\mathbf{1}$, $\mathbf{2}$ and $\mathbf{3}$ representing 1st, 2nd, 3rd class $\mathbf{C}$, respectively. Note that these numbers in \textbf{bold}, do not represent operations of exponentiality and just a rank to the their physical behaviour and rating state (speed of trigger and implementation related to `pulses of $\beta_\gamma$'), e.g., $w_i^{3rd} x_i^{3rd}$. Symbol $\mathcal{Y}$, is the yield efficiency function of the recorded data based on these pulses relative to the quantity of 1st class pulses in terms of, weight $w_i$ for its input $x_i$. Notation $\mathcal{N}$, merely stands for a neuromatrix construct on pulse $\pi$, which defines the characteristics of pulses product in the perceptron's matrix. The delta form $\delta$, for both large and too small outputs of the perceptron in form of summation $w_i = x_i\delta$, works both ways, whereas this summation by definition is given by
\vspace{6pt}

\footnotesize $\mathrm{Change \; in \; Weight \:} i = \mathrm{Current \; Value \; of \; Input\:} i × (\mathrm{Desired \; Output} - \mathrm{Current \; Output}).$

\vspace{6pt}
\normalsize Thus, computing the matrix I/O's is applicable when we product all of the paired classes from one another, with respect to a fixed class in abstract ratio terms. This is constructed in the following neuromatrix product-ratio with a fixed finite limit on \emph{i}, such that
\[\mathcal{N}_{\kappa}:=\left(\left(\frac{w_{i}^{\mathbf{CC}} x_{i}^{\mathbf{CC}} }{w_{i}^{\mathbf{C}_{\beta } } x_{i}^{\mathbf{C}_{\beta } } } \right)_{\alpha ,\beta } \right)_{3\times 3} =\left[\begin{array}{ccc} {\frac{w_{i}^{\mathbf{3}} x_{i}^{\mathbf{3}} w_{i}^{\mathbf{2}} x_{i}^{\mathbf{2}} }{w_{i}^{\mathbf{3}} x_{i}^{\mathbf{3}} } } & {\frac{w_{i}^{\mathbf{3}} x_{i}^{\mathbf{3}} w_{i}^{\mathbf{2}} x_{i}^{\mathbf{2}} }{w_{i}^{\mathbf{2}} x_{i}^{\mathbf{2}} } } & {\frac{w_{i}^{\mathbf{3}} x_{i}^{\mathbf{3}} w_{i}^{\mathbf{2}} x_{i}^{\mathbf{2}} }{w_{i}^{\mathbf{1}} x_{i}^{\mathbf{1}} } } \\ {\frac{w_{i}^{\mathbf{3}} x_{i}^{\mathbf{3}} w_{i}^{\mathbf{1}} x_{i}^{\mathbf{1}} }{w_{i}^{\mathbf{3}} x_{i}^{\mathbf{3}} } } & {\frac{w_{i}^{\mathbf{3}} x_{i}^{\mathbf{3}} w_{i}^{\mathbf{1}} x_{i}^{\mathbf{1}} }{w_{i}^{\mathbf{2}} x_{i}^{\mathbf{2}} } } & {\frac{w_{i}^{\mathbf{3}} x_{i}^{\mathbf{3}} w_{i}^{\mathbf{1}} x_{i}^{\mathbf{1}} }{w_{i}^{\mathbf{1}} x_{i}^{\mathbf{1}} } } \\ {\frac{w_{i}^{\mathbf{2}} x_{i}^{\mathbf{2}} w_{i}^{\mathbf{1}} x_{i}^{\mathbf{1}} }{w_{i}^{\mathbf{3}} x_{i}^{\mathbf{3}} } } & {\frac{w_{i}^{\mathbf{2}} x_{i}^{\mathbf{2}} w_{i}^{\mathbf{1}} x_{i}^{\mathbf{1}} }{w_{i}^{\mathbf{2}} x_{i}^{\mathbf{2}} } } & {\frac{w_{i}^{\mathbf{2}} x_{i}^{\mathbf{2}} w_{i}^{\mathbf{1}} x_{i}^{\mathbf{1}} }{w_{i}^{\mathbf{1}} x_{i}^{\mathbf{1}} } } \end{array}\right] \ \ \ \ \ \ \ \ \ \ \ \ \ \ \ \ \ \ \ \ \ \ \ \]

\[=\left[\begin{array}{ccc} {w_{i}^{\mathbf{2}} x_{i}^{\mathbf{2}} } & {w_{i}^{\mathbf{3}} x_{i}^{\mathbf{3}} } & {\frac{w_{i}^{\mathbf{3}} x_{i}^{\mathbf{3}} w_{i}^{\mathbf{2}} x_{i}^{\mathbf{2}} }{w_{i}^{\mathbf{1}} x_{i}^{\mathbf{1}} } } \\ {w_{i}^{\mathbf{1}} x_{i}^{\mathbf{1}} } & {\frac{w_{i}^{\mathbf{1}} x_{i}^{\mathbf{1}} w_{i}^{\mathbf{3}} x_{i}^{\mathbf{3}} }{w_{i}^{\mathbf{2}} x_{i}^{\mathbf{2}} } } & {w_{i}^{\mathbf{3}} x_{i}^{\mathbf{3}} } \\ {\frac{w_{i}^{\mathbf{2}} x_{i}^{\mathbf{2}} w_{i}^{\mathbf{1}} x_{i}^{\mathbf{1}} }{w_{i}^{\mathbf{3}} x_{i}^{\mathbf{3}} } } & {w_{i}^{\mathbf{1}} x_{i}^{\mathbf{1}} } & {w_{i}^{\mathbf{2}} x_{i}^{\mathbf{2}} } \end{array}\right] \equiv \left[ {\begin{array}{*{20}c}
   {w_i^\mathbf{2} x_i^\mathbf{2} } & {w_i^\mathbf{3} x_i^\mathbf{3} } & {\kappa _i^\mathbf{1} }  \\
   {w_i^\mathbf{1} x_i^\mathbf{1} } & {\kappa _i^\mathbf{2} } & {w_i^\mathbf{3} x_i^\mathbf{3} }  \\
   {\kappa _i^\mathbf{3} } & {w_i^\mathbf{1} x_i^\mathbf{1} } & {w_i^\mathbf{2} x_i^\mathbf{2} }  \\
\end{array}} \right]  \ \ \ \ \ \ \ \ \ \  (4.4) \]

\noindent which is for the pulses' learning algorithm and processor, based on fastest and efficient trigger points of a full-activation state when a product of pulse classes are experimented. The equivalence result of matrix construction denoted by symbol $\kappa$ relative to the denominator of diagonally achieved matrix elements, would be in form of the computed weighted input values in product ratio terms. That is, conducting the experiment in terms of the best outcome on factors of efficiency, project performance and the above-mentioned pulses in combination. Section \S\ref{section5} promotes this combination of pulses with variant nature to the \emph{biological vision} and its phenomena, supporting radical outcomes relevant to biovielectroluminescence, its visual technology and future peripherals advancing its applications. The \emph{database construction and analysis} on a theoretical basis of representation for the product of pulses, is submitted in Appendix II.

\section{Operation mechanism and the biovielectroluminescence phenomenon}
\label{section5}
\vspace{6pt}
As we proceed, one should not confuse luminance with the terms, luminescence and biovielectroluminescence, whereas the following equations would reflect upon a mere representation of the biovielectroluminescence phenomenon when one equation is referenced to another, correspondingly.

\vspace{8pt}
Let by notion to the phenomenon's wavelength product, luminance contrast per time sequence or luminance frequency be
\begin{flushleft}
\small $${\rm Luminance\; frequency}= \frac{{\rm Luminance\; difference}}{{\rm Average\; luminance}\times {\rm average\; time\; sequence\; \; \; }}$$
\end{flushleft} \begin{flushright}
\small $$ \; \; \equiv \frac{{\rm Luminance\; contrast}}{{\rm average\; time\; }} \; , \ \ \ \ \ $$
\end{flushright} \normalsize \begin{flushleft} or,
$$\nu _{L} =\frac{\Delta L}{\frac{1}{n} \sum _{i=1}^{n}L \times \frac{1}{m} \sum _{j=1}^{m}t_{i}  } =\frac{L^{*} }{\left\langle t\right\rangle } ,{\rm \; }n,m\in {\mathbb N} \ ,$$  such that, \end{flushleft}\begin{flushright} \vspace{-44pt}
(5.1a)\end{flushright}\vspace{12pt}$$L=\frac{\mathrm{d}^{2} F}{\mathrm{d}A\:\mathrm{d}\Omega \cos \theta }\; \; \; ,$$
\begin{flushright} \vspace{-18pt}
(5.1b)\end{flushright}

\noindent where $L$ or its formal form $L_v$ (we ignore the latter to prevent future notational confusions), is the luminance measured in candela per square meter, $F$, is the luminous flux or luminous power (lm), $\theta$, is the angle between the surface normal and the specified direction, $A$, is the area of the surface (m$^{2}$) and, $\Omega$, is the solid angle (sr) in the context of photometry (optics).

The main operation concentrates on the \emph{biovielectro-imagery state data}. That is, data being gathered in terms of an array of blurred and focused sub-images in regard to biovielectro-measurements as unoccured motion (pre-kinematics) images (recall the major subjects in Sections \S\ref{section2} and \S\ref{section3}). These discrete pre-kinematical data recordings in realtime sequences, can be satisfied by integrating appropriate waveguides and incorporating artificial ommatidia relative to the scene's objects (see Ref.~\cite{03-Lee}). The sub-images are recorded from the scene's centre with a constant size of e.g., 98$\times$84 pixels per sub-image. The recording is done by the CCD matrix through the channels of artificial ommatidia collection apparatuses $\mathrm{A}'$ and $\mathrm{A}''$, \S\ref{section2}, where this methodology and instrumentation setup partly relates to the explanations provided in Refs.~\cite{{03-Lee},{48-Rosen}}.

For clarity of images and recognizing the stationary object in the scene from any image point of the total number of different focused and blurred images, when associating sub-images with other sub-images extracted from the same matrix, the image data frame is thus shifted toward a common centre~\cite{48-Rosen}. These sub-images are by now rectified and de-speckled into enhanced images due to `virtual/real variable lens and image matrix technique' installation mentioned in \S\ref{section2}. The rest is up to the algorithm that calculates the blurred image's centre of gravity not yielding various positions of the object in all the channels regardless of object's longitudinal position explained and well-exemplified by Rosen \& Abookasis of Ref.~\cite{48-Rosen}. The algorithm in addition, maintains the timing of events relevant to the biovielectroluminescence \textbf{E} and \textbf{B}-fields behaviour measured by a multi-meter device here, e.g., a TriField meter~\cite{29-AlphaLab}.

These above-mentioned visual types of data, would subsequently \emph{expand} and then \emph{enhanced} when from left and right corners of the scene, forming the other remaining half of images' data, say, e.g., 132 previously recorded blurred images of the centre, now from the left and right corners, forming a new total of \\

\noindent $ 132 \; \textsf{C} + \frac {132}{2}\; \textsf{L} + \frac {132}{2} \; \textsf{R} = 264 \; \mathrm{images}\: , \ \ \ \ \ \ \ \ \ \ \ \ \ \ \ \ \ \ \ \ \ \ \ \ \ \ \ \ \ \ \ \ \ \ \ \ \ \ \ \ \ \ $ \\

\noindent where $\sf C$, $\sf L$ and $\sf R$ with a value of 1, represent a scene's centre, left and right respectively. In more complex systems, the more number of scenes under experimental observation, the exact number of the populated scene defined for $\sf C$, $\sf L$, $\sf R$; in form of sequence $(n\:\textsf C$ , $n\: \sf L$ , $n\: \sf R)$ = $n\: \sf S$. Let $\sf S$ always hold the $scene$ representing observable events between points A and B, \S\ref{section2}, for all recordable image-frame observations satisfying observation-scope function $scope\left( o \right)$, \S\ref{section3}. For instantaneous frame count on the previous result whilst recording the scene,
\vspace{2pt}
$$\frac {132}{t+\mathrm{cons}(t)}\; \textsf{C} + \frac {66}{t+\mathrm{cons}(t)}\; \textsf{L}+ \frac {66}{t+\mathrm{cons}(t)} \;\textsf{R} = 264 \mathrm{\; \: images \; per \; early \;second} \rightarrow$$ $\mathrm{images}\cdot(\mathrm{seconds + consumed \; seconds})^{-1}$, $\mathrm{cons}(t)=n \;\mathrm{s}\:.$ \\

\noindent Latency or signal transmission delay issue for image recording(s) and the elimination of blurry images should have been alleviated based on the logarithmic algorithm installed onto the computer system incorporating the AI components computable with organic signals' distribution per population (see Sections \ref{section2}, \ref{section4}, and Appendix II respectively). This computability of artificial neural I/O components with I/O organic components, is devoted to the observable mobile/stationary objects within the scene on the record. The formulaic orientation of luminance frequency $\nu _{L}$, biovielectroluminescence $ \beta _{\gamma }$, and 2D-to-$n$D$\:\geq\:$4D dimensional capture of the scene for their time parameter is given by
\vspace{-2pt}
\[\begin{array}{l} {\ \ \ \ \ v_{\mathrm{fly}\left(o\right)} \to v_{\mathrm{man}\left(o\right)} \;\mathrm{as}{\rm \; }f\left(v_{\mathrm{man}\left(o\right)} \right)\approx v_{\mathrm{fly}\left(o\right)} \; , \mathrm{iff}{\rm \; }\beta _{\nu } \stackrel{\mathrm{morphism}}{\longrightarrow}L_{\nu }^{*} \: \mathrm{rel}\left(A_{\mathrm{fly}\left(\sphericalangle \right)} \right) } \ \ \ \ \end{array}\]
\[\begin{array}{l}{\ \ \ \ \ \ \ \ \ \ \ \ \ \ \ \ \ \ \ \ \ \ \ \ \ \ \ \ \ \ \ \ \ \ \ \ \ \ \ \ \ \ \ \ \ \ \ \ \ \ \ \ \ \ \ \ \ \ \ \ \ \ \ \left|A_{\mathrm{fly}\left(\sphericalangle \right)} \in \left\{{\rm {\mathcal B}}^{>0} \ll {\rm {\mathcal B}}^{<\infty } \right\}\right. } \end{array}\]
\noindent thence,
\[\begin{array}{l}{\ \ \ \ \ \:  \therefore \mathbf{E}_{\mathrm{biovi}}^{L^{*} } \equiv \beta _{\nu } \to \nu _{L} \cdot Q \; \mathrm{rel}\left(A_{\mathrm{fly}\left(\sphericalangle \right)} \right)} \; \mathrm{for} \; h\nu \in \Delta E \: \wedge\: \lambda_{\mathrm{bend}}\in v_{r,g,b}\:. \ \ \ \ (5.2) \end{array}\]
\vspace{-6pt}

\noindent This mechanism is described in essence of a coproduct of electricity from the circuit in the fly's body and the bio-organic reactions. The electrical coproduct and bio-organic reactions therefore represent the morphism of luminance frequency to its powers, flux and consequently, biovielectroluminescence when luminescence is being product by charge quantity $Q$ for individuals' vision, relativistically. In Deduction (5.2), the phenomenon (later known as function $ \beta _{\gamma }$) carrying bendable wavelengths of $\lambda_{\mathrm{bend}}$ from luminance frequency $\nu _{L}$, implies to luminance frequency $\nu _{L}$ for the phenomenon's product which is $\beta _{\nu }$ equal by definition to $\mathbf{E}_{\mathrm{biovi}}^{L^{*} }$ (symbol `$\equiv$'). This solely occurs during relativistic transformations for a photon $h\nu$ if is in conservation with $\lambda_{\mathrm{bend}}$ obeying the `\emph{transformation of wave's unit circle paradigm}' of Fig.$\,$2.3, \S\ref{section2}. The involved relativity, is carried by its function, $\mathrm{rel}$, indicating the individual's observation variable, $o$, satisfying its observation-scope function $scope\left( o \right)$. Moreover, energy change $\Delta E$ relates to work function $\phi$ not of photoelectric and just of biovielectric, for charge $Q$ between points A and B on the stationary/mobile surface of the object of the scene, satisfying unoccured events of photon $h\nu$ via $\lambda_{\mathrm{bend}}$. This function is measured in watts multiplied by early second (joules). It denotes the change of photon's energy state relative to bendable length of pre-kinematical coordinate transformations, into post-state of the same photon after these transformations of pre-kinetics here, the \emph{unoccured quantum electrodynamics} (UQED).

From a formal viewpoint to~\cite[xvi]{06-Wiki}, the formal definition to Deduction (5.2), is as follows: Let for the pulses' class \textbf{C} from \S\ref{section4}, indicate a class for $\beta _{\gamma }$, and let $X_j:j\in J$ be an indexed family of objects as pulses of $\beta _{\gamma }$ in \textbf{C}. This ensues by substitution to the formal notion of $\beta _{\gamma }$, a multifunction $\mathbf{E}_{\mathrm{biovi}}^{L^{*} }$ which carries the electric field notation $\mathbf{E}$ via charge $Q$ and thus the \textbf{E}-field's function originated from Coulomb's law, possessing limits of ${L^{*}}$ for luminance frequency $\nu _{L}$ in Eq.$\,$(5.1a), and `\emph{biovi-}' of `biovielectroluminescence' (see Introduction, \S\ref{section1}). The coproduct of the set $X_j$ is an object $X$ together with a collection of morphisms $i_j:X_j\longrightarrow X$ which satisfy a universal property, in this case: for any object $Y$ and any collection of morphisms $f_j : X_j \longrightarrow Y$, there exists a unique morphism $f$ from $X$ to $Y$ such that $f_j = f \circ i_j$. That is, the following diagram commutes (for each $j$):

\[f=\coprod_{j \in J} f_j: \coprod_{j \in J} X_j \to Y \; \mathrm{for} \; X_1\oplus X_2 \sim  \mathbf{E}_{\mathrm{biovi}}^{L^{*}} \oplus \mathbf{E}_{\mathrm{biovi}} \; \mathrm{iff} \; \forall \: v_{r,g,b} \vee \Delta E \in \mathsf{S} :: \mathrm{T}\] \vspace{-16pt}\[\ \ \ \ \ \ \ \ \ \ \ \ \ \ \ \ \ \ \ \ \ \ \ \ \ \ \ \ \ \ \ \ \ \ \ \ \ \ \ \ \ \ \ \ \ \ \ \ \ \ \ \ \ \ \ \ \ \ \ \ \ \ \ \ \ \ \ \ \ \ \ \ \wedge \; \mathrm{man}(o) \stackrel{\lambda_{\mathrm{bend}}\:,\: \gamma }{\longleftrightarrow} \mathrm{fly}(o) :: \mathrm{T}\]
\vspace{-6pt}
\[\xymatrix{
& Y &\\
X_1 \ar[r]_{i_1}\ar[ur]^{f_1} & X_1\coprod X_2\ar@{-->}[u]^f & X_2\ar[l]^{i_2}\ar[ul]_{f_2}
\sim }\xymatrix{
& {\beta_\gamma} &\\
{\mathbf{E}_{\mathrm{biovi}}^{L^{*} }} \ar[r]_{{\nu_L}\wedge Q  \ \ \ \ }\ar[ur]^{\phi} & {\mathbf{E}_{\mathrm{biovi}}^{L^{*} }}\coprod {\mathbf{E}_{\mathrm{biovi}} }\ar@{-->}[u]^f & {\mathbf{E}_{\mathrm{biovi}} } \ar[l]^{{\ \ \ \ \ \ \  \beta _{\nu }}\wedge L^{*}}\ar[ul]_{\phi}
}\]

\noindent{\footnotesize{\textbf{Figure 5.1. }A commutative diagram and its unique morphism for the coproduct of the pulses' product relevant to biovielectroluminescence conversions which is done via relativity and bendable wavelengths. The diagram in proposition, indicates a man's observation scope relative to the small subject here, the fly's observation dependent to the conversions of the phenomenon as they take place. }
\smallskip \vspace{4pt}

\normalsize \noindent It is conceivable that this conversion behaviour via morphism on functions of $\beta _{\gamma }$, is in importance of the commutativity between substitutable components of biovielectroluminescence equations inclusive to observation-scope function $scope\left( o \right)$ from relation (3.7), \S\ref{section3}. This holds firm in terms of commutatively SR by GR into UR for all bodies observing space-time events, situating photon $h\nu$ for unoccured motion-stretch of the scene object's $k$-curvature, which could therefore be derived as follows
\vspace{1pt}
\[\therefore \mathbf{E}_{\mathrm{biovi}}^{L^{*} } =\frac{\nu _{L} Q}{A_{\mathrm{fly}\left(\sphericalangle \right)} } =\frac{L^{*} It}{\left\langle t\right\rangle A_{\mathrm{fly}\left(\sphericalangle \right)} } =\frac{L^{*} I}{\left(t+\mathrm{cons}\left(t\right)\right)A_{\mathrm{fly}\left(\sphericalangle \right)} } =\frac{\Delta E}{\tau _{\mathrm{fly}\left(\sphericalangle \right)} A_{\mathrm{fly}\left(\sphericalangle \right)} }\]
\[  \ \ \ \ \ \ \ \ \  \ \ \ \ \ \ \ \ \; =\frac{h\nu }{\tau _{\mathrm{fly}\left(\sphericalangle \right)} A_{\mathrm{fly}\left(\sphericalangle \right)} } =\frac{VI}{A_{\mathrm{fly}\left(\sphericalangle \right)} } {\rm \ in\; W}\:\mathrm{m}^{-2} \; \mathrm{or} \; {\rm \; kg}\:\mathrm{s}^{-3} {\rm \; } \ \ \ \ \ \ \ \ \  \ \ \ \ \ \ \ \ \ \ \ \ \ \ (5.3) \] \\
\vspace{-10pt}

\noindent This conjectures a bendable wavelength ${\lambda_{\mathrm{bend}}}$ for an unoccured event in scene $\sf S$, to be weighted by the luminosity function that models human brightness sensitivity for a certain density of biovielectroluminescence effect. So, instead of $\mathrm{kg}\:{\mathrm{m}^{-3}}$ the effect is now measured into $\mathrm{kg\:s}^{-3}$, since space-time for the latter dimensional analysis is in continuum in respect to the contents of \S\ref{section3} and \S\ref{section4}, which is thereby due to the existential notion of bendable wavelengths of $\lambda_{\mathrm{bend}}$ from Eqs.$\,$(2.2a) and (2.4b). Ergo,
\[\therefore \beta _{\gamma } \equiv \mathbf{E}_{\mathrm{biovi}} =\beta _{\nu } L_{\nu }^{*} \; \mathrm{rel}\left(A_{\mathrm{fly}\left(\sphericalangle \right)} \right)=\nu _{L} \cdot Q \; \mathrm{rel}\left(A_{\mathrm{fly}\left(\sphericalangle \right)} \right){\rm {\mathcal V}}=\frac{VI\Delta {\rm {\mathcal V}}}{A_{\mathrm{fly}\left(\sphericalangle \right)} } \int \gamma \: \mathrm{d}k \] \[ \ \ \ \ \ \ \  \ \ \ \ \ \ \ \ \ \ \ \ \ \ =h\nu v_{r,g,b} {\rm \; in\; kg}\: \mathrm{m}^{3} {\rm s}^{-3} \; \mathrm{or} \; {\rm \; J}\mathrm{m \: s}^{-1} . \ \ \ \ \ \ \ \ \  \ \ \ \ \ \ \ \ \ \ \ \ \ \ \ \ \ \ \ \ \ \ \ (5.4) \]

\noindent For typical circumstances where change of volume $\mathcal{V}$ is conditioned into the objects within the field of scene $\sf S$, one unequivocally elicits
\[\therefore \nu _{\beta } =\frac{c}{\lambda _{{\rm {\mathcal B}}} } {\rm \; }\left|\lambda _{{\rm {\mathcal B}}} \in \left({\rm {\mathcal B}}^{>0} ,{\rm {\mathcal V}}_{\sf S} \vee \Delta {\rm {\mathcal V}}\right)\right. ,\Delta {\rm {\mathcal V}}=\left({\rm {\mathcal V}}_{\sf S} -{\rm {\mathcal V}}_{\mathrm{fly}} \right)+A_{\mathrm{fly}\left(\sphericalangle \right)} \left(\Delta \lambda \right)_{\sf S\vee \mathrm{fly}} \]
\noindent and thus via general assumptions made upon bodies ${\mathcal B}^{>0}$ and ${\mathcal B}^{<\infty }$, educe

\[\begin{array}{l} {\therefore f\left(v_{\mathrm{man}\left(o\right)} \right)\approx v_{\mathrm{fly}\left(o\right)} \because v_{\mathrm{man}\left(o\right)} \to c=\nu _{\beta } \lambda _{{\rm {\mathcal B}}} \left|\lambda _{{\rm {\mathcal B}}} \in \left({\rm {\mathcal B}}^{>0} \wedge {\rm {\mathcal B}}^{<\infty } ,{\rm {\mathcal V}}_{\sf S} \wedge \Delta {\rm {\mathcal V}}\right)\right.} \end{array}\] where, \[\begin{array}{l} \; {\left\{{\rm \; {\mathcal V}}_{\sf S} \wedge \Delta {\rm {\mathcal V}}=\left({\rm {\mathcal V}}_{\sf S} -\left({\rm {\mathcal V}}_{\mathrm{fly}} +{\rm {\mathcal V}}_{\mathrm{man}} \right)\right)\times \frac{A_{\mathrm{fly}\left(\sphericalangle \right)} \left(\lambda _{2} -\lambda _{1} \right)_{\sf S\wedge \mathrm{fly}} }{A_{\mathrm{man}\left(\sphericalangle \right)} \left(\lambda _{2} -\lambda _{1} \right)_{\sf S\wedge \mathrm{man}} } \right\} \rightarrow }\end{array}\ \ \ \ \ \ \ \ \ \ \ \ \ \ \ \ \ \ \ \ \ \ \ \ \ \ \ \ \ \ \ \ \ \ \ \ \ \ \ \ \ \ \ \] \[ \ \ \ \min \left({\rm {\mathcal V}}{}_{{\rm {\mathfrak V}}\to {\rm man}\left(o\right)  } \right) = \mathfrak{R}\mathcal{V}_{{\rm man}(o)} \; \left| \: \mathfrak{R} \in \mathbb{R}^{n+1} \; , \; n\in \mathbb{Z} \; . \right. \ \ \ \ \ \ \ \ \ \ \ \ \ \ \ \ \ \ \ \ \ \ \ \ \ \ \ \ \ \ \ \ \ \ \ \ \ \ \ \ \ \  (5.5)\]\vspace{-12pt}
\smallskip

\noindent Ergo, the gradient of some visual observation for a fly works the opposite way around for the given ratio multiplied by the change of scene's volume $\mathcal{V}$, and its main subjects here, human and the fly. Hence, the change of the man's body wavelength against fly (held by symbol $\lambda _{{\rm {\mathcal B}}}$), would indicate the possibility of minimum visual condition for the man's observation, if, values are stated in the denominator as given above. Otherwise, indicates the possibility of minimum visual condition for the fly which itself corroborates with the basic assumptions made in the previous Chapters including the main Hyp.$\,$3.1, \S\ref{section3}. The result of relation (5.5), is of a product defining either bodies' visual condition dependant to ratio number $\mathfrak{R}$, which represents the multiplied division course (righthand attribute) between the two involved bodies.

With resemblance to electro-luminescence (EL) phenomenon, the `biovielectroluminescence' in its mechanism deals with radiative combination of electrons and holes in the organic material, in regard to its luminance frequency, as proposed in this passed Section. Bear in mind to the Section's conclusion, \\

\noindent ----------- Luminance frequency $\nu_L$, encompasses the measurement of density and contrast and not the bio-optical and biovielectrical phenomena, where the questioned material emits light in response to an electric current $I$ as a function of time. The responsivity is of bidirectional between source and destination satisfying pixel velocity $v_{r,g,b}$, whilst observing the phenomenon itself, it is in the form of energy $E$ multiplied by the former, simultaneously. ----------- \\

\noindent----------- Mostly used limited notations and their description in this Section were:\\

\footnotesize \begin{tabular}{p{0.9in}p{3.4in}}
$\sf S $  & scene \\
$ {} $ \\
$A_{\mathrm{fly}\left(\sphericalangle \right)} $  & surface area on the eye of the fly\\
$ {} $ \\
$\mathrm{rel}\left(A_{\mathrm{fly}\left(\sphericalangle \right)} \right)$ & relative function of $A_{\mathrm{fly}\left(\sphericalangle \right)}$ in satisfaction of synchronized observation between human and fly for each of the latter's eye lens and surface element \\
$ {} $ \\
$\tau_{\mathrm{fly}\left(\sphericalangle \right)}$ & proper time devoted to the point charge quantity from/on the fly fly's eye surface \\
$ {} $ \\
$ v_{\mathrm{fly}(o)}$ & the speed of observation devoted to fly's vision \\
$ {} $ \\
$ v_{\mathrm{man}(o)}$ & the speed of observation devoted to human vision \\
$ {} $ \\
$\mathcal{V}$ &  Volume of some quantity. Here were some types of $\mathcal{V}$ in use: volume of visual system, denoted by $\mathcal{V}_{\mathfrak{V}}$ ; volume of a human body, denoted by  $\mathcal{V}_{\mathrm{man}}$; volume of a fly's body, denoted by $\mathcal{V}_{\mathrm{fly}}$; volume transformation of visual system into human observation system i.e. human vision, expressed via $\mathcal{V}_{\mathfrak{V}\to {\rm man}(o)}$; volume of scene \textsf{S}, denoted by $V_\textsf{S}$  \\
$ {} $
\end{tabular}
\normalsize whereas more on the general notations for all Sections are given in Appendix III. -----------

\section{Conclusion and future remarks}
\vspace{4pt}
\label{section6}
The basics of optimizing graphical images including interpolation, constructing images automatically whilst picture frame anticipation of near future events due to the geometry of the fly's eye and its vision, is explained on this level of the research. However, it is believed that the theory would finalize and vindicate its basic claims/objectives experimentally. Thus examining the resultants regarding time factor, latency/anti-latency issues, pixel integration and etc. comes into focus. However, the main concentration of all of the outlined tasks shall be invested into computing the efficiency factor for the progressing levels of the project as an industrial product-frame accomplishment, \S\ref{section5} and Appendix II.

In the concept of advanced photography such as applications of holography techniques, advanced light imaging techniques, upgrading immensely the scope of image capture and reconstructing computer synthesized $360^{\mathrm{o}}$ 2D-to-5D virtual images for optically variable security features(OVD), are the other promising applications of this project. This is also applicable to metascan e.g., medical image post-processing applications analysis via MRI as Dynamic Contrast-Enhanced MRI (DCE-MRI), in an advanced manner~\cite{46-Metascan}.

Aspects of liquid crystal display (LCD) systems are treated in form of incorporating biovielectroluminescence effect as \emph{bright monochrome pixels} and for its projection unit conducting the phenomenon, used as a filter amongst LCD filters~\cite{06-Wiki}. This substitution for LCD filters in specification as perfect organic eye filters, reduces the complexity of luminance techniques concerning the level of light emission with the attainment of simplified channeling of pixel resolution due to the filter's shape, structure and application. In this project, the lifetime parameter regular to OLEDs as a problematic issue, could be similarly measured using a system for the apparatus's lifetime testing. This measurement could be employed during experimentation when loss of functionality on one or more of the device components under development are confronted with.

Early cancer detection in medical applications, and security purposes including military applications are the other potentials of the project. For instance, determining rocket's trajectory of its future motion or unoccured motion (pre-kinematics) regardless of the rocket's installed programming coordinates for a hit in self-defense military applications could be of use; in specialization of \emph{bendable wavelengths}, Section \S\ref{section2} via Fig.$\,$2.3 and Eqs.$\,$(2.2a), (2.4b). The previous aspect requires a thorough investigation coming up with an ultimate simulation model and thus being tested under laboratorial conditions as a separate issue.

In conclusion, the project's aim is to conclude in major, the 8th Objective, \S\ref{section1} of the project which should be based on powerful deductions on Objectives \#3 and \#6, \S\ref{section1}, of the UR theory, as a provable theory and empirically correct in \emph{n}-dimensional physical systems and applications. In the minor, the experimental results will speak for themselves leading to the 8th Objective of the project.

\begin{flushleft}
\noindent \textbf{Statement of conclusion:} \end{flushleft}
\vspace{4pt}
\noindent --------- \emph{Not being overprotective upon the well-known notions submitted, evaluated, proven and thus scientifically confirmed, yet, there does exist an unquestionable doubt on discovering oddity and occult beyond one's knowledge and intelligence. I think, this is what we seek for, conceptually, visually, cognitively and beyond any system of realities in our universe which hence in notion, reflects upon the pragmatics of n-dimensional visionary projects. In the case of my own visionary prospect to any dimension, is defined in terms of `the theory of unified relativity for a biovielectroluminescence phenomenon via some small body's n-dimensional visual and imaging system linked to giant systems.'} ---------

\begin{flushright}
The Author \end{flushright} \vspace{-12pt}

\section*{Acknowledgments}
\vspace{4pt}
The work described in this paper was supported personally as an extension to the basis of the author's scientific book, research project license no: TXU001347562, USA (2007). The author thanks, G. Goodwin \& M. Dickinson \emph{of} the Faculty \textit{of }Technology, Department \textit{of} Computing, University \textit{of }Lincoln, UK (2005), for their departmental written confirmation on behalf of the author's personalized scientific activities. It is highly appreciated for Dr. H. Alipour \& Dominic S. Zandi's help, their moral support and basic costs of the project in release.
\vspace{-2pt}
\textbf{  }

\section*{Appendix I. {\large \emph{Sample problems and solutions }}}
\markboth{}{Appendices}
\vspace{6pt}
The following problems are in conjunction with Sections \S\ref{section2}, \S\ref{section5}, and similar issues that may relate to Eqs.$\,$(5.4), (2.2-2.6), and Subhyp.$\,$3.1.2, \S\ref{section3}.

\vspace{6pt}
\noindent \textbf{Sample problem 2.1. }It is desired to find for one object in its frame of curvature motion, a velocity of $0.1 \mathrm{m\:s}^{-1}$ relative to a stationary body conducting light and projecting it onto the moving object in a $95^\mathrm{o}$ angle. \emph{\textbf{a}}) What is the anticipated length as a consumed worldline in pixel size $x = 0.023 \: \mathrm{cm}$, while $x' = 0.01 \: \mathrm{m}$ as a displaced length to its parallel aligned distance, $y$= 0.4318 mm as another pixel dimension for the frame of motion off of the scene recorded relative to $x$ and $x'$? Give this in a Newtonian time frame of 1s, in the process of a visually utilized body beyond our three-dimensional system; \emph{\textbf{b}}) What is the total length inclusive to its anticipated release denoting circuital travel of the length's data and the area of scene's space, accumulatively? Construct this in an \emph{n}-dimensional matrix of the cross product in $\mathbb{R}^{n+1}$ via the notion of wedge product in exterior algebraic form (mathematics, Ref.~\cite[xviii]{06-Wiki}). Rationalize why this form of construction is appropriate? What is the constraint to the matrix in terms of its interdependency with the object in question of the problem's face? \\

\noindent \textbf{Solution to }\emph{\textbf{a}}). Using Eqs.$\,$(2.5) and (2.3), due to the inputs given, $x_k$, appears as a transformable length into pixel length displayed on screen, thus $x_{k \leftrightarrow r,g,b}$. We then compute the anticipated length as follows
$$x_{k \leftrightarrow r,g,b}=\frac{{v_{k} \: \Delta xy_{r,g,b} \cos \theta  }}{{c^2 u\left( t \right)}}  = \frac {(0.1 \: \rm ms^{-1})(0.023 \mathrm{cm} - 0.01 \mathrm{m})(0.4318 \mathrm{mm})\cos(95^\mathrm{o}) } {(c^2) (1 \mathrm{s})} \ \ \ \ \ \ \ \ $$  $$ \ \ \ \ \ \ \ \ \ \ \ \ \approx 4.09102 × 10^{-25} \: \mathrm{m} \ \ \ \ \ \ \ \ \ \ \ \ \ \ \ \ \ \ \ \ \ \ \ \ \ \ \ \ \ \ \ \ \ \ \ \ \ \ \ \ \ \ \ \ \ \ \ \ \ \ \ \ \ \ \ \ \ \ \ \ \ \ \ (\rm I.1)$$ \\
\smallskip
\noindent \textbf{Solution to }\emph{\textbf{b}}). Benefiting from the basics of vector calculus, the length component for data for a $k$-curve in a tangential course of data construction in triangular mappings, appears as
\[\left\{\mathrm{hypotenuse\longrightarrow  }\Delta x_{r,g,b\longrightarrow k}\right\}= \frac {\mathrm{adjacent}}{\cos \theta}= \frac {(0.023 \: \mathrm{cm}) - (0.01 \mathrm{m})}{\cos(95^\mathrm{o})} \] \[ \ \ \ \ \ \ \ \ \ \  \ \ \ \ \ \ \ \ \ \ \  \approx 0.11209 \: \mathrm{m} \ \ \ \ \ \ \ \ \ \ \ \ \ \ \ \ \ \ \ \ \ \ \ \ \ \ \ \ \ \ \ \ \ \ \ \ \ \ \ \ \ \ \ \ \ \ \ \ \ \ \ \ \ \ \ \ \ \ \ \ \ \ (\mathrm{I}.2)\]
\noindent therefore, the total length inclusive to its anticipated release would be

\[   0.11209 + 4.09102 × 10^{-25} = 0.11209(25) \: \mathrm{meters},\] \noindent giving out for its discrete area of the space of scene accumulated with the previous addition,
\[ A = \Delta xy_{r,g,b} \cos \theta =  (-9.77 \: \mathrm{mm}) (0.4318 \: \mathrm{mm})  \cos(95^\mathrm{o}) \] \[ \ \ \ \ \ \ \ \ \ \ \ \ \ \ \ \ \ \ \ \ \ \ \ \ \ \ \ \ \ \ \ \ \ \ \ \ \ \ \  = 3.67682 × 10^{-7} \mathrm{m}^2 \parallel 0.11209(25) \: \mathrm{m} \ \ \ \ \ \ \ \ \ \   (\rm I.3) \]

\noindent whence by closure, comes the higher dimensional construction in $\mathbb{R}^{{n+1}\geq3}$ as
\vspace{-4pt}
\[\bigwedge(\mathbf{v}_1,\cdots,\mathbf{v}_n)= \ \ \ \ \ \ \ \ \ \ \ \ \ \ \ \ \ \ \ \ \ \ \ \ \ \ \ \ \ \ \ \ \ \ \ \ \ \ \ \ \ \ \  \ \ \ \ \ \ \ \ \ \ \ \ \ \ \ \ \ \ \ \  \ \ \ \ \ \ \ \ \ \ \ \ \ \ \]
\[ \left| {\begin{array}{*{20}c}
   {v_1 ^1 } &  \ldots  & {v_1 ^{n + 1} }  \\
    \vdots  &  \ddots  &  \vdots   \\
   {\begin{array}{*{20}c}
   {v_n ^1 }  \\
   {{\bf e}_1 }  \\
\end{array}} & {\begin{array}{*{20}c}
    \cdots   \\
    \cdots   \\
\end{array}} & {\begin{array}{*{20}c}
   {v_n ^{n + 1} }  \\
   {{\bf e}_{n + 1} }  \\
\end{array}}  \\
\end{array}} \right| \Rightarrow \left| {\begin{array}{*{20}c}
   {{\rm 0}{\rm .11209} \ldots _{\rm 1} {\rm m}} &  \cdots  & {{\rm 3}{\rm .67682  \times  10}^{ - 7} _1 {\rm m}^2 }  \\
    \vdots  &  \ddots  &  \vdots   \\
   {\begin{array}{*{20}c}
   {{\rm 0}{\rm .11209} \ldots _n {\rm m}}  \\
   {\left( {{\rm 1,0}} \right)}  \\
\end{array}} & {\begin{array}{*{20}c}
    \cdots   \\
    \cdots   \\
\end{array}} & {\begin{array}{*{20}c}
   {{\rm 3}{\rm .67682   \times  10}^{ - 7} _n {\rm m}^2 }  \\
   {\left( {{\rm 1,1}} \right)}  \\
\end{array}}  \\
\end{array}} \right| \]
\vspace{4pt}
\noindent which is well defined for the articulated problem. $\ \ \ \ \ \ \ \ \  \ \ \ \ \ \ \ \ \ \ \ \ \ \ \ \ \ \ \ \ \ \ \ (\mathrm{I}.4)$

\vspace{4pt}
\textbf{Rationale:} The established higher dimensional construction, is identical in structure to the determinant formula for the normal cross product in $\mathbb{R}^3$ except that the row of basis vectors is the last row in the determinant rather than the first. The reason for this is to ensure that the ordered vectors $(\mathbf{v}_1,...,\mathbf{v}_n,\wedge$ $(\mathbf{v}_1,...,\mathbf{v}_n))$ have a positive orientation with respect to $(\mathbf{e}_1,...,\mathbf{e}_{n+1})$. If $n$ is even, this modification leaves the value unchanged, so this convention agrees with the normal definition of the binary product. In the case that $n$ is odd, however, the distinction must be kept~\cite[xviii]{06-Wiki}. Perceivably, all of these essentials are appreciated in the constructed higher dimensional matrix. In this problem, one must realize $\mathbf{e}_{i}$ represents orthogonal vectors defining spacial and timelike components for the accumulated dimensions of space.

\vspace{4pt}
\textbf{Constraint:} From the problem's face, the object is strongly intact to its curvature in terms of its course of motion. Thus, the matrix constructed above, is of instantaneous state anticipated on the length competent and must be updated when frame of curvature motion, a velocity of $0.01 \mathrm{m\:s}^{-1}$ changes before the object's space-time event occurrence. That is, factors of pixel velocity and indeed the photon being conducting by the stationary body onto the object with the specific angle against non-relativistic moments of motion-stretch in practice. This pre-moment of determination is plausible in-between the dimensions of $\mathbf{v}_i^1$  and $\mathbf{v}_n^2$ with respect to $\mathbf{e}_1$ and $\mathbf{e}_2$, which are the dotted parts between them in the matrix, accordingly (say, 1.5-dimensional approaching 2-dimensional in this problem). \\

\noindent \textbf{Sample problem 2.2.} Using the concept of anticipated phase velocity $|v_\mathrm{p}|$ from Section \S\ref{section2}, \emph{\textbf{a}}) give a solution to the object's speed as in Sample problem 2.1, and its frame of curvature motion travelling at the speed of light representing the projected light onto it pre-kinematically. Suppose the pixel frame is a fixed frame in square dimensions of 0.377 mm, and the speed of light $c= 299,792,458 \: \mathrm{m\:s}^{-1}$. \emph{\textbf{b}}) For a collapsing dimension beyond the subatomic levels under forces of quantum gravity, imagine the beam of light exhibits quantum states in a square slice of a cavity with a diameter $d=1.6264 × 10^{-35} $m in a 10 degree course covering a magnitudal area of quantum trap at one unit of Planck time regardless of its phase velocity $v_\mathrm{p}$. Coordinates are ($\varrho$,0,0) for $\bf x$, and (0,$\varrho$,0) for $\bf y$, where $\varrho$ is the random radius of quantum trap.  What is the cavity's length in consumption defining the dimensions of its curvature in space? Does this cavity \emph{at the moment of pre-kinematics} sustain an amorphous geometry for its dimension, otherwise a stable one such as a crystalline structure? \textbf{\emph{c}}) What would be the recordable biovielectroluminescent data on the experimenter's  monitor concerning part 2.2\emph{a} after 1s, once the photon is released from its cavity for the acquired cosmic frequency on the object in part 2.2\emph{b}, whilst propagates in space pre-kinematically regardless of its direction. Also give the observation's hypothetical coordinates from the scene of event occurrence and exemplify a post-kinematical sample of the event proving states of $\lambda_{\mathrm{bend}}$. Planck constant $h=6.626068 × 10^{-34} \mathrm{J}\:\mathrm{s}$, and Planck time $t_P \approx 5.39121\left( {40} \right) \times 10^{ -44} {\rm s}$. \\

\noindent \textbf{Solution to }\emph{\textbf{a}}). Based on Eqs.$\,$(2.4b) and (2.4e), we obtain for the anticipated phase velocity
\vspace{-4pt}
\[|v_\mathrm{p}| = \frac{c^2}{v_{r,g,b\rightarrow k}} = \frac {c^2}{0.1 \: \mathrm{ms}^{-1}} = 8.98755179 × 10^{14} \: \mathrm{km}\:\mathrm{s}^{-1} \:, \]
\[ \ \ \ \ \ \ \ \ \ \ \ \ \ \ \ \ \ \ \ \ \ \ \ v_{\mathrm{p}} = |v_{\mathrm{p}}| + c = \xi c + c = 8.98755179\ldots × 10^{14} \: \mathrm{km\:s}^{-1} \ \ \ \ \ \ \ \ \ \ \ \ \ \ \ \ \ \ \ \ \  (\mathrm{I}.5) \]

\noindent the fixed pixel frame has no use at this time until the attainment of resultants on Solution 2.2\emph{c}. The propagation of wave faster than light (FTL) does not prohibit the particle performing superluminal velocities since the notion is observed pre-kinematically and not as it occurs after the event. Thus, this velocity for phases of $k$, is of the velocity in a pre-kinematical manner faster than the projected photon onto the object as its unoccured information or matter on the curvature motion. \\

\noindent \textbf{Solution to }\emph{\textbf{b}}). The direction of the photon is in favour of the sine of the penetrated angle in cross product magnitude for the area of quantum trap whereby,
\[|\mathbf{A}| =|\mathbf{x} × \mathbf{y}| =|\mathbf{x}| \:|\mathbf{y}| \: \mathrm{sin} \theta \: , \]
\[\ \ \  \ \ \ \ \ \ \ \ \   \ \ \ \ \ \ \ \ \ \ \ \ \ \ \ \ \  x_k = \frac{(|v_{\mathrm{p}}| + c) |\mathbf{x}|_{x^{\mu},\lambda_{\mathrm{bend}}} |\mathbf{y}|_{x^{\mu},\lambda_{\mathrm{bend}}} \sin \theta  }{ c^2 \: t_{P}} \: , \ \ \ \ \ \ \ \ \ \ \ \ \ \ \ \ \ \ \ \ \ \ \ \ \ (\mathrm{I}.6)\]
\[x_k=  \frac {v_{\mathrm{p}}\sqrt{\left(\frac{1.6264 × 10^{-35} \: \mathrm{m}}{2}\right)^2} \sqrt{\left(\frac{1.6264 × 10^{-35} \: \mathrm{m}}{2}\right)^2}\; \sin(10^\mathrm{o})} {c^2 (5.3912 × 10^{-44} \: \mathrm{s})} \approx 2.1299 × 10^{-30}\:  \mathrm{km} \]

\noindent To determine whether the structure is of amorphus type or not, one must compute the frequency of a second photon (assume it to exist) in projection to the material here, the cavity while having its values of pre-kinematics calculated from before. Hence, $\vspace{-6pt}$
\[\ \ \  \ \ \ \ \ \ \ \ \ \ \ \ \ \ \ \ \ \ \ \ \ \ \ \ \ \ \ \ \ \nu = \frac{c}{x_k}= \frac {299792458 \: \mathrm{m\:s}^{-1}}{2.1299 × 10^{-27} \:\mathrm{m}}=1.40747658 × 10^{35} \: \mathrm{Hz} \ \ \ \ \ \ \ \ \ \    \ \ \ \ \  (\mathrm{I}.7) \]

\noindent The obtained result is in a frequency range above the Yottahertz class ($10^{24}$)~\cite{06-Wiki}, at the moment of pre-kinematics. Thus, interpretably one may deduce
$$\therefore x_k \longrightarrow  \lambda_{\mathrm{bend}}\;, \ \mathrm{where} \ \lambda_{\mathrm{bend}} = \frac{v_w t}{\nu t_{\mathrm{cons}}}\;,$$

\noindent which is of Eq.$\,$(2.2a), and the producible energy from the material (Solution 2.2\emph{c}) is most probably of an amorphous structure, since it satisfies measurements of diffuse gamma-ray emission~\cite{28-Schlessingerman}. This type of cavitation in periods of $t_P$, is of an extreme cosmic spectra class of the electromagnetic waves in the universe, despite of secondary photons covering magnitudes of $d$, the Planck time holds a multi-dimensional collapse on matter for $x_k$. One could state it imaginatively: \emph{A cavity where all matter collapses and any release of it, is in form of pulses holding the bombardment of catastrophic rays or the birth of new matter!}

\vspace{6pt}
\noindent \textbf{Solution to }\emph{\textbf{c}}). Benefiting from Eqs.$\,$(2.6), solution frame (I.5), and de Broglie's wave equation on part 2.2\emph{b}'s solution result, the previous data is recorded and displayed after computing the amount of photon's cosmic energy $E_{h\nu}$, into the object in terms of
\vspace{-4pt}
\[\nu = \frac {c} {x_k} = \frac{ c^3 \: t_{P}}{v_{\mathrm{p}} x_{d}y_{d}  }\: ,  \]
\[ \nu = \frac  { c^3 (5.3912 \times 10^{-44} \: \mathrm{s})}{(8.98755179 × 10^{14} \: \mathrm{km} \: \mathrm{s}^{-1}) (1.6264 \times 10^{-35}\: \mathrm{m})^2} \approx 6.11014357 × 10^{33} \: \mathrm{Hz} \: , \]
\[\ \ \  \ \   \ \ \ \ \ \ \ \ \ \ \ \ \ \ \ \ \ \ \ \ \ \ \ \ \ \ \ \ \ \ \ \therefore E_{h\nu}=h\nu \approx 4.04862268 \:\mathrm{ J} \: , \ \ \ \ \ \ \ \ \ \ \ \ \ \ \ \ \ \ \ \ \ \ \ \ \ \ \ \ \ \   \ (\mathrm{I}.8)\]
\vspace{2pt}
\noindent which is at the point of photon's release (already traversed a full diameter of $d$, of the cavity) regardless of its direction. The amount of energy exhibits values much greater than a visible light given for an energy carried by a single photon around a tiny $4 × 10^{-19}$ joules. Now after the previous metric occurrence $(r,r,0,t_P)$, for 1s as $(r,r,0,1)$, the transformations appear in form of
\[ \mathrm{\mathrm{firstly}, \; \mathrm{let}} \ f(x_d)= x_{\varrho\rightarrow r,g,b} = \ \sqrt{\left(\frac{1.6264 × 10^{-35} \: \mathrm{m}}{2}\right)^2 + (0.377 \: \mathrm{mm})^2} = \mathfrak{X} \ \ \ \ \ \ \ \ \]secondly,\[ \; \mathrm{let} \ f(y_d)= y_{\varrho\rightarrow r,g,b} = \frac {x_{d\rightarrow r,g,b}}{2}= \mathfrak{Y} \stackrel{\mathrm{dis}}{\sim} \mathfrak{X} \ \because d= g(\varrho) \; \mathrm{for} \; {\bf x}\; \mathrm{as}\; x^{\mu},\lambda_{\mathrm{bend}}  \]\[ \  \ \ \ \ \ \ \ \ \ \ \ \ \ \ \ \ \ \ \ \ \ \ \ \ \ \ \ \ \ \ \ \ \ \ \ \ \ \ \ \ \ \ \ \ \ \ \ \ \ \ \ \ \ \ \ \ \; \wedge \; d = g(\varrho) \; \mathrm{for} \; {\bf y}\; \mathrm{as} \; x^{\mu},\lambda_{\mathrm{bend}}  \ .  \]
Thus
\[ \therefore \nu = \frac {c} {x_k} = \frac{ 2c^3 \: u(t)}{v_{\mathrm{p}} x_{\varphi\rightarrow r,g,b}x_{d\rightarrow r,g,b}  } = \frac{ 2c^3 \: u(t)}{v_{\mathrm{p}} \mathfrak{XY} \stackrel{\mathrm{dis}}{\sim} \mathfrak{X}  } = \frac{ c^3 \: u(t)}{v_{\mathrm{p}} \mathfrak{X}^2  }\: , \]
\[\ \ \ \ \ \ \ \ \therefore  \nu = \frac  { c^3 (1) \: \mathrm{s}}{8.98755179 × 10^{14} \: \mathrm{km} \:\mathrm{s}^{-1} (3.77 × 10^{-7}) \: \mathrm{m}^2 } \approx 7.95205 × 10^{13}  \: \mathrm{Hz} \ \ \ \ \   (\mathrm{I}.9) \] \vspace{4pt}

\noindent which is in the infrared range of electromagnetic spectrum, even in cases where direction is considered. So, the biovielectroluminescence $\beta_\gamma$ exhibits in virtue of
\[\therefore \beta_\gamma \equiv Ev_{r,g,b}=h\nu v_{r,g,b} = \frac { 5.26908545 × 10^{-20} \: \mathrm{J}} {10 \: \mathrm{m}^{-1}\:\mathrm{s}} \: , \] \vspace{-1pt}
\[\ \ \ \ \ \ \ \ \ \ \ \ \ \ \ \ \ \ \ \ \ \ \ \  \ \ \ \ \ \ \ \ \ \ \ \ \ \ \ \ \ \ \ \ \therefore \beta_\gamma  \approx 5.26908 × 10^{-21}  \: \rm kg \: \rm {m}^3 \rm{s}^{-3}  \ \ \ \ \ \ \ \ \ \ \ \ \   \ \ \ \ \ \  \ \  \ \ \ \ \ \ \ \ \ \ \ \ \ \ \ \ \ \ (\mathrm{I}.10)\]

\noindent which is from the scene to the monitor. The amount of energy is of a safe kind lesser then the amount needed to excite a single molecule in a photoreceptor cell of an eye thus contributing to vision, Refs.~\cite{50-Vimal et al.},~\cite[xix]{06-Wiki}.

The location of observation is deemed to be far from the scene since the outcome is of Doppler effect as the length of wavelength is increased a `red-shift' maintaining $(r,r,0,t_P)$, becomes $( v_\rho u(t)-\varrho,0,0,1-t_P)$ for the observer perceiving events with the infrared frequency, pre-kinematically. The post-kinematics may refer to frequencies when one tries a value of $c$ instead of $8.98755179 × 10^{14} \: \mathrm{km} \:\mathrm{s}^{-1}$ for $v_P$, giving out $2.38396599 × 10^{23}$ Hz, or simply appreciating this formulaic frame
\[\ \ \ \ \ \ \ \ \ \ \ \ \ \ \ \ \ \ \ \ \ \ \ \ \ \ \ \ \ \ \ \ \ \ \nu_{\mathrm{post}} = \frac{ c^3 \: u(t)}{v_{\mathrm{p}}\mid\lim_{v_{\mathrm{p}}\rightarrow c}  \mathfrak{X}^2  }=\frac{ c^2 \: u(t)}{ \mathfrak{X}^2} \ . \ \ \ \ \ \ \ \ \ \ \ \ \ \ \ \ \ \ \ \ \ \ \ \ \ \ \  (\rm I.11)\]

\noindent Let $\nu_{\mathrm{post}}$ represent frequencies of post-kinematical events which occur, \emph{after, the effect of cavitation event}, expectably. The notation $u(t)$, is the unit of time function in accordance with Eq.$\,$(2.5), and $v_\rho$ is radial velocity as the measurement of the speed at which stars and galaxies (in this case, the amorphus cavity), are approaching toward or receding from the hypothetical observer.

So in general, (I.11), is an after-effect event equation. This shows the event has occurred and approaching the observer compared with the pre-kinematical state of observation. In fact the difference between pre-kinematical events and post-kinematical events as a function $\mathcal{K}$ say,
\[\Delta \mathcal{K} = \mathcal{K}_ \mathrm{post} -  \mathcal{K}_ \mathrm{pre}\]
\noindent if implied to $\nu$,  then
\[\ \ \ \ \ \ \ \ \  \ \  \ \ \ \ \ \ \ \ \ \ \ \ \ \ \ \ \ \ \ \ \ \ \ \ \ \ \ \ \ \ \ \mathcal{K}(\nu)\equiv \nu_{\mathrm{post}} - \nu \ \ \ \ \ \ \ \ \ \ \ \ \ \ \ \ \ \ \ \ \ \ \ \ \ \ \ \ \ \ \ \ \  (\rm I.12)\]

\noindent would basically comply with wavelength $\lambda$ as being bendable or, $\lambda_{\mathrm{bend}}$, in consequence to the observer's vision. \\

\section*{Appendix II. {\large \emph{Database construction and analysis}}}
\vspace{6pt}
Resultants would be based upon the gathered data in terms of an array of blurred and focused sub-images recorded from the scene's \textsf{C}, \textsf{L} and \textsf{R}, explained in Section \S\ref{section4}. The database shall also contain in its attributes' identity, the neuromatrix data values on I/O's of the realtime recorded pulses. The physical information, however, must be in form of basic data types such as categorized in \emph{real}, \emph{integer} and \emph{natural numbers}; through which, not altering the actual content of the submitted formulae and necessary calculus hereon stated, prior to the experiment. Here is an example of such a formal class of neuromatrix representation: \vspace{-18pt}
\begin{center} \vspace{6pt}$\!$---------------------------------------------------------------------------------
\footnotesize \begin{tabular}{p{0.3in}p{0.5in}p{0.3in}p{0.3in}p{0.3in}p{0.3in}p{0.3in}p{0.3in}}
\textit{O} \ \ \ ------\vspace{1pt} & \textit{Dir }  \ \ \ \ \ \ \ \ \ \ --------\vspace{1pt} &  \textit{In}$^\textbf{3}$\textit{}  ------\vspace{1pt} & \textit{In}$^\textbf{2}$\textit{} ------\vspace{1pt} & \textit{In}$^\textbf{1}$\textit{} ------\vspace{1pt} & \textit{Out}$^\textbf{3}$\textit{}  ------\vspace{1pt} & \textit{Out}$^\textbf{2}$\textit{}  ------\vspace{1pt} & \textit{Out}$^\textbf{1}$\textit{}  ------\vspace{1pt} \\
\textit{}1$\times$1 & 1 & $w_{1}^{\mathbf{3}} x_{1}^{\mathbf{3}} $ & $w_{1}^{\mathbf{2}} x_{1}^{\mathbf{2}} $ & $w_{1}^{\mathbf{1}} x_{1}^{\mathbf{1}} $ & $1\vee 0$ & $1\vee 0$ & $1\vee 0$ \\
2$\times$1 & 2 & $w_{2}^{\mathbf{3}} x_{2}^{\mathbf{3}} $ & $w_{2}^{\mathbf{2}} x_{2}^{\mathbf{2}} $ & $w_{2}^{\mathbf{1}} x_{2}^{\mathbf{1}} $ & $1\vee 0$ & $1\vee 0$ & $1\vee 0$ \\
3$\times$1 & 3 & $w_{3}^{\mathbf{3}} x_{3}^{\mathbf{3}} $ & $w_{3}^{\mathbf{2}} x_{3}^{\mathbf{2}} $ & $w_{3}^{\mathbf{1}} x_{3}^{\mathbf{1}} $ & $1\vee 0$ & $1\vee 0$ & $1\vee 0$ \\
$\vdots $ & $\vdots $ & $\vdots $ & $\vdots $ & $\vdots $ & $\vdots $ & $\vdots $ & $\vdots $ \\
\textit{n}$\times$1\textit{} & \textit{n} & $w_{n}^{\mathbf{3}} x_{n}^{\mathbf{3}} $ & $w_{n}^{\mathbf{2}} x_{n}^{\mathbf{2}} $ & $w_{n}^{\mathbf{1}} x_{n}^{\mathbf{1}} $ & $1\vee 0$ & $1\vee 0$ & $1\vee 0$ \\ ------\vspace{1pt}
$\sum$  --------\vspace{1pt} &  --------\vspace{1pt} $\sum dir_{i}$  --------\vspace{1pt} &  --------\vspace{1pt} $\ge \theta ^{\mathbf{3}} $  --------\vspace{1pt} &  --------\vspace{1pt} $\ge \theta ^{\mathbf{2}} $ --------\vspace{1pt} & --------\vspace{1pt} $\ge \theta ^{\mathbf{1}} $ --------\vspace{1pt} & --------\vspace{1pt} $\forall$T  --------\vspace{1pt} & --------\vspace{1pt} $\forall$T --------\vspace{1pt} & --------\vspace{1pt} $\forall$T --------\vspace{1pt}\\
\textit{$\overline{O}$} & $\left\langle dir_{i} \right\rangle $ & n/a & n/a & n/a & $\exists$T & $\exists$T & $\exists$T \\
\end{tabular}\\
\hspace{20pt}------------------------------------------------------------------------------------------------$\ \ \ \ \ \ $ \end{center}

\noindent{\footnotesize{\textbf{Table II.1.} In this neuromatrix database, $\overline{O}$ stands for the average of all elements in the list for the given operation e.g., $w_1^\mathbf{2}x_1^\mathbf{2}$ for data index 1. Additionally, for all notations' semantical standards, $Dir\buildrel\wedge\over=$ Data index record; $O \buildrel\wedge\over=$  Operation, valued as Operation Row $\times \; Dir$ column.}
\smallskip

\normalsize The extended database exemplar in realization of the corresponding product's normal distribution density function $f(x)$, could be resumed to the one above, consisting sub-attributes of the database original fields. That is,

\begin{center}---------------------------------------------------------------------------------------------$\ \ \ \  $
\footnotesize \begin{tabular}{p{0.6in}p{1.8in}p{1.7in}}
\textit{$Out^{\mathbf{CC}} $} \ \ \ \ ------------ & ${\rm {\mathcal Y}}\left(\pi \right) \: \mathrm{to} \:    w_{i}^{\mathbf{1}} x_{i}^{\mathbf{1}} \: \mathrm{in} \: (k {\rm \% }) \ \ \ \ \ \ \ $ \ \ \ \ \ \ \ \ \ \ \ \ \ \ \ \ \ \ \ \ \ \ \ \ \ \ \  \ \ \ \ \ \ --------------------------------------------- & $\mathcal{X}_{\beta _{\gamma}} \left(\pi \right)\! \sim \mathcal{N}\left(\mu ,\sigma ^{2} \right)^{\mathbf{C}} \ \ \ \ \ \ \ \ $ \ \ \ \ \ \ \ \ \ \ \ \ \ \ \ \ \ \ \ \ \ \ \ \ \ \ --------------------------------------- \\
\textit{}$\kappa _{1}^{\mathbf{3}} ,\kappa _{1}^{\mathbf{2}} ,\kappa _{1}^{\mathbf{1}} $ & $\left\{w_{1}^{\mathbf{3}} x_{1}^{\mathbf{3}} :w_{1}^{\mathbf{1}} x_{1}^{\mathbf{1}} ,{\rm \; }w_{1}^{\mathbf{2}} x_{1}^{\mathbf{2}} :w_{1}^{\mathbf{1}} x_{1}^{\mathbf{1}} {\rm \; }\right\}$${\rm \% }$ & $\mathcal{X}_{\beta _{\gamma } } \left(\pi \right)\sim \mathcal{N}\left(\mu ,\sigma ^{2} \right)_{1}^{\mathbf{3},\mathbf{2},\mathbf{1}} $ \\
$\kappa _{2}^{\mathbf{3}} ,\kappa _{2}^{\mathbf{2}} ,\kappa _{2}^{\mathbf{1}} $ & $\left\{w_{2}^{\mathbf{3}} x_{2}^{\mathbf{3}} :w_{2}^{\mathbf{1}} x_{2}^{\mathbf{1}} ,{\rm \; }w_{2}^{\mathbf{2}} x_{2}^{\mathbf{2}} :w_{2}^{\mathbf{1}} x_{2}^{\mathbf{1}} {\rm \; }\right\}{\rm \; \% }$ & $\mathcal{X}_{\beta _{\gamma } } \left(\pi \right)\sim \mathcal{N}\left(\mu ,\sigma ^{2} \right)_{2}^{\mathbf{3},\mathbf{2},\mathbf{1}} $ \\
$\kappa _{3}^{\mathbf{3}} ,\kappa _{3}^{\mathbf{2}} ,\kappa _{3}^{\mathbf{1}} $ & $\left\{w_{3}^{\mathbf{3}} x_{3}^{\mathbf{3}} :w_{3}^{\mathbf{1}} x_{3}^{\mathbf{1}} ,{\rm \; }w_{3}^{\mathbf{2}} x_{3}^{\mathbf{2}} :w_{3}^{\mathbf{1}} x_{3}^{\mathbf{1}} {\rm \; }\right\}{\rm \% }$ & $\mathcal{X}_{\beta _{\gamma } } \left(\pi \right)\sim \mathcal{N}\left(\mu ,\sigma ^{2} \right)_{3}^{\mathbf{3},\mathbf{2},\mathbf{1}} $ \\
$\vdots $ & $\vdots $ & $\vdots $ \\
$\kappa _{n}^{\mathbf{3}} ,\kappa _{n}^{\mathbf{2}} ,\kappa _{n}^{\mathbf{1}} \ \ \ \ \ \ \  $ ------------ & $\left\{w_{n}^{\mathbf{3}} x_{n}^{\mathbf{3}} :w_{n}^{1} x_{n}^{\mathbf{1}} ,{\rm \; }w_{n}^{\mathbf{2}} x_{n}^{\mathbf{2}} :w_{n}^{\mathbf{1}} x_{n}^{\mathbf{1}} {\rm \; }\right\}{\rm \; \% }$ --------------------------------------------- & $\mathcal{X}_{\beta _{\gamma } } \left(\pi \right)\sim \mathcal{N}\left(\mu ,\sigma ^{2} \right)_{n}^{\mathbf{3},\mathbf{2},\mathbf{1}} \ \ \ \ \ \ \ \ \ \ \ $ --------------------------------------- \\
$\sum \kappa _{i}^{\mathbf{C}} \ \ \ \ \ \ \  $ ------------ & $\sum {\rm {\mathcal Y}}\left(\pi \right) $ to $w_{i}^{\mathbf{1}} x_{i}^{\mathbf{1}} $ in (\textit{k}${\rm \% })\ \ \ \ \ \ \ \ \ \ \ \ \ \ \ \ \ \ $   --------------------------------------------- & $\sum \mathcal{X}_{\beta _{\gamma } } \left(\pi \right)\sim \mathcal{N}\left(\mu ,\sigma ^{2} \right)^{\mathbf{C}}  \ \ \ \ \ \ \ \ \ \ \ $ ---------------------------------------  \\
$\left\langle \kappa _{i}^{\mathbf{C}} \right\rangle $ & $\left\langle {\rm {\mathcal Y}}\left(\pi \right)\right\rangle $ to $w_{i}^{\mathbf{1}} x_{i}^{\mathbf{1}} $ in (\textit{k}${\rm \% }$) & $\left\langle \mathcal{X}_{\beta _{\gamma } } \left(\pi \right)\sim \mathcal{N}\left(\mu ,\sigma^{2} \right)^{\mathbf{C}} \right\rangle $ \\
\end{tabular}\\  \vspace{1pt}
------------------------------------------------------------------------------------------------------------$\ \ \ \ \  $\end{center}

\normalsize \noindent \noindent{\footnotesize{\textbf{Table II.2.} The yield efficiency percentage via its function $\mathcal{Y}$, is committed to $w_i^\mathbf{1}x_i^\mathbf{1}$ for $\pi$-product comparisons. One must consider a more complex format than this table for the real circuit I/O's. However, the data type representation and substructuring method used in form of this table, suffices the neuromatrices concept and the pulses' product-based decisions, concerning their efficiency performance prior to their normal distribution (rightmost column).}

\smallskip
\vspace{4pt}
\normalsize In Tab.$\,$II.2, we have simplified the neuromatrix data format into sequences of lists reflecting the relevant matrix situated within the context of the neuromatrix as introduced in Section \S\ref{section4}. This is to prevent further convolution to the apparent structure of the complex form of the database where the actual one in practice, relies upon computational elementary factors for data value comparisons based on the learning circuit and computational decisions. However, mathematical convolution is evaluated between normal distribution function via the sum operation in the previous database. This allows precise analysis and error report generation for best pulses' product choice (see why this is preferred, and thus conditioned for the functions' evaluation, Ref.~\cite[xv]{06-Wiki} for more details). The graph outlining the characteristics of the neuromatrix is plotted after the data atoms and their statistical values actually being stored in the laboratory.

These are the necessary elements on recording pulses/images data for demonstrating and analyzing the theoretical concept, which leads to a successful experiment in exhibiting the biovielectroluminescence phenomenon accompanying its effectively efficient product of its pulses. Thus, the notion of time-data interpretation holding the visual aspects of some tiny being such as a fly's eye committed to human's vision, is formal to the practical construction of the neuromatrix, the learning circuit and thereby data analysis in one utilizable package.

\section*{Appendix III. {\large \emph{Symbols, expressions and abbreviations}}}
\markboth{}{Appendices}
\vspace{6pt}

\textbf{General list of symbols and expressions} \\
\vspace{6pt}

\footnotesize \begin{tabular}{p{0.9in}p{3.4in}}
$body\left({\rm {\mathcal B}}\right)$ & General body function for non-subatomic bodies \\
$ {} $\\
${\rm {\mathcal B}}$ & Specific body function in a defined range of general body function; body state or simply, body \\
$ {} $\\
$scope\left( o \right)$ & observation-scope function for body ${\rm {\mathcal B}}$ \\
$ {} $\\
${\rm {\mathcal B}}\left(\ell \right)$ & Function of body length or overall body size function \\
$ {} $\\
rel & Relativistic function in special relativity (SR); function of relativity \\
$ {} $\\
$\nu _{L} $ & Luminance frequency \\
$ {} $\\
$\beta _{\nu } $ & Biovielectroluminescence quantity \\
$ {} $\\
$\lambda _{{\rm bend}} $ & Bendable wavelength in UR theory \\
$ {} $\\
$v$ & Velocity from classical mechanics \\
$ {} $\\
$c$ & The speed of light \\
$ {} $\\
$\lambda $ & Wavelength in quantum wave theory \\
$ {} $\\
$\nu $ & Frequency in quantum wave theory \\
$ {} $\\
\textbf{C} & Product class e.g., class of pulses product \\
$ {} $\\
$C_i $ & Element of unit circle; unit circle index; contour index  \\
$ {} $\\
$\mathrm{C}$ & Circuit, given in the form of e.g., $x_\mathrm{C}$ a `circuital travel distance'    \\
$ {} $\\
$x_{{\rm cons}} $ & Distance in consumption; anticipated distance in UR theory  \\
$ {} $\\
$t_{{\rm cons}} $ & Newtonian time in consumption; maximally/minimally time ahead parameter in UR theory \\
$ {} $\\
$v_{r,g,b} $ & Pixel velocity in UR theory \\
$ {} $\\
$\tau $ & Proper time in SR and GR theories \\

\end{tabular}

\begin{tabular}{p{0.9in}p{3.4in}}
$t_{P} $ & Planck time \\
$ {} $\\
$E$ & Energy quantity \\
$ {} $\\
$E$$v_{r,g,b} $ & Quantity of energy-pixel-velocity or, biovielectroluminescence \\
$ {} $\\
$\gamma $ & Relativistic gamma; Lorentz factor in SR theory \\
$ {} $\\
$x_{k} $ & Distance with space-time curvature $k$ in the context of GR to SR as UR theory \\
$ {} $\\
$\left|v_{{\rm p}} \right|$ & Anticipated phase velocity; virtual part of phase velocity $v_{{\rm p}} $ in UR \\
$ {} $\\
$\emph{g}$ & Space-time coefficient as functions of $t,x,y,z$ in GR theory. Not to be confused with ordinary function, $\textmd{g}$, in logic \\
$ {} $\\
$Q$ & Charge quantity in the context of electromagnetism \\
$ {} $\\
$q$ & Specific charge in the context of electromagnetism \\
$ {} $\\
$t_{q} $ & Relativistic time interval for specific charge $q$ of elementary particle (electron) frame of reference in the time dilation concept of SR, now to GR and ergo to SR (see deductive relations of Eq.$\,$3.5c) \\
$ {} $\\
$\mathbf{E}$ & Electric field in the context of electromagnetism. \\
$ {} $\\
$\mathbf{B}$ & Magnetic field in the context of electromagnetism. \\
$ {} $\\
$\mathbf{E}_{{\rm biovi}}^{L^{{\rm *}} } $ & Multifunction and product of biovielectroluminescence $\beta _{\nu } $; Carrier of electric field $\mathbf{E}$ and luminance contrast, $L^{*} $, from luminance frequency, $\nu _{L} $, in product terms of biovielectroluminescence  \\
$ {} $\\
$\rho _{{\rm {\mathcal B}}\left(o\right)} $ & Observation density of relative body in the context of UR theory \\
$ {} $\\
${\rm {\mathfrak V}}$ & Visual system satisfying an observation $o$, of some body e.g., `man' as a giant body or `fly' as a small body, in terms of their $o$'s volume dimensions or hyper dimensions beyond the vision's time limit \\
$ {} $\\
${\rm {\mathcal V}}$ & Volume representing an observation $o$, of some body in terms of length dimensions in the frame of some time limit (in successful scenarios, defined in the frame of space-time limit denoted by interval $s$ from GR theory) \\
$ {} $\\
${\rm {\mathcal K}}$ & Function of pre and post-kinematics. Applicable to frequency $\nu $ or any related physical parameter describing the kinematical concept of the UR system. See solution frames (I.11), (I.12), Appendix I \\
$ {} $\\
$\mathcal{Y}$ & Yield efficiency function. Subject of \S\ref{section4} and Appendix II \\
$ {} $\\
$\mathcal{N}_{\pi}$ & Neuromatrix product. Subject of \S\ref{section4}, Eq.$\,$(4.4) \\
$ {} $\\
\end{tabular}

\begin{flushleft}
\normalsize \textbf{Main list of operators and some elementary notations in logic}
\end{flushleft}
\vspace{7pt}

\footnotesize \begin{tabular}{p{0.9in}p{3.4in}}
$\equiv $ & Equivalence or, `identically equal by definition'  \\
$ {} $\\
$\buildrel\wedge\over= $ & Stands for \dots ; represents. Relevant to the theory of semantics.  \newline Not to be confused with `estimates\dots ' in operation \\
\end{tabular}

\begin{tabular}{p{0.9in}p{3.4in}}

$ {} $\\
$\stackrel{\propto}{=}$ & Proportionally equal to... Sustains a frame of equality between both sides, whilst one side variably proportional to the other side \\
$ {} $\\
$\mathbf{o}$ & Abstract algebraic operator defining laws of commutative, associative, satisfying `distributivity' and `product', in the theory of UR. Study relation (3.7) and its commencing predicate, its differential form in (3.8)     \\
$ {} $\\
$\stackrel{}{\longrightarrow}$ & Implication; implies to... Sometimes denotes a transformation of function via some variable or constant otherwise merely a transformation or implication. See for instance, `morphism' (Eq.$\,$5.2) \\
$ {} $\\
$\in $ & Elements of \dots  in set theory  \\
$ {} $\\
$\left|\right. $ & Such that\dots ; restricted to\dots   in logic, algebraic limits, and in integration expressed as integral restriction(s)  \\
$ {} $\\
$\because $ & Because\dots ; with reason to\dots   \\
$ {} $\\
$\therefore $ & Therefore\dots  in set theory, physical statements and deductive equations \\
$ {} $\\
$\cong $ & Congruent to\dots  , geometrically satisfying some physical parameter of the system \\
$ {} $\\
$\ll $ & Much lesser than\dots  , other relational operators are: `much greater than\dots ', $\gg $; `much-much lesser than\dots ', $\ll \ll $; `much-much greater than \dots ', $\gg \gg $. The last two are considered between colossal differences of a physical body's size dimension. See Pred.$\,$(3.1) \\
$ {} $\\
$\mathrm{T}$   &  `True state' in logic. Substitutable for `1' in boolean logic.    \\
$ {} $\\
$\mathrm{F}$   &  `False state' in logic. Substitutable by `0' in boolean logic    \\
$ {} $\\
$\wedge$   &  `AND logic', in boolean algebraic statements (predicates), related to physical `interceptive' conditions \\
$ {} $\\
$\vee$ &  `OR logic', in boolean algebraic statements (predicates), related to physical `alternative' or `union' conditions\\
$ {} $\\
$\Leftrightarrow$   & If and only if; iff; logical equivalence  \\
$ {} $\\
$\bigcap\nolimits_a^b$   & Intersection closure, suited for discrete mathematical and algebraic problems for any uncertainty behaviour on `unitary physical vectors'. Relations $\,$(3.7) and (3.8), \S\ref{section3}  \\
$ {} $\\
$\bigcup\nolimits_{a'}^{b'} $ & Discrete union closure, suited for discrete mathematical and algebraic problems related to any uncertainty behaviour on `a set of unitary physical vectors'. Relations $\,$(3.7) and (3.8), \S\ref{section3}   \\
$ {} $\\
$\prod  $ & Product of \dots  , relevant to deductive equations of Eq.$\,$(3.7) \\
$ {} $\\
$\coprod  $ & Coproduct of \dots  , relevant to Fig.$\,$5.1 and Eq.$\,$(5.2)   \\
$ {} $\\
\end{tabular}


\begin{thebibliography}{99}
\markboth{}{References}
\vspace{6pt}
\footnotesize
\bibitem{01-Alipour} P. B. Alipour, `Logic, Design and Organization of Prallel Time Varying Data and Time Super-helical Memory As; PTVD-SHAM', Article-id. arXiv:\,0707.1151, \textit{CompSci. Ar. ArXiv.org.}, pp. 1-33, 2007.

\bibitem{02-Alipour} P. B. Alipour, `Theoretical Engineering and Satellite Comlink $\emph{of}$ a PTVD-SHAM System', Article-id. arXiv:\,0710.0244, \textit{CompSci. Ar. ArXiv.org.}, pp. 1-50, 2007.

\bibitem{03-Lee} L. P. Lee, J. Kim and K. H. Jeong, `Biologically Inspired Artificial Compound Eyes', \textit{J. Sci.}, {\textit{Vol.}} \textbf{312} (5773), pp. 557-561, 2006 (\htmladdnormallink {\textcolor{MyBrown}{doi:10.1126/science.1123053}}{http://dx.doi.org/10.1126/science.1123053}), or see for instance: \\ \htmladdnormallink {http://www.berkeley.edu/news/media/releases/2006/04/27\_compoundeye.shtml}{http://www.berkeley.edu/news/media/releases/2006/04/27_compoundeye.shtml}, \emph{Berkeley news California Press, Berkeley University of California}, accessed on 28 August 2007.

\bibitem{04-Fischer} H. Fischer-Nagel and A. Fischer-Nagel, \textit{The housefly}, Carolrhoda Books: Minneapolis, (1990) ISBN 9780876143742.

\bibitem{05-Biro} L. P. Biró \textit{et al.}, `Scanning tunneling microscopy of tightly wound, single-wall coiled carbon nanotubes', \textit{Europhys. Lett.}, {\textit{Vol.}} \textbf{50 }(4), pp. 494-500, 2000.

\bibitem{06-Wiki} \textit{\htmladdnormallink{Online encyclopedia} {http://en.wikipedia.org/wiki/Main_Page}} (2007) \textbf{i-} Scanning tunneling microscope, see for instance: \htmladdnormallink{Scanning tunneling microscope}{http://en.wikipedia.org/wiki/Scanning_tunneling_microscope}; \textbf{ii-} Complex plane, see for instance: \htmladdnormallink{Complex plane}{http://en.wikipedia.org/wiki/Complex_plane}; \textbf{iii-} Visual Basic, see for instance: \htmladdnormallink{Visual Basic(VB)}{http://en.wikipedia.org/wiki/Visual_Basic}; \textbf{iv-} probability distribution, see for instance: \htmladdnormallink{Gaussian distribution}{http://en.wikipedia.org/wiki/Gaussian_distribution} and \htmladdnormallink{Continuous probability distribution} {http://en.wikipedia.org/wiki/Continuous_probability_distribution}; \textbf{v-} Compound eye, see for instance: \htmladdnormallink{Compound eye}{http://en.wikipedia.org/wiki/Compound_eye } and for \emph{vision}, dragonfly, see, \htmladdnormallink{Dragonfly}{http://en.wikipedia.org/wiki/Dragonfly }; \textbf{vi-} Fisheye lens, see for instance: \htmladdnormallink{Fisheye lens}{http://en.wikipedia.org/wiki/Fisheye_lens}; \textbf{vii-} Organic material, see for instance: \htmladdnormallink{Organic matter}{http://en.wikipedia.org/wiki/Organic_material}; \textbf{viii-} \emph{Physiology and genetics}, human biology, see for instance: \htmladdnormallink{Human}{http://en.wikipedia.org/wiki/Human}; \textbf{ix-} \emph{Space-time intervals}, spacetime, see for instance: \htmladdnormallink{Spacetime}{http://en.wikipedia.org/wiki/Spacetime}; \textbf{x-} Field emission and Fowler-Nordheim equation, see for instance: \htmladdnormallink{Fowler-Nordheim equation}{http://en.wikipedia.org/wiki/Fowler-Nordheim_equation}; \textbf{xi-} Planck time and Planck length, see for instance: \htmladdnormallink{Planck time}{http://en.wikipedia.org/wiki/Planck_time}, see also, \emph{future} on:  \htmladdnormallink{Future}{http://en.wikipedia.org/wiki/Future}; \textbf{xii-} \emph{Types of perception}, perception, see for instance: \htmladdnormallink{Color perception}{http://en.wikipedia.org/wiki/Color_perception}; \textbf{xiii-} \emph{Lorentz force in special relativity}, Lorentz force, see: \htmladdnormallink{Lorentz force}{http://en.wikipedia.org/wiki/Lorentz_force}; \textbf{xiv-} \emph{Proper time mathematical formalism}, proper time, see: \htmladdnormallink{Proper time}{http://en.wikipedia.org/wiki/Proper_time}; \textbf{xv-} \emph{Functional analysis}, mathematical convolution, see: \htmladdnormallink {Convolution}{http://en.wikipedia.org/wiki/Convolution}; \textbf{xvi-} \emph{Definition of coproduct}, coproduct, see for instance: \htmladdnormallink {Coproduct}{http://en.wikipedia.org/wiki/Coproduct}; \textbf{xvii-} \emph{Matter wave phase}, phase velocity, see: \htmladdnormallink {Phase velocity}{http://en.wikipedia.org/wiki/Phase_velocity}, see also, \emph{matter wave group velocity}: \htmladdnormallink {Group velocity}{http://en.wikipedia.org/wiki/Group_velocity} and \htmladdnormallink {Wavenumber}{http://en.wikipedia.org/wiki/Wavenumber} ; \textbf{xviii-} \emph{Wedge product}, exterior algebra, see: \htmladdnormallink {Exterior algebra}{http://en.wikipedia.org/wiki/Wedge_product}, see also, \emph{higher dimensions}: \htmladdnormallink {Cross product}{http://en.wikipedia.org/wiki/Cross_product}; \textbf{xix-} Photon, see: \htmladdnormallink {Photon}{http://en.wikipedia.org/wiki/Photon}; \textbf{xx-} \emph{Structure}, Minkowski space, see: \htmladdnormallink {Minkowski space}{http://en.wikipedia.org/wiki/Minkowski_metric}; \textbf{xxi-} \emph{Complex analysis}, line integral, see: \htmladdnormallink {Line integral}{http://en.wikipedia.org/wiki/Line_integral} \textbf{ weblink}.

\bibitem{07-Braitenberg} V. Braitenberg and P. Debbage, A regular net of reciprocal synapses in the visual system of the fly, Musca domestica, \textit{Comp. Physiol. A}, {\textit{Vol.}} \textbf{90}, pp. 25-31, 1974 (\htmladdnormallink{\textcolor{MyBrown}{doi:10.1007/BF00698364}}{http://dx.doi.org/10.1007/BF00698364})

\bibitem{08-Popowicz} Z. Popowicz, `Integrable System Constructed out of Two Interacting Superconformal Fields'. \textit{Inst. Theor. Phys. University of Wroclaw, Poland, }pp. 7-8 (1997), \textit{Quant. Phys. ArXiv.org.}

\bibitem{09-Pucknell} D. A. Pucknell and K. Eshraghian, \textit{Basic VLSI Design} (3rd Ed.) Prentice-Hall Silicon Systems Engineering, Chaps. 4, 9 \& 11, sect. 4-2, 4-2-1; 9-1, 9-2-1-2, 9-2-3, 9-2-3-4; 11-6-1, 11-6-1-1, 11-6-1-2, 11-6-1-3, 1994.

\bibitem{10-Barber} S. Barber, AI: An introduction into Neural Networks, see, for instance: \htmladdnormallink {Introduction into neural networks} {http://www.codeproject.com/cs/algorithms/NeuralNetwork_1.asp}, Sussex University, \textit{Comp. Sci. \& AI}, 2007; see also \emph{Perceptrons: Basic Neural Networking}, AI Horizon: \emph{Comp. Sci. and AI Programming} on \htmladdnormallink{http://www.aihorizon.com/essays/generalai/perceptrons.htm}{http://www.aihorizon.com/essays/generalai/perceptrons.htm}

\bibitem{11-Dickinson} M. Dickinson (Ph.D.) and G. Goodwin, `Mathematics I \& II Lecture Notes on \textit{Discrete Mathematics}', University of Lincoln, UK. (2002-2005).

\bibitem{12-Senior} J. M. Senior, \textit{Optical fiber communications,} (2nd Ed.), Prentice Hall International (UK), Ltd., Chaps. 7 and 10, 1992.

\bibitem{13-Saathoff} G. Saathoff \textit{et al.}\textbf{ }`Relatavisic Doppler effect in the theory of special relativity', \textit{Presentation of the Guido Saathoff modern reenactment of the Ives-Stilwell experiment,} \textit{Max-Planck-Inst}., 2007.

\bibitem{14-Talantsev} A. D. Talantsev, `on the analysis and synthesis of certain electrical circuits by means of special logical operators', \textit{Autom. and Telemech}. {\textit{Vol.}} \textbf{20}, 895-907, 1959.

\bibitem{15-Wang} X. Wang, Y. Liu and D. Zhu, `Pillar-shaped structures and patterns of three-dimensional carbon nanotube alignments', \textit{R. Soc. Chem. Commun.,} pp. 751-752, 2001~(\htmladdnormallink{\textcolor{MyBrown}{doi:10.1039/b100326g}}{http://dx.doi.org/10.1039/b100326g})

\bibitem{16-Joshi} A. Joshi and P. Rheingans, `Illustration-inspired techniques for visualizing time-varying data' on, \htmladdnormallink {http://www.cs.umbc.edu/$\sim$alark1/illustration.pdf}{http://www.cs.umbc.edu/~alark1/illustration.pdf}, University of Maryland Baltimore County, \textit{Proceedings of IEEE Visualization,} pp. 679-686, 2005.

\bibitem{17-Kapany} Kapany and Burke entitled `Optical Waveguides'; Academic Press. N.Y. and London, pp. 233-257, 1972.

\bibitem{18-Gaten} E. Gaten, `Optics and phylogeny: is there an insight? The evolution of superposition eyes in the Decapoda (Crustacea)', \textit{Contributions to Zoology}, {\textit{Vol.}} \textbf{67} (4), 223-236, 1998.

\bibitem{19-Einstein et al.} A. Einstein, H. A. Lorentz, H. Weyl, and Minkowski, \textit{The principle of relativity}, pp. 83-91, 1952.

\bibitem{20-David} D. Barwacz, \emph{time vector}, see for instance: \htmladdnormallink {Time Vector} {http://members.triton.net/daveb/} on, \\ \textcolor{blue}{http://members.triton.net/daveb}, \textit{J. Gen. Sci.}, 2003.

\bibitem{21-Egmond} W. Egmond, \emph{A microscopic survey of the ultimate arthropod machine}, see for instance: \htmladdnormallink {The Eye of the Fly}{http://www.microscopy-uk.org.uk/mag/indexmag.html?http://www.microscopy-uk.org.uk/mag/artfly/eye.html} on, \textcolor{blue}{http://www.microscopy-uk.org.uk}, \textit{Micr. UK, Pub. }1999, accessed on 24 August 2007.

\bibitem{22-Grossberg} M. D. Grossberg  and S. K. Nayar, `A general imaging model and a method for finding its parameters', in: \emph{IEEE Inter. Conf. Comp. Vis.}, {\textit{Vol.}} \textbf{II.}, pp. 108–115, 2001.

\bibitem{23-Orghidan} R. Orghidan, European Ph.D. Thesis, `\emph{Calibration of a structured light-based stereo catadioptric sensor}', Supervisors: J. Salvi, and E. Mouaddib Dr. J. Batlle, Submitted in Girona, University of Girona, pp. 1-5, 2005.

\bibitem{24-Shwartz} M. Shwartz,  \emph{Casanova to caveman}, `Casanova or caveman: Scientists isolate nerve cells that choreograph male fly's courtship behavior', Standford University Press., see for instance: \textcolor{blue}{http://news-service.stanford.edu/pr/2004/flysex-84.html}, \emph{ Pub. J. Nature paper}, 2004.

\bibitem{25-Einstein} A. Einstein, `on the electrodynamics of moving bodies', \emph{English translation of his original 1905 German-language paper published as Zur Elektrodynamik bewegter Körper, J. Annalen der Physik}, {\textit{Vol.}} \textbf{17} (891), pp. 1-24, 1905.

\bibitem{26-Starman} L. Starman, Micro-Electro-Mechanical Systems (MEMS), see for instance: \htmladdnormallink{http://www.cs.wright.edu/people/faculty/kxue/mems/MEMS\_1IntroductionM06.pdf}{http://www.cs.wright.edu/people/faculty/kxue/mems/MEMS_1IntroductionM06.pdf}, \\ Wright State University, pp. 1-15, 2006.

\bibitem{27-Kippers} V. Kippers, `PHYSICAL DEVELOPMENT, GROWTH, MATURATION, AND AGEING', see for instance: \htmladdnormallink{http://www.uq.edu.au/$\sim$anvkippe/an229/growth.html}{http://www.uq.edu.au/~anvkippe/an229/growth.html}, \\ \emph{Dept. of Anat. Sci.}, University of Queensland, Australia, 1999.

\bibitem{28-Schlessingerman} A. Schlessingerman, \emph{The Physics Factbook, An Encyclopedia of Scientific Essays},\\ \textbf{i-} `Mass of an Adult', see for instance: \\ \textcolor{blue}{http://hypertextbook.com/facts/2003/AlexSchlessingerman.shtml}; \\ \textbf{ii-} `Frequency of the Highest Electromagnetic Waves', see for instance:  \\ \textcolor{blue}{http://hypertextbook.com/facts/2000/ElaineLo.shtml}, \emph{Affil., Phys. Sci. Dept.}, Midwood High School, NY, USA, 2000-2003.

\bibitem{29-AlphaLab} Alpha Lab Inc., TriField® meter: `measures electric field, magnetic field and radio/microwave', see for instance:\htmladdnormallink{http://www.trifield.com}{http://www.trifield.com/}, Salt Lake City UT, USA, no. 10060, 2005.

\bibitem{30-Stallings} W. Stallings, \emph{Data and computer communications}, Pearson Education, Inc., pp. 247-248, 2007.

\bibitem{31-Senior} J. M. Senior, \emph{Optical fibers communications}, (2nd ed.), Prentice Hall International (UK), Ltd., pp. 262-270, 1992.

\bibitem{32-Kaushik et al.} B. K. Kaushik, S. Goel and G. Rauthan, Future VLSI interconnects: optical fiber or carbon nanotube - a review. \emph{J. Micro. Electro. Inter.}, {\textit{Vol.}} \textbf{24} Issue: 2 pp. 53-63, 2007~(\htmladdnormallink {\textcolor{MyBrown}{doi:10.1108/13565360710745601}}{http://www.emeraldinsight.com/10.1108/13565360710745601})

\bibitem{33-Ren} H. Ren and S. T. Wu, Variable-focus liquid lens College of Optics and Photonics, University of Central Florida, \emph{J. Opt. Soc. Amer.}, {\textit{Vol.}} \textbf{15} no. 10/OPTICS EXPRESS, pp. 5931-5936, 2007.

\bibitem{34-Philips} Philips Research Team, `FluidFocus Variable Focus Lens System', see for instance: \textcolor{blue}{http://www.research.philips.com/newscenter/archive/2004/fluidfocus.html}, \emph{Royal Philips Electronics, Koninklijke Philips Electronics}, 2004.

\bibitem{35-Einstein} A. Einstein, `Die Grundlage der allgemeinen Relativitaetstheorie', \emph{Annalen der Physik}, {\textit{Vol.}} \textbf{49}, 769-822, 1916; translated without p.769 as `The Foundation of the General Theory of Relativity', pp. 111-164 in H. A.Lorentz \emph{et al.}, \emph{The Principle of Relativity}, Dover, 1952.

\bibitem{36-Sniffen} D. Schaner and M. Sniffen, `Adult Brain Dissection Protocol', see for instance: \htmladdnormallink{http://anatomy.ucsf.edu/heberlein/Protocols/Adult\_CNS\_dissection.pdf}{http://anatomy.ucsf.edu/heberlein/Protocols/Adult_CNS_dissection.pdf}, \emph{Dept. Anat.}, University of California, San Francisco, USA, 2005.

\bibitem{37-Katayama} M. Katayama \emph{et al.}, `Ultra-Low-Threshold Field Electron Emission from Pillar Array of Aligned Carbon Nanotube Bundles', \textit{Jap. J. Appl. Phys.}, {\textit{Vol.}} \textbf{43} no. 6B, pp. L774-L776, 2004~(\htmladdnormallink {\textcolor{MyBrown}{doi:10.1143/JJAP.43.L774}}{http://jjap.ipap.jp/link?JJAP/43/L774/})

\bibitem{38-Beaty} B. Beaty, `Speed of electricity', see for instance: \htmladdnormallink{http://amasci.com/miscon/speed.html}{http://amasci.com/miscon/speed.html}, \emph{Dept. Chem.}, University of Washington, USA, 1996; \\ see also, \htmladdnormallink{http://www4.ncsu.edu/$~\sim$rwchabay/mi/circuit.pdf}{http://www4.ncsu.edu/~rwchabay/mi/circuit.pdf}, of B. A. Sherwood and R. W. Chabay, `A unified treatment of electrostatics and circuits', \emph{Dept. of Phys.}, Carnegie Mellon University, 1999.

\bibitem{39-AlphaLab} Botest Systems GmbH, `Measurement and test systems for OLEDs (organic light emitting devices) and OPVs (organic photovoltaic cells)', on,  \htmladdnormallink{http://www.botest.com/data/en/index.htm}{http://www.botest.com/data/en/index.htm}, Reichenäcker 11, 97877 Wertheim, Germany, accessed on 03 Sep. 2007.

\bibitem{40-Behounek} L. Behounek, `On the difference between traditional and deductive fuzzy logic' on, {http://www.volny.cz/behounek/logic/papers/FLd.pdf}, \emph{Inst. Comp. Sci., Academy of Sciences of the Czech Republic, Preprint:} pp. 14-21, 2007.

\bibitem{41-Bezanilla} F. Bezanilla, `Voltage-Gated Ion Channels', University of Maryland Baltimore County, \textit{IEEE TRANSACTIONS ON NANOBIOSCIENCE, } \textbf{4}, no. 1, pp. 34-48, March 2005.

\bibitem{42-Wolfram} E. W. Weisstein, \textbf{i-} `\htmladdnormallink{Euclidean Metric}{http://mathworld.wolfram.com/EuclideanMetric.html}'; \textbf{ii-} `\htmladdnormallink{Minkowski Metric}{http://mathworld.wolfram.com/MinkowskiMetric.html}', From MathWorld--A Wolfram Web Resource on, \htmladdnormallink{http://mathworld.wolfram.com}{http://mathworld.wolfram.com}, 1999; accessed on 06 Sep. 2007.

\bibitem{43-Mojahedi} M. Mojahedi and J. Fleck (\textit{Journal Staff Writer}), 1st paragraph, Uncovering a Loophole in Einstein's Law, Article from the \textit{Albuquerque Journal}, see for instance: \htmladdnormallink{http://www.chtm.unm.edu}{http://www.chtm.unm.edu/AA_NEWS_STORIES_DATA/NEWS_ITEMS/mojahedi_story.html} \textbf{weblink}, \emph{Pub.} on Sunday, 19 Nov. 2000.

\bibitem{44-Fante} R. L. Fante, Theory of Propagation of Electromagnetic Waves in Space-Time Varying Media, \emph{Subj. Categ.} \emph{Radiofrequency Wave Propagation}, \textit{Phys. Sci. Res. Papers}, \emph{Corp. Author.} AIR FORCE CAMBRIDGE RESEARCH LABS L G HANSCOM FIELD MASS, Accession No. AD0746676, 1972.

\bibitem{45-Whites} K. W. Whites, `Propagating Electromagnetic Waves in Lossless Media', \emph{A MATHCAD ELECTRONIC BOOK}, see for instance: \\ \htmladdnormallink{http://www.adeptscience.co.uk/products/mathsim/mathcad/add-ons/free\_ebooks}{http://www.adeptscience.co.uk/products/mathsim/mathcad/add-ons/free_ebooks/vis_elec_samp.htm} \\ \textbf{weblink}, The McGraw-Hill Companies, 1998.

\bibitem{46-Metascan} \textbf{Terminology}: `\emph{Dynamic Contrast-Enhanced MRI Analysis}', for further example: Metascan image analysis centre ltd., \emph{Solution via DCE MRI image analysis}, `Medical metascan imaging', see for instance, \htmladdnormallink{http://www.metascan-centre.com}{http://www.metascan-centre.com/Services/solution.html} \textbf{weblink}, Cyprus, accessed on 08, Sep. 2007.

\bibitem{47-Krane} K. S. Krane, \emph{Modern Physics}, 2nd ed. p. 491-494, (1996) ISBN 0471828726.

\bibitem{48-Rosen} J. Rosen and D. Abookasis, `Seeing through biological tissues using the fly eye principle', \emph{Opt. Soc. Amer.}, \textit{Vol.} \textbf{11}, no. 26 /OPTICS EXPRESS, pp. 3605-3611, 2003.

\bibitem{49-Borchardt} J. K. Borchardt, `Developments in organic displays', \emph{Elsevier Ltd.}, \emph{Vol.} \textbf{7}, Issue 9, pp. 42-46, September 2004~(\htmladdnormallink {\textcolor{MyBrown}{doi:10.1016/S1369-7021(04)00401-8}}{http://dx.doi.org/10.1016/S1369-7021(04)00401-8})

\bibitem{50-Vimal et al.} R. L. P. Vimal, J. Pokorny, V. C. Smith, and S. K. Shevell, `Foveal cone thresholds', \emph{Vision Res.}, 29(1), 61-78, 1989.

\bibitem{51-Asma et al.} E. Asma, T. E. Nichols, Qi Jinyi and R. M. Leahy, `4D PET image reconstruction from list mode data', \emph{Nuc. Sci. Symp. Conf. Rec., IEEE}, \emph{Vol.} \textbf{2}, Issue 2000, pp. 57-65, 2000.

\bibitem{52-Kautz} J. Kautz and H. Seidel, `Hardware accelerated displacement mapping for image based rendering', \emph{Can. Info. Proc. Soc.}, pp. 61-70, (2001) ISBN$\sim$ISSN: 0713-5424, 0968880800.

\bibitem{53-Howard} W. E. Howard and O. F. Prache, `Microdisplays based upon OLEDs', \emph{IBM J. Res. \& Dev.}, \emph{Vol.} \textbf{45}, no. 1, pp. 115-127, 2001~(\htmladdnormallink {\textcolor{MyBrown}{doi:10.1147/rd.451.0115}}{http://dx.doi.org/10.1147/rd.451.0115})

\bibitem{54-Tan et al.} L. W. Tan, X. T. Hao, K. S. Ong, Y. Q. Li and F. R. Zhu, `An efficient top-emitting electroluminescent device on metal-laminated plastic substrate', \emph{Inst. Mat. Res. \& Eng.}, Singapore, paper no. DD11.12, 2004; see also, `Top-emitting Organic Light-Emitting Device (TOLED)-with video' on \htmladdnormallink{http://www.a-star.edu.sg}{http://www.a-star.edu.sg/astar/sciengr/action/sciengr_project_details.do?id=0f531e799chR}, \emph{Sci. \& Eng.}, 2007; `Flexible top-emitting electroluminescent devices on polyethylene terephthalate substrates', \emph{Amer. Inst. Phys.}, \emph{Appl. Phys. Lett.} 86, 153508 (2005) (3 pages),~(\htmladdnormallink {\textcolor{MyBrown}{doi:10.1063/1.1900940}}{http://dx.doi.org/10.1063/1.1900940})

\bibitem{55-Prosper} Prof. H. B. Prosper, `Introduction to General Relativity' on \\ \htmladdnormallink{http://www.physics.fsu.edu/Courses/Spring98/AST3033/Relativity}{http://www.physics.fsu.edu/Courses/Spring98/AST3033/Relativity/GeneralRelativity.htm}, \emph{Astr. \& Phys., General Relativity Course Material}, AST 3033, Florida State University, USA, 1998.


\end{thebibliography}
\end{document}